\documentclass[aps,prl,10pt,twocolumn,superscriptaddress,balancelastpage,showpacs,reprint,english]{revtex4-2}
\usepackage[colorlinks,bookmarks=false,citecolor=blue,linkcolor=blue,urlcolor=black]{hyperref}
\usepackage[T1]{fontenc}
\usepackage[latin9]{inputenc}
\usepackage{amsmath}
\usepackage{amssymb}
\usepackage{graphicx}
\usepackage{color}
\usepackage{bm}

\makeatletter

\newcommand{\qq}{\begin{eqnarray}}
\newcommand{\qqq}{\end{eqnarray}}

\newcommand{\p}{\partial}

\newcommand{\bfx}{\mathbf{x}}

\newcommand{\bfr}{\mathbf{r}}
\newcommand{\bfp}{\mathbf{p}}

\newcommand{\bfm}{{\bf m}}
\newcommand{\bfu}{{\bf u}}
\newcommand{\bfv}{{\bf v}}
\newcommand{\bfJ}{{\bf J}}
\newcommand{\bfq}{{\bf q}}

\pdfpageheight\paperheight
\pdfpagewidth\paperwidth

\makeatother

\usepackage{babel}

\begin{document}

\title{Active phase separation: new phenomenology from non-equilibrium physics}
\author{M. E. Cates}
\affiliation{DAMTP, Centre for Mathematical Sciences, University of Cambridge, Wilberforce Road, Cambridge CB3 0WA, UK}
\author{C. Nardini}
\affiliation{Service de Physique de l'Etat Condens\'e, CEA, CNRS Universit\'e Paris-Saclay, CEA-Saclay, 91191 Gif-sur-Yvette, France}
\affiliation{Sorbonne Universit\'e, CNRS, Laboratoire de Physique Th\'eorique de la Mati\`ere Condens\'ee, 75005 Paris, France}

\date{\today}
\begin{abstract}
In active systems, whose constituents have non-equilibrium dynamics at local level, fluid-fluid phase separation is widely observed. Examples include the formation of membraneless organelles within cells; the clustering of self-propelled colloidal particles in the absence of attractive forces, and some types of ecological segregation. A schematic understanding of such active phase separation was initially borrowed from what is known for the equilibrium case, in which detailed balance holds at microscopic level. However it has recently become clear that in active systems the absence of detailed balance, although it leave phase separation qualitatively unchanged in some regimes
(for example domain growth driven by interfacial tension via Ostwald ripening),
can in other regimes
radically alter its phenomenology at mechanistic level. For example, microphase separation can be caused by reverse Ostwald ripening, a process that is hard to imagine from an equilibrium perspective. This and other new phenomena arise because, instead of having a single, positive interfacial tension like their equilibrium counterparts, the fluid-fluid interfaces created by active phase separation can have several distinct interfacial tensions governing different properties, some of which can be negative. These phenomena can be broadly understood by studying continuum field theories for a single conserved scalar order parameter (the fluid density), supplemented with a velocity field in cases where momentum conservation is also present. More complex regimes arise in systems described by multiple scalar order parameters (especially with nonreciprocal interactions between these); or when an order parameter undergoes both conserved and non-conserved dynamics (such that the combination breaks detailed balance); or in systems that support orientational long-range order in one or more of the coexisting phases.  In this Review, we survey recent progress in understanding the specific role of activity in phase separation, drawing attention to many open questions. We focus primarily on continuum theories, especially those with a single scalar order parameter, reviewing both analytical and numerical work. We compare their predictions with particle-based models, which have mostly been studied numerically although a few have been explicitly coarse-grained to continuum level. We also compare, where possible, with experimental results. In the latter case, qualitative comparisons are broadly encouraging whereas quantitative ones are hindered by the dynamical complexity of most experimental systems relative that of simplified (particle-level or continuum) models of active matter.
\end{abstract}

\maketitle

\tableofcontents
\section{Introduction}\label{sec:intro}
Phase separation is commonly seen in systems far from equilibrium. Examples arise in suspensions of self-propelled particles~\cite{Cates:15}; biological tissues~\cite{steinberg1963reconstruction,foty2005differential}; biomolecular condensates in cells~\cite{hyman2014liquid,banani2017,Weber2019review,berry2018physical}; chromatin organisation in the nucleus~\cite{erdel2018formation,cook2018transcription,mirny2019two}; vibrated granular materials~\cite{oyarte2013phase,clerc2008liquid,prevost2004nonequilibrium}; social dynamics~\cite{schelling1971dynamic,schelling1969models,sen2014sociophysics}, 
and ecology~\cite{rietkerk2008regular}. A crude understanding of such phase separations is often possible by modelling the interaction between particles as a conservative force field, and supposing an effective thermal equilibrium to be reached: in this case, attractions cause phase separation via a mechanism first understood by van der Waals~\cite{chaikin2000principles}. But on closer inspection, phenomena are often seen that are impossible in thermal equilibrium -- such as continuous fluxes of material or spatially moving patterns in the statistical steady state. In other cases the observations would require implausible choices of equilibrium model -- for example, by invoking long-ranged interactions to explain microphase-separated patterns, defined as those in which phase-separated domains grow to a finite length scale and then stop. 

 \begin{figure*}
\begin{centering}
\includegraphics[width=2\columnwidth]{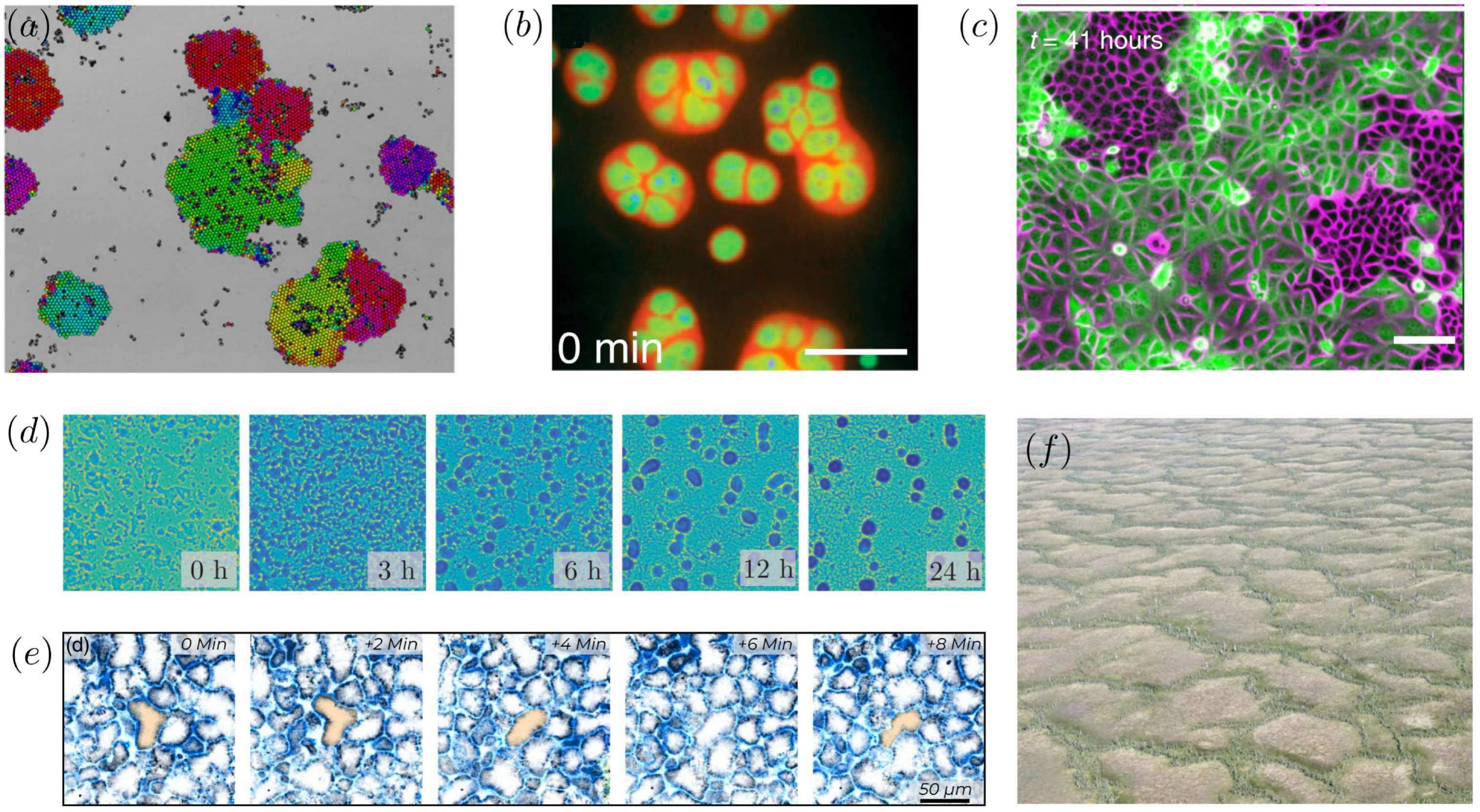}
\par\end{centering}
\caption{Experimental observation of phase separation in active systems. $(a)$ Cluster phases in Janus colloids activated by electrophoresis. The clusters form a microphase separated state in which dense clusters display strong hexatic order (the color-code denotes its orientation) and do not coarsen to the system-size scale. Figure adapted with permission from~\cite{van2019interrupted}. 
$(b)$ Biomolecular condensates (nucleoli) in {\em X. laevis} nuclei (in vivo) display an hierarchical structure; the granular component is visualised in red, dense fibrillar component in green and fibrillar centre in blue; scale bar is $20\mu m$. 
Figure adapted with permission from~\cite{feric2016coexisting}.
$(c)$ Cell-sorting or microphase separation in a monolayer of two populations of cells that are effectively contractile (green) or extensile (magenta) and exert only weak differential adhesion.  Figure adapted with permission from~\cite{balasubramaniam2021investigating}. 
$(d)$ Bacteria ({\em Myxococcus xanthus}) forming fruiting bodies when deprived of nutrients, attributable to phase separation induced by a modification in motility. Figure adapted with permission from~\cite{liu2019self}.  
$(e)$ Mixtures of filaments and molecular motors forming a three-dimensional active foam state. Highlighted is a foam cell that undergoes topological rearrangements. Kinesin motors are represented in blue and microtubules in black.
Figure adapted from~\cite{lemma2022active}.  
$(f)$ Irregular spatial patterns in vegetal ecosystems (here, peatlands in Siberia)~\cite{siteur2023phase}, attributed to a phase separation mechanism~\cite{liu2016phase}. Image credit: Maarten B. Eppinga.}
\label{fig:intro}
\end{figure*}

Some of the recent experimental and fieldwork observations of phase separation in active systems are reported in Fig.~\ref{fig:intro}.  In all cases, detailed balance, or equivalently time-reversal symmetry (and more formally also charge-parity-time reversal symmetry) is broken microscopically, because of dissipative energy fluxes at the scale of the individual constituents  or  `active particles'. This is the defining property of active matter~\cite{Marchetti2013RMP,gompper20202020}. It contrasts with types of nonequilibrium system more often studied in the past, where driving is imposed at the boundary while the bulk dynamics respects detailed balance in its absence. (Examples include, a sheared liquid, or a thermally conducting wire coupled to thermostats at different temperatures at its ends.) Comprising a large class of non-equilibrium systems, active matter has developed its own theoretical framework, and this Review focuses on recent progress in the theoretical description of phase separation in active systems. We start by surveying in more detail some of the physical and biological systems in which active phase separation emerges (Sec.~\ref{sec:instances}), and then in Sec.~\ref{sec:scope} delineate the scope of this Review.

\subsection{Instances of active phase separation}\label{sec:instances}
The following survey is not comprehensive; our main focus is on studies that identify phase separation or microphase separation as a mechanism for the observed patterns. Note that many papers report `aggregation' or `coagulation'  which, at least in the physics literature, differ from true (micro-)phase separation. In the latter, the coexisting phases regions are generally in dynamic equilibrium with each other rather than formed by an irreversible process leading to local arrest. However in passive colloidal gels, for instance, it is widely understood that phase separation can itself lead to arrest~\cite{poon2002physics}. Moreover, in areas such as ecology where dynamical fluctuation times may exceed observation times, such distinctions may not be helpful.

Given the complexity of many active systems, a rather general question to ask, when spatial patterns are seen, is whether these are due to some sort of phase separation, or due to other instability mechanisms. One of the best known alternatives is the Turing mechanism~\cite{turing1990chemical,cross1993pattern,cross2009pattern}: the instability of a spatially homogeneous state formed by (at least) two species, an activator and a inhibitor, with very different diffusivities. This is quite different mechanistically from phase separation~\cite{siteur2023phase}, and the ensuing phenomenology at large scale is also different. For example, a feature of phase separation is that the identity of the two coexisting phases does not change as the global composition is swept from one to the other along a straight `tie-line' in composition space. There is no analogue to this among Turing patterns. Also, at dynamical level, Turing patterns are stabilised by local nonlinear effects, while phase separating systems reach the steady state by large-scale fluxes of the order parameter. Thus, although ambiguity might arise in practice, there is a clear distinction in principle between active (micro-)phase separation and Turing instabilities. An interesting bridge between the two arises in the study of reaction-diffusion models with global but not local particle conservation~\cite{brauns2020phase,freyfoam2024}. This can closely resemble true phase separation but lies beyond the scope of this Review.

One class of active systems routinely displaying phase separation comprises self-propelled colloidal particles~\cite{Cates:15}. Several experiments on self-propelled colloids that are homogeneous in the absence of activity have reported clustering upon making them self-propelled~\cite{Palacci:12,palacci2013living,Speck:13,van2019interrupted,thutupalli2017boundaries,zhang2021active,vutukuri2020light,ginot2018aggregation}. Intriguingly, while some of these experiments found complete phase separation into two system-spanning domains~\cite{Speck:13,thutupalli2017boundaries,zhang2021active}, many reported microphase-separated clusters on a characteristic scale that is independent of system size~\cite{Palacci:12,palacci2013living,Speck:13,van2019interrupted,vutukuri2020light,ginot2018aggregation}. Several of these colloidal studies report crystalline or hexatic order within the dense phase ({\em e.g.},~\cite{Palacci:12,van2019interrupted}, Fig.~\ref{fig:intro}(a)). Although this ordering is important for colloids, it is less relevant for other active systems, and we discuss it only in passing below.

Phase separation mechanisms have also been proposed for clustering and pattern formation in bacteria~\cite{liu2019self,schwarz2012phase,catesPNAS2010,liu2011sequential}, in
mixtures of microtubules and molecular motors~\cite{lemma2022active}, and in
colonies of worms or other animals~\cite{demir2020dynamics,schellinck2011review,anderson2022social}. It was further suggested that phase separation is a crucial mechanism leading to biofilm formation~\cite{grobas2021swarming,worlitzer2022biophysical}. In these cases, as in many colloids, the phase separation is between dense and dilute amorphous (fluid) phases of the same species of particle. However phase separation by demixing of species in bacterial colonies was also recently studied~\cite{curatolo2020cooperative}.

Another class of active systems in which demixing arises are biological tissues, where it is often called compartmentalization or cell sorting.
This is crucial for several biological phenomena, such as anterior-posterior stripe  formation, tissue patterning and body-axis formation in development~\cite{akam1989making,krens2011cell,christian1991xwnt,heller2015tissue}, and is also implicated in tumor invasion~\cite{lowengrub2009nonlinear,wise2008three,byrne2003modelling,kang2021novel,ilina2020cell,szabo2012invasion,basan2009homeostatic}. Cell sorting was classically explained in terms of the hypothesis that the adhesion between two cells depends on their type~\cite{steinberg1963reconstruction,foty2005differential}, resembling differential attractions between species in passive fluids~\cite{steinberg1963reconstruction}. It was however soon proposed that more complex physics is at play~\cite{brodland2002differential,krens2011cell,harris1976cell}. Building on early models for cell-sorting in confluent tissues~\cite{glazier1993simulation,nagai2001dynamic,szabo2013cellular}, theorists are now incorporating the effect of activity to understand whether this can drive phase separation even in the absence of differential adhesion, and whether it affects the properties of the phase separated state~\cite{barton2017active,krajnc2020solid,ai2021large,sahu2020small,sussman2018soft,mccarthy2024demixing,balasubramaniam2021investigating,matoz2017cell}. A distinct, but related, line of research has developed models for tumor growth by analogy with binary fluid phase separation~\cite{lowengrub2009nonlinear,wise2008three,byrne2003modelling}; here the search for activity-specific effects is in its infancy~\cite{delarue2014compressive,montel2012isotropic,waclaw2015spatial,sinha2020self,sinha2020spatially,kang2021novel}.

Turning to the subcellular domain, a major paradigm shift in recent years has promoted liquid-liquid phase separation into the dominant explanation for membraneless organelles, also known as biomolecular condensates~\cite{berry2018physical,Weber2019review,Jacobs2017biophysical,banani2017}. These aggregates, forming in the interior of cells (including within the cell nucleus), are composed of specific proteins and RNA molecules that are much more abundant in the condensate than in the surrounding cytoplasm. Many appear to be self-limiting in size (akin to microphase separation). 
Some of them, such as the nucleolus~\cite{pederson2011nucleolus} of eukaryotic cells, or Cajal bodies~\cite{gall2000cajal,morris2008cajal,machyna2013cajal}, were discovered more than a century ago. It is nowadays fully established~\cite{hyman2014liquid, berry2018physical, Jacobs2017biophysical, Weber2019review, hyman2011, banani2017, brangwynne2015, nott2015phase}, that these condensates can have liquid-like mechanical properties~\cite{wolf1983segregation,strome1983generation}. (However in other cases solid or elastic properties were observed ~\cite{michaels2023amyloid,law2023bending,garaizar2022aging,folkmann2021regulation,farr2021nucleosome}.) Intense recent research has exposed complex substructure in some such condensates~\cite{lafontaine2021nucleolus,erkamp2023spatially}. 

Although the biological function of membraneless organelles is still not fully settled~\cite{pederson2011nucleolus,morris2008cajal,machyna2013cajal}, it seems likely that physics as well as biology plays a role in their function in health and disease~\cite{shin2017liquid,cohen2013proliferation,jain2017rna,michaels2023amyloid,derenzini2009nucleolus,buchwalter2017nucleolar}. 
While the subcellular environment is undoubtedly an active medium, the {\em specific role of activity} within the phase separation paradigm for organelles is largely unexplored. Although many authors have stressed similarities with passive phase separation~\cite{Weber2019review,tanaka2022viscoelastic,rosowski2020elastic,style2018liquid,pappu2023phase,li2024predicting}, the main driving force could be either of thermodynamic or of active origin depending on the specific condensate~\cite{donau2020active,boulon2010nucleolus,ganai2014chromosome}. For instance, protein-chromatin interactions could provide a generic mechanism for the formation of condensates in the cell nucleus~\cite{brackley2013nonspecific,erdel2018formation}, with specific active interactions ensuring that condensates remain of finite size and highly dynamic in character~\cite{brackley2017ephemeral,brackley2020bridging}.  Notably also, subcellular molecules have a turnover time determined by active synthesis and degradation rates. The combination of such chemical kinetics with phase separation generically leads to far-from-equilibrium structures ~\cite{zwicker2017growth,catesPNAS2010,li2021hierarchical}
that are likely implicated in many subcellular phase separations. Conversely, the resulting condensates likely have a role in organising how reactions take place within cells~\cite{Weber2019review,hyman2014liquid}. 
All in all, a clear understanding of the specific roles of activity in the formation and function biomolecular condensates remains an important future goal. 

A somewhat distinct type of subcellular phase separation occurs in the organization of the genome, which supports a hierarchy of spatial structures, ranging downwards from chromosome territories, through compartments and topologically-associating domains, to contact domains~\cite{cook2018transcription}. Here too, the precise role of activity remains under investigation. For example, segregation between high-activity and low-activity regions of the genome was proposed as a mechanism for spatially organising chromosome territories~\cite{ganai2014chromosome,agrawal2017chromatin,adame2023regulation,smrek2017small,goychuk2023polymer,mahajan2022euchromatin,adame2023regulation,bajpai2020mesoscale}. 

Above we have considered phase separation in colloids, microbes, cells, and tissues; the corresponding length scales span microns to centimetres. Phase separation is also implicated in structuring at much larger length scales, for instance in ecology~\cite{okubo2001diffusion}. The spatial self-organization of vegetation into large-scale patterns is observed in many ecosystems, from arid bush lands~\cite{klausmeier1999regular} to marine corals~\cite{jokiel2004hawai}, peatlands~\cite{eppinga2009nutrients} and during invasion by alien species~\cite{invasive}.
In some cases the formation of regular vegetation patterns has been linked to Turing instabilities~\cite{klausmeier1999regular,rietkerk2008regular}, via the effect of one species to the growth rate of others. In other cases, however, such as grazing systems, biogeomorphological systems and nutrient-poor ecologies, the emergent patterns are irregular and a Turing-like mechanism seems unlikely~\cite{klausmeier1999regular,siteur2023phase}. It has been proposed that some of these cases involve a phase separation mechanism~\cite{rietkerk2004putative,liu2016phase,liu2013phase,siteur2023phase,couteron2023conservative}. Some closely related mechanistic concepts such as nucleation and coarsening were also introduced ~\cite{michaels2020nucleation,liu2016phase,gandhi1999nucleation}. Understanding the patterning mechanisms of plant ecosystems is not only a scientific puzzle, but has also implications for their resilience to external perturbations including climate change~\cite{liu2016phase,rietkerk2004self,siteur2023phase,couteron2023conservative,rietkerk2021evasion,vallespir2019structured}.The specific role of activity in ecological phase separation remains to be fully explored.

Similar remarks apply to studies of human demographics, addressing segregation by ethnicity or other characteristics. It was first realized by Schelling~\cite{schelling1969models} that a mild preference of agents to have neighbours of their own type can lead to strongly segregated regions~\cite{sakoda1971checkerboard,schelling1971dynamic,schelling1969models,sen2014sociophysics,hegselmann2017thomas}, which can be viewed as phase separated domains~\cite{vinkovic2006physical,grauwin2009competition}. While such models were initially equilibrium-like, recent versions include more realistic utility functions that break detailed balance~\cite{grauwin2009competition,seara2023sociohydrodynamics,zakine2023socioeconomic}, leading to active phase separation in the sense of this article. (Indeed an explicit mapping to `Active Model B+', introduced in Sec.~\ref{sec:AMB+} below, was proposed in~\cite{zakine2023socioeconomic,seara2023sociohydrodynamics}.) Schelling-type models have been used to help shape the thinking of policymakers~\cite{hegselmann2017thomas,clark2008understanding}, and further lessons might in future be drawn from their active counterparts, especially if these give improved quantitative accounts of the empirical data. 

Finally, and also at a supra-organism scale, phase separation is implicated in the collective dynamics of flocking animals such as birds and wildebeest. Flocking was arguably the first area of active matter to gain the attention of theoretical physicists ~\cite{vicsek1995novel,toner1995long,toner2005hydrodynamics}, but only relatively recently was it understood that flocks are generically prone to spatiotemporal patterning via a type of microphase separation~\cite{gregoire2004onset,mishra2010fluctuations,caussin2014emergent,Solon:2015:PRL,geyer2019freezing,chate2019dry,martin2021fluctuation,anderson2024traveling}.

\subsection{Scope of this Review}\label{sec:scope}
Although the above survey of physical candidates for active phase separation is far from complete, one might already wonder if, echoing Tolstoy \footnote{Happy families are all alike; every unhappy family is unhappy in its own way. (Tolstoy, Anna Karenina).} ``passive phase separations are all alike; every active phase separation is active in its own way''.  If this was the case, there could be no generic theory for active phase separation and each specific instance would need to be understood separately. In this Review we will argue instead that general principles do exist, so that there is much to learn by addressing active phase separation by studying, in depth, a small number of minimalist theories. 

Just as in passive systems~\cite{hohenberg1977theory,Bray} such theories are best formulated phenomenologically at a continuum level, where much microscopic detail falls away. Those we discuss will mostly take the form of `active field theories' in which minimal detailed-balance violations are added to pre-existing theories of passive matter.
For the main problems we address, which primarily involve phase separation of a scalar order parameter, this approach was laid out in two earlier reviews~\cite{CatesJFM:2018,cates2022active} (since when the field has evolved considerably). However, the underlying `top-down' philosophy dates back to the birth of active matter physics as an identifiable subdiscipline in the 1990s; for reviews see \cite{Marchetti2013RMP,ramaswamy2010mechanics,toner2005hydrodynamics}.

The widespread occurrence of phase separation in systems lacking detailed balance then motivates the main questions that this Review addresses: $i)$ what \emph{generic, qualitative} phenomenology, independent of system details, should we expect? $ii)$ What \emph{universal, quantitative} properties, likewise independent of system details, should we expect? $iii)$ What are the \emph{minimal microscopic ingredients} needed to obtain these non-equilibrium properties, and in which physical systems should they arise? 

This top-down approach has an inbuilt limitation: it cannot relate the generic or universal phenomena to the underlying microscopic interactions. Doing that requires a bottom-up strategy which is challenging in active systems, whose complexity extends down to the microscopic scale of a single constituent `particle'. This is obvious when that particle is a living object such as an animal, micro-organism, or cell. However it is also true for synthetic cases, such as colloids that self-propel by phoretic mechanisms. Here, mechanistic theories for a single colloid initially failed to predict not only the magnitude but even the sign of the self-propulsion~\cite{golestanian2005propulsion,brown2014ionic,boniface2019self} as well as the interactions among the colloids~\cite{carrasco2024quantitative}; and where predictive first-principle descriptions have been obtained, these are highly system-specific~\cite{bricard2013emergence}. 

Minimalist particle models with simple propulsion rules, such as ABPs (Active Brownian Particles), RTPs (Run-and-Tumble Particles) and AOUPs (Active Ornstein-Uhlenbeck Particles) have accordingly played a key role in developing theories of active phase separation~\cite{Cates:15}. Once coarse-grained and rewritten at Landau-Ginzburg level, these models can, with suitable interactions, generate the leading-order active terms that are invoked \emph{a priori} in the top-down approach~\cite{tjhung2018cluster}. Also these models are, via simulation, a crucial source of empirical data against which to test continuum theories, and have been heavily relied upon in this capacity as we review in Sec.~\ref {sec:particle-models} below. Nonetheless, the same continuum theories should in principle describe much wider classes of system than represented by minimal particle models. (An example is their recent use to describe demographic segregation in human populations~\cite{zakine2023socioeconomic,seara2023sociohydrodynamics}.)

The developments reviewed here parallel those of the 1970s--1990s for equilibrium systems, where the phenomenology of phase separation was profoundly illuminated by top-down field-theories, constrained as usual by symmetry arguments and conservation laws. Such theories were named Models A-J in the influential Hohenberg-Halperin classification~\cite{hohenberg1977theory}, differing in the number and types of order parameters present; whether these have a conserved or non-conserved dynamics; and whether momentum (and in some cases energy) is or is not conserved. This top-down approach not only elucidated universal properties near the critical point~\cite{Onuki}, as was its original intention, but subsequently also uncovered many other generic properties of phase separating systems, including  nucleation rates, coarsening exponents, and droplet size distributions for non-critical phase separation~\cite{Bray,CatesJFM:2018}. These aspects are governed by their own universalities, linked to a `zero-temperature' fixed point of the dynamical field theory~\cite{Bray}. Although less universal, top-down continuum theories have also proved essential in understanding microphase separation, which arises in equilibrium either from long-ranged interactions~\cite{Muratov2002}, or from free energies that in effect depend on surface-to-volume ratios of phase-separated domains due to the presence of fixed amounts of amphipile~\cite{gompper1994phase}.

A much richer physics is present in active phase separating systems than in their equilibrium counterparts. As we shall see, extending a small subset of Models A-J to active systems (primarily, Models B and H) already reveals various new generic and/or universal phenomena arising specifically from activity. Indeed, activity can induce a reversal of the Ostwald process, cause capillary waves to become unstable, and induce the spontaneous elongation and splitting of droplets. These effects lead to qualitatively altered stationary states, generally involving continuous entropy production, and sometimes matter fluxes, that have no equilibrium counterpart. Such states include bubbly phase separation, microphase separated states, active foam states, hierarchical microphase separation, and travelling waves patterns, all of which we will discuss later in this Review. 

Note also that activity can cause phase separation in systems where it would not otherwise occur at all, for example liquid-liquid phase separation among repulsive self-propelled particles, as reviewed at length in~\cite{Cates:15}. However, from a top-down perspective, the resulting phase separation is not distinctively active until terms breaking detailed balance (which are not essential for phase separation itself) are added to the coarse-grained description. Until that point, activity has merely changed the parameters of the equilibrium model from values that do not show phase separation, to values that do.

This Review aims to provide a relatively self-contained overview of the theoretical understanding of active phase separation that has developed during the last decade, highlighting where nonequilibrium physics leads to distinctively altered phenomenology. In some areas, especially those covered by the simplest scalar ($\phi^4$) field theory, we present what is known in some detail. We think this most minimal of active field theories has special pedagogical value -- just as the scalar $\phi^4$ theory does in equilibrium statistical physics. In some of the other areas covered, we instead give a birds-eye view which is primarily intended as a guide to the literature.

We thus start in Sec.~\ref{sec:AMB+} with the simplest possible case of an active system described by a single, conserved, density field $\phi$. Broken detailed balance at the microscale leads to introduce a generalisation of Model B for phase separation that includes leading-order terms to break detailed balance locally. The resulting field theory, Active Model B+, describes so-called `dry' systems, in which momentum is not conserved. (In many biological or sociological settings, and also in self-propelled colloids that often form quasi-two-dimensional layers in proximity to a basal wall, this is the appropriate picture.) We then discuss the effects of adding momentum conservation, as is relevant for active particles suspended in a bulk fluid, whose role is minimally described by Active Model H.

Sec.~\ref{sec:particle-models} connects the phenomenology predicted at continuum level in Sec.~\ref{sec:AMB+} with the calculated and/or simulated behaviour of microscopic models of self-propelled particles. This creates a connection between top-down and bottom-up physics. Where appropriate we also refer to experimental systems, although here such connections remain far from complete, for reasons already described above. 
We show that at least some of the predictions of Sec.~\ref{sec:AMB+} are borne out by particle-baseds models, and discuss future directions for both dry and wet systems.

Sec.~\ref{sec:two-order-parameters} extends the continuum description of active phase separation to systems that either have multiple conserved order parameters, or break mass conservation. The first case brings novel physics because, unlike in equilibrium systems, effective interactions among different types of active particles are generically non-reciprocal, which can result in waves and travelling patterns. We will again see how a minimal generalisation of passive Model B (the non-reciprocal Cahn-Hilliard model) predicts such novel phenomenology, and discuss in which physical systems it can be expected. We will then discuss cases in which particle number is not conserved, a case relevant to the formation of both microbial colonies and membraneless organelles, which can be addressed by a different minimal model (Model AB).
We conclude in Sec.~\ref{sec:Conclusions} with a brief outlook of future research directions involving active phase separation.

\section{Active phase separation of a conserved scalar order parameter}\label{sec:AMB+}

As already indicated, activity can have a profound impact on the phenomenology of phase separating systems. This happens even in the simplest case of `dry' active systems (without momentum conservation) governed by a single conserved order parameter $\phi({\bf r},t)$. This order parameter can be the coarse-grained particle density, or a composition variable in a binary fluid mixture. In either case, it is a conserved quantity if particles are not created or destroyed; this we assume throughout Section~\ref{sec:AMB+}. Bulk orientational ordering of the particles is excluded as this would require vector (polar) or tensor (nematic) order parameters. 

In fact, active particles generally do have an orientation, such as a swimming direction, but this can be ignored at large scales (in the absence of Goldstone modes arising from long range orientational order), because the corresponding order parameters are not conserved and therefore relax quickly compared to the diffusive relaxation of the scalar field~\cite{Cates:15}. On the other hand, although not treated explicitly in our theory, the orientational order need not vanish locally. For instance in the phase separation of self-propelled particles, the particles around the edges of a dense droplet, where $\nabla\phi$ is large, are oriented on average up the density gradient, because any that swim in the other direction will leave the droplet immediately. This and other `bottom-up' connections to particle models are discussed in Sec.~\ref{sec:particle-models} below. 

A crucial concept for phase separating systems is the interfacial tension. In equilibrium fluid-fluid systems, a single interfacial tension describes multiple phenomena, such as the interfacial free energy cost~\cite{Onuki}; the effects of interfacial curvature on domain growth via the Ostwald process~\cite{Bray}; and the amplitudes and relaxation rates of thermal capillary waves~\cite{rowlinson2013molecular}. Activity lifts this degeneracy in a remarkable way: while the Ostwald process and capillary waves each remain governed by an interfacial tension, these need not be equal. Moreover, each can turn negative at sufficiently strong activity; surprisingly, this can happen without destroying phase separation entirely. It should be further observed that these effects are not related to the ambiguity in defining the interfacial tensions that is present even in equilibrium systems due to how the boundary between the two phases is identified~\cite{rowlinson2013molecular}.

\begin{figure*}
\begin{centering}
\includegraphics[width=2\columnwidth]{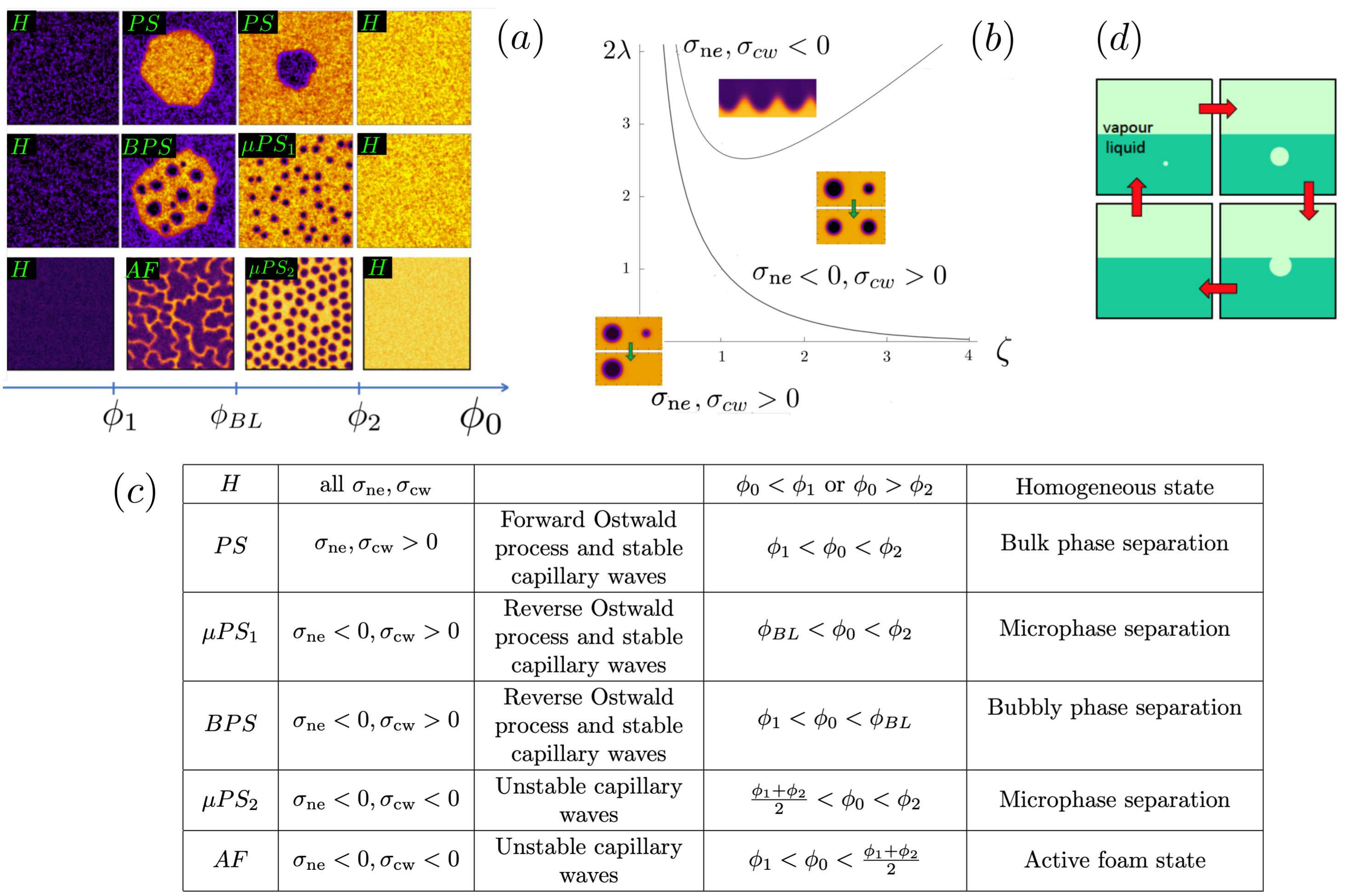}
\par\end{centering}
\caption{Figure adapted from~\cite{tjhung2018cluster,fausti2021capillary}.
Panel (a): Phases  of Active Model B+ found by numerical simulations on varying the global density $\phi_0$. Phase separations arise when this lies between the binodals, $\phi_1,\phi_2$. The colour scale is dark for positive $\phi$ (`liquid') and light for negative $\phi$ (`vapour'), or {\em vice versa} depending on the sign of the activity parameters (see text for discussion).
Each row exemplifies a different regime governed by the signs of the interfacial tensions $\sigma_{\rm ne}$ and $\sigma_{\rm cw}$ defined in the text.  Top row, both tensions positive and the system undergoes bulk phase separation ($PS$). Middle row, $\sigma_{\rm ne}$ negative, $\sigma_{\rm cw}$ positive; depending on the global density, the system undergoes either microphase separation ($\mu PS_1$) or bubbly phase separation ($BPS$), corresponding to the coexistence between a vapour-bubble-in-liquid microphase separation and bulk vapour (or with liquid and vapour exchanged for opposite values of activity). Bottom row, both tensions negative; depending on the global density the system is either microphase separated ($\mu PS_2$) or in an active foam state ($AF$).
Panel (b): 
Regimes of positive and negative $\sigma_{{\rm ne}, \rm cw}$ for varying activity parameters $\lambda,\zeta$ that enter into Active Model B+ and exemplification of the physical mechanism leading to the novel forms of phase separation. Insets show (top) an instability of capillary waves of the liquid-vapour interface; (middle) initial and final states of a two droplet system in the reverse Ostwald regime without noise; (bottom) likewise for the forward Ostwald regime. Panel (c): Summarizes the various phases found in Active Model B+ when $\lambda,\zeta>0$. For $\lambda,\zeta<0$ the phase diagram is obtained exchanging the liquid with the vapour and viceversa due to the symmetry of the model under $(\phi,\lambda,\zeta)\to -(\phi,\lambda,\zeta)$. The boundaries in densities where various phases are observed should be considered indicative as they depend on the noise level (see Sec. \ref{subsec:statistical} for a detailed discussion and the definition of $\phi_{BL}$, for which it was found $\phi_{BL}\gtrsim (\phi_1+\phi_2)/2$ in all simulations so far performed).
Panel (d): Schematic of a circulating phase space current seen in steady state within the regime of bubbly phase separation~\cite{tjhung2018cluster}. Vapour bubbles are repeatedly nucleated within the liquid and ejected to join the bulk vapour, while the time reversed process is not seen. 
}
\label{fig:phi+-}
\end{figure*}

In the rest of Section~\ref{sec:AMB+} we show how such predictions emerge from Active Model B+, a minimal  model introduced in Sec.~\ref{subsec:AMB+}, whose phase diagram is summarized in Fig.~\ref{fig:phi+-}.
We review the effect of activity on the binodals in Secs.~\ref{subsec:AMB+binodals},\ref{subsec:pseudo-variables} and on the interfacial profile in Sec.~\ref{Sec:interfacial-shape}.  In Sec.~\ref{subsec:AMB+Ostwald} we show that activity can send the Ostwald process into reverse, thereby providing a generic route to microphase separation and bubbly phase separation. Sec.~\ref{subsec:AMB+coarsening} addresses coarsening dynamics, while
Section~\ref{subsec:AMB+capillary} describes how activity modifies the dynamics of capillary waves, giving a second route to microphase separation via an active foam state. After reviewing the properties of these various states in Sec.~\ref{subsec:statistical}, we discuss the impact of activity on nucleation in Sec.~\ref{subsec:AMB+nucleation} and on interfacial roughening in Sec.~\ref{subsec:AMB+roughening}. 
Sec.~\ref{subsec:critical} concerns the universality class of the critical point for active phase separation. Finally, in Sec.~\ref{subsec:AMH} we turn from dry systems to the wet case, where in addition to a conserved scalar density, there is a conserved fluid momentum. This is described by a second minimal model, Active Model H.

\subsection{Active Model B+}\label{subsec:AMB+}
An equilibrium system undergoing diffusive fluid-fluid phase separation can be modelled on continuum scales by `Model B', according to the Hohenberg-Halperin classification~\cite{hohenberg1977theory}. Model B describes the dynamics of a scalar field $\phi(\bfr,t)$ which, for the liquid-vapour transition, is a linear transform of the particle density $\rho(\bfr,t)$, chosen such that $\phi = 0$ at the mean-field critical point density $\rho_c$ of the model \cite{chaikin2000principles}. For a symmetric incompressible binary fluid mixture $\phi$ is a compositional order parameter for which the terminology of vapour ($\phi<0$) and liquid ($\phi>0$) will still be used in this Review.
The conserved dynamics of $\phi$ obeys:
\qq
\p_t\phi &=& -\nabla \cdot \left( \bfJ_{\rm eq}+\sqrt{2D M[\phi]}\mathbf{\Lambda} \right) \,,  \label{eq:MB}\\
\bfJ_{\rm eq}       &=& -M[\phi]\nabla\mu_{\rm eq} \equiv -M[\phi]\nabla\frac{\delta \mathcal{F}}{\delta \phi} \,,  \label{eq:MBJ}
\qqq
where $\mathbf{\Lambda}(\mathbf{r},t)$ is a Gaussian white vector noise with zero mean and (component-wise) unit variance. 
$M[\phi]$, which is positive definite, is the collective mobility and $D$ is a `noise temperature' (which for equilibrium systems is the actual temperature $T$ in units where $k_B = 1$).  Model B has an equilibrium structure, meaning that the deterministic current $\bfJ_{\rm eq}$ is proportional to the negative gradient of a chemical potential $\mu_{\rm eq}$,
which itself derives from a free energy functional, $\mu_{\rm eq} ={\delta \mathcal{F}}/{\delta\mathcal{\phi}}$. This is taken of square-gradient, $\phi^4$ form:
\begin{equation}
\mathcal{F}[\phi] = \int \bigg\{ \underbrace{\frac{a}{2}\phi^2 + \frac{b}{4}\phi^4}_{f(\phi)} + \frac{K}{2}|\nabla\phi|^2 \bigg\} d\mathbf{r} \,, \label{eq:F}
\end{equation}
where $f(\phi)$ is the bulk free energy density and $b,K>0$.  Any linear term in $f(\phi)$ contributes a constant to $\mu_{\rm eq}$ and therefore has no effect, while any cubic term can be absorbed by shifting $\phi$ and $a$ to make these both vanish at the mean-field critical point. Then \eqref{eq:F} is applicable even if the microscopic interactions are asymmetric, as it is generally the case. For $|\phi| < \phi_s =\sqrt{-a/3b}$ the dynamics is linearly unstable towards phase separation; this defines the spinodal regime. For $\phi_s < |\phi| <  \phi_b = \sqrt{-a/b}$ phase separation requires a noise-induced nucleation event. The binodals $\pm\phi_b$ are, for all equilibrium models,  the points of common tangency to the $f(\phi)$ curve, between which the bulk free energy can be reduced by phase separation into coexisting states at these densities, at the (sub-extensive) cost of creating an interface. In equilibrium Model B that cost is, per unit area, $\sigma_{\rm eq} = (-8a^3K/9b^2)^{1/2}$~\cite{Bray}. 

Model B neglects terms in $f$ beyond quartic order which would quantitatively but not qualitatively change the above results for $\phi_{s,b}$. Moreover it is normally assumed that $M[\phi]$ is $\phi$-independent -- gaining significant technical advantage from the fact that multiplicative noise is then avoided. These approximations are valid in principle only when $\phi$ remains small in the two coexisting phases. In practice however, Model B remains qualitatively valid even in the strongly separated regime, reflecting the universality of the zero-temperature fixed point governing phase ordering~\cite{Bray}. (Exceptions to this arise for singular choices of $M[\phi]$, not relevant here~\cite{bray1995lifshitz}.)

With the choice of noise made in \eqref{eq:MB}, Model B obeys the principle of detailed balance. Put differently, the noise competes with the deterministic current (which causes a gradient flow on the free energy landscape $\mathcal{F}$) in such a way that, in steady state, the Boltzmann distribution $\exp[-\mathcal{F}/D]$ is recovered. The statistics of forward and backward trajectories is then identical: in steady state, all movies look statistically the same running backwards.
Detailed balance is broken in active matter, so that this time-reversal symmetry (TRS) should not be built into a continuum model. A minimal route to describe phase separation of an active scalar order parameter is thus to extend Model B to include terms that break TRS at the lowest order possible in $\phi$ and $\nabla$. 

Retaining constant $M$ and non-multiplicative noise, and in absence of external fields coupling to gradients of $\phi$ (that would break rotational symmetry), such terms first arise in $\partial_t\phi$ at order ${\mathcal O}(\nabla^4,\phi^2)$. This is because odd powers of $\nabla$ are eliminated by isotropy, while terms $\nabla^2\phi^n$ can be absorbed into ${\cal F}$, and so cannot break TRS in (\ref{eq:MB},\ref{eq:MBJ}).
The minimal field theory that emerges is called Active Model B+ (AMB+) ~\cite{Wittkowski14,nardini2017entropy,tjhung2018cluster}. It is obtained from Model B by adding all terms to order $\mathcal{O}(\nabla^4\phi^2)$, while retaining  constant $M$ and $D$. AMB+ can be written
\qq
\p_t\phi &=& -\nabla\cdot\left(\mathbf{J}+\sqrt{2D M}\mathbf{\Lambda}\right) \,,  \label{eq:AMB+}\\
\bfJ/M       &=& -\nabla \mu_\lambda + \zeta (\nabla^2\phi)\nabla\phi   \label{eq:AMB+J}\\
\mu_\lambda &=& \frac{\delta \mathcal{F}}{\delta\phi} +\lambda|\nabla\phi|^2\,.\label{eq:AMB+L}
\qqq
There are two separate TRS-breaking terms, in $\lambda$ and $\zeta$. A third term arising at the same order does not break TRS, because it can be absorbed into $\mathcal{F}$ via $K\to K+K'\phi$  in (\ref{eq:F}).
With constant $M$ and $D$, any other term at order $\mathcal{O}(\nabla^4\phi^2)$ can be written as a linear combination of the $\lambda,\zeta$ and $K'$ terms. The special case of $\zeta=0$ was studied first, and is called Acitve Model B (AMB)~\cite{Wittkowski14}. The phenomenology obtained when $\zeta=0$ is qualitatively similar to the one of Model B, although quantitative differences are present. Entirely new phenomenology is instead found when both $\lambda$ and $\zeta$ are present. This is discussed in detail below and summarised in Fig. 2.
For a full discussion of TRS breakdown in these field theories, and the resulting steady-state entropy production (which of course vanishes in the passive limit), see~\cite{nardini2017entropy,cates2022stochastic,markovich2021thermodynamics}. 

Note that the new $\lambda$ and $\zeta$ terms are directly suggested by the coarse-graining of microscopic models of active phase separation~\cite{stenhammar2013continuum,tjhung2018cluster}; see Sec.~\ref{sec:particle-models} below. There are, however, other interesting ways to break TRS that lie beyond our scope here, including choosing colored noise in \eqref{eq:MB} ~\cite{paoluzzi2016critical,maggi2022critical,paoluzzi2023noise,paoluzzi2024noise}. However, for mobilities that are expandable in $\phi$ and its gradients, and/or for finite noise correlation times, one can expect similar results to those found below.

Although for simplicity we choose below a symmetric potential $f(\phi) = f(-\phi)$ as in Model B, all the results of this section can be easily generalised to the case of a generic double-well $f$. With this choice, although the active terms break $\phi\to-\phi$ symmetry, a residual symmetry $(\phi,\lambda,\zeta)\to -(\phi,\lambda,\zeta)$ is retained. (This reduces the parameter space by half, and we exploit it extensively below.) Also, because 
variations in $K$ do not break detailed balance, we focus hereafter on constant $K$, although most of the results below can be extended to any $K(\phi)>0$~\cite{tjhung2018cluster,fausti2021capillary,cates2023classical}. 
Note further that a specific extension of AMB+ that avoids expanding in powers of the density will be discussed in Sec.~\ref{subsec:particles-towards-quantitative-continuum} below.

\subsection{Coexistence conditions}\label{subsec:AMB+binodals}
Observe from \eqref{eq:AMB+} that the condition for linear instability is unaffected by the active nonlinearities $\lambda$ and $\zeta$, because these are quadratic in derivatives of the field $\phi$. The spinodals at $|\phi| = \phi_{s}=\sqrt{-a/3b}$ from passive Model B are therefore unaffected by these terms. (The subsequent nonlinear evolution in the spinodal regime is affected, as discussed further in Sec.~\ref{subsec:AMB+coarsening} below.)

The binodals are, in contrast, altered by activity; denoting them by $\phi_{1,2}$, we no longer have $\phi_{1,2} = \mp\phi_b$.
The new binodals can be found by generalising the common tangent construction on $f(\phi)$ that holds for equilibrium models, but this requires non-straightforward steps as we shall explain next. The shift in binodals is at first sight surprising because the standard construction does not mention gradient terms, yet only those terms have been changed in going from Model B to AMB+. As we shall see, the usual construction holds only when all such terms derive from a free energy functional.

We consider at mean-field level coexisting bulk phases separated by a globally flat interface. Clearly (without noise) the density profile, which we denote  $\varphi(x)$, depends only on the normal coordinate $x$ and the problem is effectively one-dimensional. 
It follows that the active terms $\lambda,\zeta$ enter only through the specific combination
\begin{equation}
\bar\lambda \equiv \lambda -\frac{\zeta}{2}. \label{eq:lambdabar}
\end{equation}
To see this, note that in $d=1$ only, $\nabla|\nabla\phi|^2 = 2(\nabla^2\phi)\nabla\phi$ so the terms in $(\lambda,\zeta)$ are not independent. It follows that all further effects of the $\zeta$ term, beyond those captured in $\bar\lambda$, will arise only in situations of interfacial curvature. (These effects are explored in Sec.~\ref{subsec:AMB+Ostwald} below.)
Indeed, in $d=1$, the deterministic current in (\ref{eq:AMB+J}) is simply $J=-\p_x\mu$, with
\qq\label{eq:stationary-AMB+explicit-flat}
\mu = f'(\varphi) -K \varphi'' + \bar\lambda\varphi'^2\,
\qqq
where $f'(\varphi)$ again denotes $df/d\varphi$, while $\varphi'$ denotes $\p_x\varphi$.

We now look for the stationary solution $\varphi(x)$ subject to boundary conditions $\varphi(-\infty)=\phi_1$ and $\varphi(+\infty)=\phi_2$ with a smooth interface around $x=0$. (The binodal values $\phi_{1,2}$ are not yet specified, but will emerge below.)
Setting $J=0$ implies that the chemical potential $\mu(x)$ takes a constant value $\tilde\mu$. For this to hold in the bulk phases requires
\qq
\tilde\mu = f'(\phi_1) = f'(\phi_2) \,. \label{eq:mu-flat}
\qqq
In equilibrium systems, alongside this `common slope' requirement one has a second condition of equal thermodynamic pressure $P_{\rm eq}(\phi)=\phi f'(\phi) - f(\phi)$ in the two bulk phases. This implies a `common intercept', thus completing the familiar common tangent construction.

Notably though, this second condition can also be derived, without reference to thermodynamics, by multiplying (\ref{eq:mu-flat}) by $\varphi'$ and integrating across the interface. Doing the same for AMB+, the pressure equality is replaced by 
\qq\label{eq:pressure-eq-equality-AMB+}
P_{\rm eq}(\phi_1) = P_{\rm eq}(\phi_2) + \bar\lambda\int_{-\infty}^\infty \varphi'^3(x)\,.
\qqq
Note that the coefficients of the equilibrium-like gradient terms in the dynamics do not enter here because, thanks to the underlying free energy structure, they are exact derivatives and vanish upon integration.
Eqs.~(\ref{eq:mu-flat},\ref{eq:pressure-eq-equality-AMB+})  fix the binodals of AMB+ in principle~\cite{Wittkowski14}. To find them explicitly, however, we need the integral on the right hand side of (\ref{eq:pressure-eq-equality-AMB+}), which depends on the unknown interfacial profile $\varphi(x)$, seemingly creating an {\em impasse}. 

\subsection{Transformation of fields and potentials}\label{subsec:pseudo-variables}
The {\em impasse} can be swerved by a nonlinear change of variables, $\varphi\to\psi(\varphi)$, which allows the binodals to be found without first knowing $\varphi(x)$. The approach follows lines long known for equilibrium systems~\cite{aifantis1983mechanical,aifantis1983equilibrium}, but was first deployed for active systems in the contexts of quorum-sensing particles~\cite{solon2018generalized} (see Sec.~\ref{subsec:QS} below) and AMB+~\cite{tjhung2018cluster}. The reader should note the new variables do not necessarily have a useful physical interpretation in their own right. They do however allow a century of expertise in equilibrium thermodynamics to be applied to the nonequilibrium problems under study here. 

We first multiply (\ref{eq:stationary-AMB+explicit-flat}) by a function $\psi'$ such that $\int dx \psi' \left[-K\varphi''+\bar\lambda\varphi'^2\right] =0 $.
Now defining $g$ as the solution to $\p g/\p \psi = \p f/\p \phi$ (so that $g=f$ for the passive case), the pressure relation \eqref{eq:pressure-eq-equality-AMB+} is transformed into
\qq\label{eq:AMB+pseudo-pressure-equality}
(\mu\psi-g)_1 = (\mu\psi-g)_2
\qqq
with subscripts $1,2$ denoting the two bulk phases. (In~\cite{tjhung2018cluster},  $\mu\psi-g$ is called the `pseudo-pressure'.)

It remains to determine $\psi(x)$. First observe that 
\qq
&\int dx \, \psi' \left[-K\varphi''+\bar\lambda\varphi'^2\right] \\
&=
\int dx \,
(\varphi')^3
\left[\frac{K}{2}\frac{\p^2\psi}{\p\varphi^2}  +\bar\lambda\frac{\p\psi}{\p\varphi} \right],
\qqq
where we integrated by parts in the first term and used the fact that $\varphi'(x\to\pm\infty)=0$. 
We conclude that $\psi$ obeys
\qq\label{eq:AMB+psi-equation}
\frac{K}{2}\frac{\p^2\psi}{\p\varphi^2}  +\bar\lambda\frac{\p\psi}{\p\varphi}=0.
\qqq
Without loss of generality, we may now choose a boundary condition on $\psi$ so that the transformation leaves the passive case unaffected: $\psi\to\varphi$ for $\bar\lambda \to 0$. This gives the explicit transformation for AMB+ as 
\qq 
\psi &=&\frac{K}{2\bar\lambda} 
\left(1-e^{\frac{-2\bar\lambda}{K}\varphi}\right)=\frac{K}{\zeta-2\lambda} 
\left(e^{\frac{\zeta-2\lambda}{K}\varphi}-1\right)\label{eq:AMB+psi},
\qqq
where the second equality uses \eqref{eq:lambdabar}.

Meanwhile $g(\phi)$ can be found explicitly as
\qq 
   g(\phi) = \int^\phi d\varphi \frac{\p f(\varphi)}{\p \varphi} \frac{\p \psi(\varphi)}{\p \varphi}.
   \label{eq:AMB+g}
   \qqq
The binodals are thus the solutions of (\ref{eq:mu-flat}) and (\ref{eq:AMB+pseudo-pressure-equality}) wherein $\psi$ and $g$ respectively obey (\ref{eq:AMB+psi}) and (\ref{eq:AMB+g}); finding explicit values becomes a straightforward numerical problem. In the limit $\bar\lambda = 0$ the binodals of course reduce to those of Model B,  $\phi_{1,2}=\mp \phi_b$ with $\phi_b = \sqrt{-a/b}$, since (\ref{eq:mu-flat}) and (\ref{eq:AMB+pseudo-pressure-equality}) then reduce to $\psi=\phi$ and $g=f$. This means that on a special line $\bar\lambda = 0$ in the $(\lambda,\zeta)$ activity plane, the {\em only} effects of activity on phase separation, at mean field level, arise through interfacial curvature.

\subsection{ Interfacial profile and its stability}\label{Sec:interfacial-shape}
While the transformed variables allow binodals to be found without reference to the profile $\varphi(x)$, the same is not true of the interfacial tension, whose physics will be addressed extensively in what follows. To obtain  $\varphi$ we may solve (\ref{eq:stationary-AMB+explicit-flat}) with boundary conditions $\varphi(\pm\infty)=\phi_{1,2}$ (or $\phi_{2,1}$). For $\bar\lambda = 0$ this can be done explicitly and gives $\varphi(x) =\varphi_{\rm eq}(x) =\pm \phi_b\tanh(x/\xi_{\rm eq})$, where $\xi_{\rm eq}= \sqrt{-2K/a}$ is the interfacial width \cite{Bray}. For $\bar\lambda\neq0$, no explicit solution is known, but a parametric one is found via the function $w(\varphi) \equiv (\p_x \varphi)^2$~\cite{Wittkowski14,cates2023classical}. This is obtained by multiplying (\ref{eq:stationary-AMB+explicit-flat}) by $\p_x\psi$ and integrating on $(-\infty,x)$, to give
\qq\label{app:eq-w}
w(\varphi)
=
\frac{2}{K} \left[
g(\varphi)-g_{-} -\mu\psi(\varphi)+\mu\psi_{-}
\right]\left(\frac{\p \varphi}{\p \psi}\right)
\qqq
where $\mu=f'(\phi_{-\infty})$, $g_{-}=g(\phi_{-\infty})$ and $\psi_{-}=\psi(\phi_{-\infty})$, and the transformed variables obey (\ref{eq:AMB+psi}, \ref{eq:AMB+g}).

One may ask whether the resulting interfacial profile is dynamically stable against variations that leave the interface strictly one dimensional. (Stability against higher-dimensional perturbations, such as height fluctuations, will considered in Sec.~\ref{subsec:AMB+capillary}.)   Because $\int\phi(x,t)dx$ is conserved, such perturbations can be written as $\phi(x,t)=\varphi(x) + \p_x \epsilon(x,t)$. The linear operator ruling the relaxation of $\epsilon$ has a continuous spectrum whose characterisation establishes stability for Model B~\cite{bricmont1999stability}.  
For AMB+, it has been found numerically that the profile is indeed stable against normal perturbations~\cite{fausti2021capillary}, but showing this analytically remains an open problem.

\subsection{The Ostwald process and its reversal}\label{subsec:AMB+Ostwald}
In equilibrium systems governed by Model B, bulk phase separation is driven by the Ostwald process~\cite{Bray,CatesJFM:2018}. For such systems, local thermodynamic equilibrium requires, in the sharp interface limit, a discontinuity in pressure across any curved interface, called the Laplace pressure: $\Delta P_{\rm eq} = \sigma \kappa$, with $\kappa$ the mean curvature. In particular, $\kappa = (d-1)/R$ for a spherical droplet of radius $R$, with  $\Delta P_{\rm eq}$ positive in the droplet interior. This allows the Laplace force on a hemisphere to be balanced by the interfacial tension around the equator. (For the same reason, $\Delta P_{\rm eq}$ is also positive in the interior of a vapour bubble surrounded by liquid.)
 
In contrast to the pressure, the chemical potential at a curved interface remains continuous,  but the Laplace effect shifts it by an increment that in Model B reads $\Delta\mu = \pm\sigma\kappa/2\phi_b +\mathcal{O}(\kappa^2)$. The sign is now such that $\mu$ is {\em increased} at the surface of a liquid-in-vapour droplet. The result is a compositional excess $\delta$ outside each droplet that decreases with increasing droplet size: in fact, $\delta = \sigma/(R\phi_b f''(\phi_b))$~\cite{Bray,CatesJFM:2018}. This local excess in turn sets boundary conditions on a quasi-static solution of a (linearized) diffusion equation, which slowly drives material down the density gradients from smaller to larger droplets. Small droplets evapourate, and large ones grow; the mean droplet size increases forever. This is the Ostwald route to complete phase separation.  Noting that $\mu$ is similarly {\em decreased} for a vapour-in-liquid bubble, exactly similar arguments apply to the growth of vapour bubbles on the opposite side of the phase diagram. Moreover, although precise calculations are not possible for the aspherical and/or bicontinous domains that often result from spinodal decomposition, exactly the same physical mechanism acts to coarsen such domains. In all cases, sharply curved regions are smoothed out by diffusive transfer of $\phi$ from regions of strong to weak curvature. This reduces the interfacial area and hence the free energy.

The effect of activity on the Ostwald process in AMB+ can be addressed using very similar arguments to those summarised above for the passive case~\cite{tjhung2018cluster}. Here we take a different route, following~\cite{cates2023classical}. We choose this line of analysis because it can also be used to address capillary waves (Sec.~\ref{subsec:AMB+capillary}) and nucleation phenomena (Sec.~\ref{subsec:AMB+nucleation}). The results concerning the Ostwald process are valid in any dimension $d\geq 2$.

For the Ostwald process, consider a liquid droplet in a supersaturated vapour environment at density $\phi=\phi_1+\epsilon$. Ignoring noise (setting $D=0$), and considering droplets whose radius $R$ is much larger than the interfacial width $\xi$, we argue that the evolution of the droplet is quasi-stationary in the sense that the interfacial profile has relaxed on the time-scale in which $R$ changes. This holds because the profile relaxes on a fixed diffusive time-scale set by the length-scale $\xi$, while $R$ itself evolves on a time-scale that increases indefinitely at large $R$. This time-scale separation can be formalised in the {\em ansatz}  
\qq\label{eq:Ansatz}
\phi(\bfx, t) &= &\varphi_{R}(r- R)\nonumber\\
&=&
\varphi(r-R)+\frac{\varphi_1(r-R)}{R}+\mathcal{O}(R^{-2})\,
\qqq
where $\varphi$ is the stationary density profile of a flat interface and $\varphi_1$ a correction due to curvature.

We then inject (\ref{eq:Ansatz}) in (\ref{eq:AMB+}) (without noise), multiply by $\psi'$ and integrate across the interface. A straightforward calculation~\cite{cates2023classical} leads to 
\qq\label{eq:AMB+Ostwald}
\dot{R} = 
 \frac{(d-1)M\sigma_{\rm ne}}{R\,(\Delta\phi)^2}  
\left[ \frac{1}{R_s} -\frac{1}{R} \right] +\mathcal{O}\left(\frac{1}{R^3}\right)
\qqq
where we have now defined $\Delta\phi=\phi_2-\phi_1$ (analogously for $\Delta\psi$ and $\Delta g$ appearing below) and $R_s$ is given by 
\qq\label{eq:Rc}
R_s=\frac{ (d-1)\sigma_{\rm ne}}{f''(\phi_1)\epsilon\Delta\phi}\,.
\qqq
Here, $\sigma_{\rm ne}$ is a nonequilibrium interfacial tension found using (\ref{app:eq-w})  as
\qq\label{eq:AMB+sigma}
\sigma_{\rm ne} 
= 
\frac{\Delta\phi}{\Delta\psi}\int_{\phi_1}^{\phi_2} d\phi \, \left\{ 
K\frac{\p \psi}{\p \phi}
-\zeta \left[\psi(\phi) -\bar\psi\right]\right\}\sqrt{w(\phi)}\,,\nonumber\\
\qqq
where $\bar\psi$ is evaluated at the droplet centre (so that here, $\psi = \psi_2+\mathcal{O}(1/R)$).
Note that the $\mathcal{O}(1/R)$ profile correction $\varphi_1$ in (\ref{eq:Ansatz}) does not enter the results for $\sigma_{\rm ne}$ or $R_s$ as calculated here to leading order. Note also that \eqref{eq:AMB+sigma} differs by the front factor $\Delta\phi/\Delta\psi$ from a quantity denoted $\sigma$ in the previous literature~\cite{tjhung2018cluster,fausti2021capillary,fausti2023,cates2023classical}. It can be further shown that, as happens in passive systems, $\sigma_{\rm ne}$ determines the deviations of the coexisting densities from the binodals in liquid droplets or vapour bubbles due to the curvature of the interface~\cite{tjhung2018cluster}. 

Up to the replacement of $\sigma_{\rm eq}$ by $\sigma_{\rm ne}$, the results (\ref{eq:AMB+Ostwald}, \ref{eq:Rc}) governing droplet dynamics coincide with those for passive Model B. It follows that, so long as $\sigma_{\rm ne}$ is positive (see below), $R_s(\epsilon)$ is an unstable fixed point that separates growing from shrinking droplets. As time evolves and small droplets are eliminated, the mean droplet size $\bar R$ increases as $\bar R(t)^3 \propto M\sigma_{\rm ne} t/(\Delta \phi)^2$, so that $R\sim t^{1/3}$, and accordingly the ambient supersaturation $\epsilon$ (which obeys $\epsilon^{-1}\simeq f''(\phi_1)\bar R\Delta\phi/\sigma_{\rm ne}$) decreases as $t^{-1/3}$~\cite{Bray,CatesJFM:2018}. This is the standard Ostwald ripening scenario, albeit with $\sigma_{\rm eq}\to\sigma_{\rm ne}$, which also admits a full theory of the droplet size distribution (the Lifshitz-Slyozov-Wagner theory)~\cite{lifshitz1961kinetics,wagner1961theory}. As previously mentioned, although precise calculations are not possible, the same physical mechanism of curvature-driven diffusion applies beyond the regime of spherical droplets, so that even for bicontinuous, interpenetrating domains the typical domain scale obeys $L(t)\sim (\sigma_{\rm ne}t)^{1/3}$. (For a fuller discussion of this coarsening law, see Section~\ref{subsec:AMB+coarsening} below.)

Closer inspection shows two important new features that distinguish the Ostwald process from its purely passive counterpart. First, unlike $\sigma_{\rm eq}$, the nonequilibrium tension $\sigma_{\rm ne}$ depends via \eqref{eq:AMB+sigma}
on $\bar\psi$, the value of $\psi$ at $r=0$.  
Hence there are in fact two {\em different} nonequilibrium  tensions: one found above for a liquid droplet surrounded by vapour, hereafter denoted $\sigma_D$, and another found by the same calculation applied to a vapour bubble  in a liquid environment. The latter has $\sigma_{\rm ne} = \sigma_B$, the value obtained by setting $\bar\psi=\psi_1$ in  \eqref{eq:AMB+sigma}.

A second major surprise emerges from evaluating these two quantities explicitly. As shown in~\cite{tjhung2018cluster} and in Fig.~\ref{fig:phi+-}, the nonequilibrium tension turns negative for sufficiently large negative values of $\bar\lambda,\zeta$ (for liquid droplets) or sufficiently large positive values of these (for vapour bubbles). 

Whenever $\sigma_B<0$, two bubbles of different sizes deterministically equalise their volume as time proceeds; this is `reverse Ostwald' behavior (see Fig.~\ref{fig:phi+-}). 
In AMB+ at nonzero noise, $D\neq 0$, two novel types of phase separation emerge in the reverse Ostwald regime defined by $\sigma_B<0$ and a second condition $\sigma_{\rm cw}>0$ (see Sec.~\ref{subsec:AMB+capillary}). These are microphase separation at high global density $\phi_0$, and so-called `bubbly phase separation'~\cite{tjhung2018cluster} at low $\phi_0$, as portrayed in the left panel of Fig.~\ref{fig:phi+-}. (Recall that the global density is defined as $\phi_0=V^{-1}\int\phi\,d{\bf r}$.) The noise-dependent statistical properties of these phases will be discussed in Sec.~\ref{subsec:statistical}. 
The reverse Ostwald scenario for droplets at $\sigma_D<0$ is identical to the one just described for bubbles, as guaranteed by the symmetry of AMB+ under $(\phi,\lambda,\zeta)\to -(\phi,\lambda,\zeta)$.  

Mathematically, one way to understand the reverse Ostwald regime is to note that
(\ref{eq:AMB+}) for $\p_t\phi$ is unchanged if \eqref{eq:AMB+J} is replaced by
\qq
{\bf J}/M = -\nabla\left(\frac{\delta \mathcal{F}}{\delta\phi} + \lambda|\nabla\phi|^2+\mu_\zeta\right)
\label{eq:AMB+nonloc1}
\qqq
where $\mu_\zeta$ obeys the nonlocal equation
\qq
\nabla^2\mu_\zeta = -\zeta \nabla\cdot\left((\nabla^2\phi)\nabla\phi\right) \label{eq:AMB+nonloc2}
\qqq
Here there is an analogy with Coulomb's equation; viewing the right hand side of
\eqref{eq:AMB+nonloc2} as a scalar charge density, one sees this to be localized at interfaces. Moreover, in the sharp interface limit, it becomes a dipole layer of strength proportional to the mean curvature  $\kappa$~\cite{tjhung2018cluster}, at least for spherical droplets. The resulting potential $\mu_\zeta$ has a discontinuity at the interface of strength $\propto\kappa$, which must be compensated by a  discontinuity in $\delta F/\delta\phi$ to ensure that ${\bf J}$ remains finite. This shifts the order-parameter increment $\delta$ adjacent to the interface; recall that the latter sets the boundary condition on diffusion and hence drives the Ostwald process. Because of the discontinuity in $\mu_\zeta$, the shift differs between the interface's two sides, creating new values $\delta_{B,D}\propto\sigma_{B,D}\kappa$. When activity is large, one or other of $\sigma_{B,D}$ can become negative, driving the Ostwald process into reverse for either bubbles, or droplets. 

Note that in AMB+ at fixed model parameters, a reverse Ostwald regime cannot arise on {\em both} sides of the phase diagram -- that is, for both positive and negative global density $\phi_0$. An exception could be engineered in principle by requiring the activity parameter $\zeta$ to vary with $\phi_0$, but such dependence is nonlocal in space and therefore unphysical. While a local dependence $\zeta(\phi)$ might achieve the same result, it also might not, since $\zeta$ would then vary sharply across every interface in the system, not just between the two sides of the phase diagram. (We are not aware of work on such a model.) Currently therefore, the mutual exclusivity of droplet and bubble phases, on opposite sides of the same phase diagram in any given system, appears to be a relatively robust prediction of AMB+. We are not so far aware of any numerical or experimental model that defies this prediction, at least among those for which a single scalar order parameter should offer a sufficient description.

Although AMB+ should describe a large class of systems with different microphysics, it is tempting to look for a generic {\em microscopic} mechanism that can explain the reverse Ostwald effect. There is so far no such explanation, even though the $\zeta$ term does arise by coarse-graining certain microscopic models~\cite{tjhung2018cluster}; see Sec.~\ref{subsec:particles-towards-quantitative-continuum}. However, we note that activity can, generically, cause pumping of material across interfaces, which in the flat case can alter the binodals (as found in Sec.~\ref{subsec:AMB+binodals}). Any curvature dependence of this pumping can deeply alter the physics of Laplace pressure, allowing a negative tension to arise in \eqref{eq:AMB+Ostwald}. This happens without inherently destabilizing the interfaces, as we show in Sec.~\ref{subsec:AMB+capillary} below.

\subsection{Spinodal decomposition and phase ordering}\label{subsec:AMB+coarsening}
We restrict attention here to the case of positive $\sigma_{\rm ne}$ for which the dynamics is qualitatively the same as in passive phase separation. (There is little analytic work so far on the corresponding time evolution when $\sigma_{\rm ne}<0$.) 

In AMB+, as in Model B, any uniform state with $|\phi_0|<\phi_s$ is linearly unstable. (Recall that $\phi_s$ is unaffected by active terms as these are nonlinear.) Accordingly, we expect no change to the early stages of spinodal decomposition, in which density fluctuations grow to form regions of local density close to $\phi_{1,2}$, separated by sharp interfaces. After this, then so long as $\sigma_{\rm ne} > 0$, the standard Ostwald mechanism takes over, moving material so as to reduce interfacial curvature. The structural length scale $L(t)$ then grows in time as $(\sigma_{\rm ne} t)^{1/3}$~\cite{Bray} in $d\geq2$ (with logarithmic corrections for the case $d=2$~\cite{marqusee1984dynamics,zheng1989theory}). The same Ostwald dynamics governs droplet growth in a metastable system whose initial population of small droplets is created by nucleation. 

A competing channel for domain growth is via diffusive coalescence, whose role depends on a power law exponent $u>0$ governing the size-dependence of the diffusivity of liquid droplets (or vapour bubbles) as $D_{ld}\sim L^{-u}$; taken alone this leads to $L(t)\sim t^{1/(u+2)}$. For Model B where momentum is not conserved, $u=d$~\cite{Onuki,kawasaki1983kinetics} so that coalescence is  subdominant for $d\geq 2$. In contrast, with a momentum-conserving fluid (Model H, Sec.~\ref{subsec:AMH} below) one has $u=1$ in $d=3$ dimensions~\cite{Onuki} so that coalescence can dominate (still giving $t^{1/3}$ scaling, but with a quite different expression for the prefactor) at small enough values of the fluid viscosity and/or low enough solubility of diffusant. 

In summary, with $\sigma_{\rm ne}>0$ and no momentum conservation, the Ostwald process should lead to the standard $(\sigma_{\rm ne} t)^{1/3}$ growth at long times, with coalescence and other noise-induced effects asymptotically negligible. (Said differently, the long time behaviour is governed by a zero-temperature fixed point of the renormalization group~\cite{Bray}.) The relevant tension $\sigma_{\rm ne}$ equates to $\sigma_{D,B}$ for droplet/bubble growth, but is some interfacially averaged combination of these in bicontinuous morphologies. Under fully symmetric conditions only, this average must reduce to $(\sigma_B+\sigma_D)/2$; see Sec.~\ref{subsec:AMB+capillary} below.  

These expectations are clear, yet their numerical verification is technically difficult, leaving room for doubt. Results compatible with $t^{1/3}$, but with slightly smaller measured exponents, were found in numerical studies of AMB+ for $\zeta=0$ both without noise~\cite{Wittkowski14} and with it~\cite{dikshit2024domain}, although in~\cite{pattanayak2021ordering} slower coarsening was reported. Whether these differences are caused by numerical limitations, or due to some unknown fundamental physics at play, is still an open question. In general though it is dangerous to estimate the coarsening exponent from any single simulation run (in which $L$ typically changes by a decade or less before finite-size effects kick in). Merging different runs onto a single scaling plot can help~\cite{kendon2001inertial}, but is numerically intensive and has not yet been attempted for active theories such as AMB+.

In both active and passive phase separation, an interesting feature is the emergence of hyperuniformity:  on length-scales much larger than $L(t)$, density fluctuations are far less than for a random pattern of domains of this size~\cite{ma2017random,tomita1991preservation,Bray,zheng2023universal,deluca2024}. More specifically, if the structure factor $S_q(t)=\langle |\phi_\bfq |^2\rangle$ is rescaled as $s_{\hat{q}}=S(\hat{q}/L(t)) / L^d(t)$, where $\hat q \simeq q L(t)$, then at $\hat q\to 0$,  $s_{\hat{q}}\sim \hat{q}^4$.
The absence of a $\hat q^0$ term (which defines hyperuniformity), and surprisingly also of a $\hat q^2$ term, was originally explained using isotropy arguments for passive Model B without noise~\cite{tomita1991preservation,Bray}. The result was recently found numerically to also hold for AMB+~\cite{zheng2023universal}. This was extended to Model B and 
AMB+ with non-constant mobility $M$, and also with finite noise $D$, in~\cite{deluca2024} where the isotropy arguments of~\cite{tomita1991preservation,Bray} were partly extended to these cases also.  
Designing hyperuniform materials is a significant agenda in materials science~\cite{torquato2018hyperuniform}, so it is intriguing that active as well as passive phase separation can achieve this without the need for long-range order. We further notice that the role of fluid flows has recently started to be characterized, although so far not in the regime where the system undergoes bulk phase-separation~\cite{padhan2025suppression}.

\subsection{Capillary waves and their stability}\label{subsec:AMB+capillary}
In passive liquids, thermal noise disturbs the liquid-vapour interface and is resisted by interfacial tension $\sigma_{\rm eq}$. In active systems we expect an interfacial tension also to resist height fluctuations but, because time-reversal symmetry is broken, there is no reason {\em a priori} to identify this with $\sigma_B,\sigma_D$ or any other particular value. It was however shown in~\cite{fausti2021capillary} that the interfacial tension that damps the noise-induced fluctuations of the interfacial height is given by the symmetric combination
\qq\label{eq:sigma-cw}
\sigma_{\rm cw} =\frac{1}{2} (\sigma_B+\sigma_D)
\qqq
where $\sigma_{B,D}$ are as defined previously. (Our notation for $\sigma_{\rm cw}$ differs from that in~\cite{fausti2021capillary} by a factor $\Delta\phi/\Delta\psi$.) Notice that (\ref{eq:sigma-cw}) is valid only for constant mobility $M$.

\begin{figure}
\begin{centering}
\includegraphics[width=0.9\columnwidth]{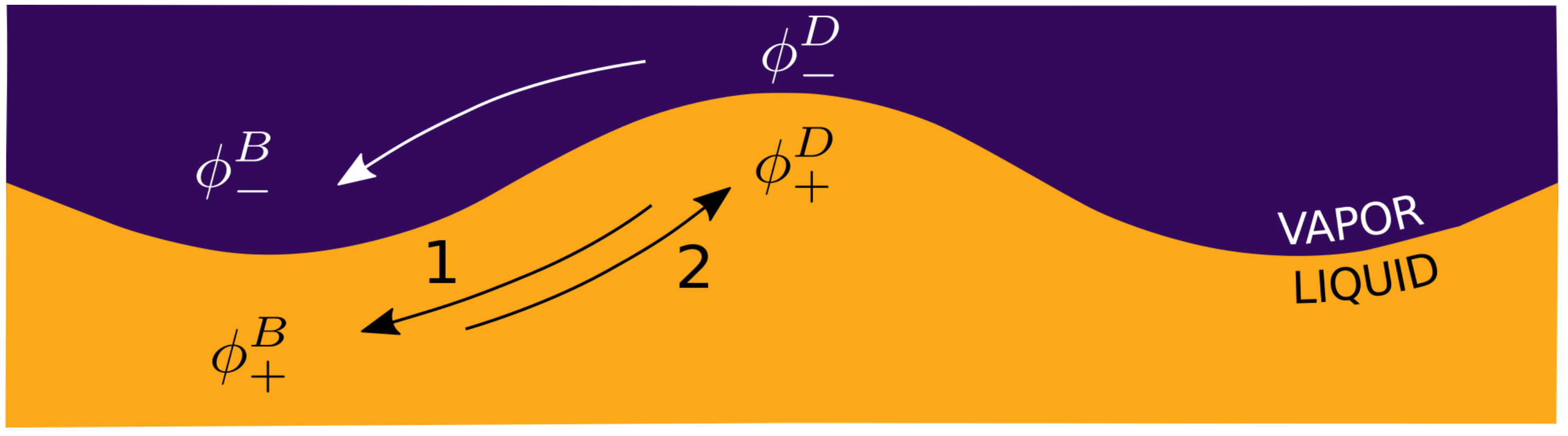}
\par\end{centering}
\caption{
Mechanism leading to the instability of capillary waves for $\lambda,\zeta>0$. Here $\phi_\pm^{B}$ and $\phi_\pm^{D}$ denote the quasi-stationary densities adjacent to the interface if the latter is curved, respectively, towards the vapour or the liquid regions. Arrows show the ensuing currents of $\phi$. The case marked (1) is stabilizing, and (2) is destabilizing. The destabilizing current outweighs the stabilizing one in the vapour phase when $\sigma_{\rm cw}<0$. }\label{fig:CW}.\end{figure}
 
Eq.~(\ref{eq:sigma-cw}) was established in~\cite{fausti2021capillary} by deriving the effective dynamics for small fluctuations in the interfacial height field, neglecting overhangs, and assuming quasi-static relaxation of the order parameter. These dynamics are nonlocal (which is also true in the equilibrium case) because of the presence of diffusive fluxes through both bulk phases. The fluxes are driven by interfacial curvature, just as they are in the Ostwald process. Bearing that in mind, \eqref{eq:sigma-cw} is easily understood in physical terms. The capillary tension can be found by considering the deterministic relaxation of a sinusoidal height perturbation via curvature driven currents: see Fig.~\ref{fig:CW}. In Model B, which has $\phi\to-\phi$ symmetry and  a single tension, the contributing currents through the two phases are exactly equal, and both are proportional in magnitude to $\sigma_{\rm eq}$. Therefore each phase contributes half the relaxation. In AMB+, the currents on the two sides of a sinusoidal interface have exactly the same spatial structure as in equilibrium, but are now proportional to $\sigma_D$ on one side of the interface and $\sigma_B$ on the other;  (\ref{eq:sigma-cw}) then follows.

\begin{figure}
\begin{centering}
	\includegraphics[width=1\linewidth]{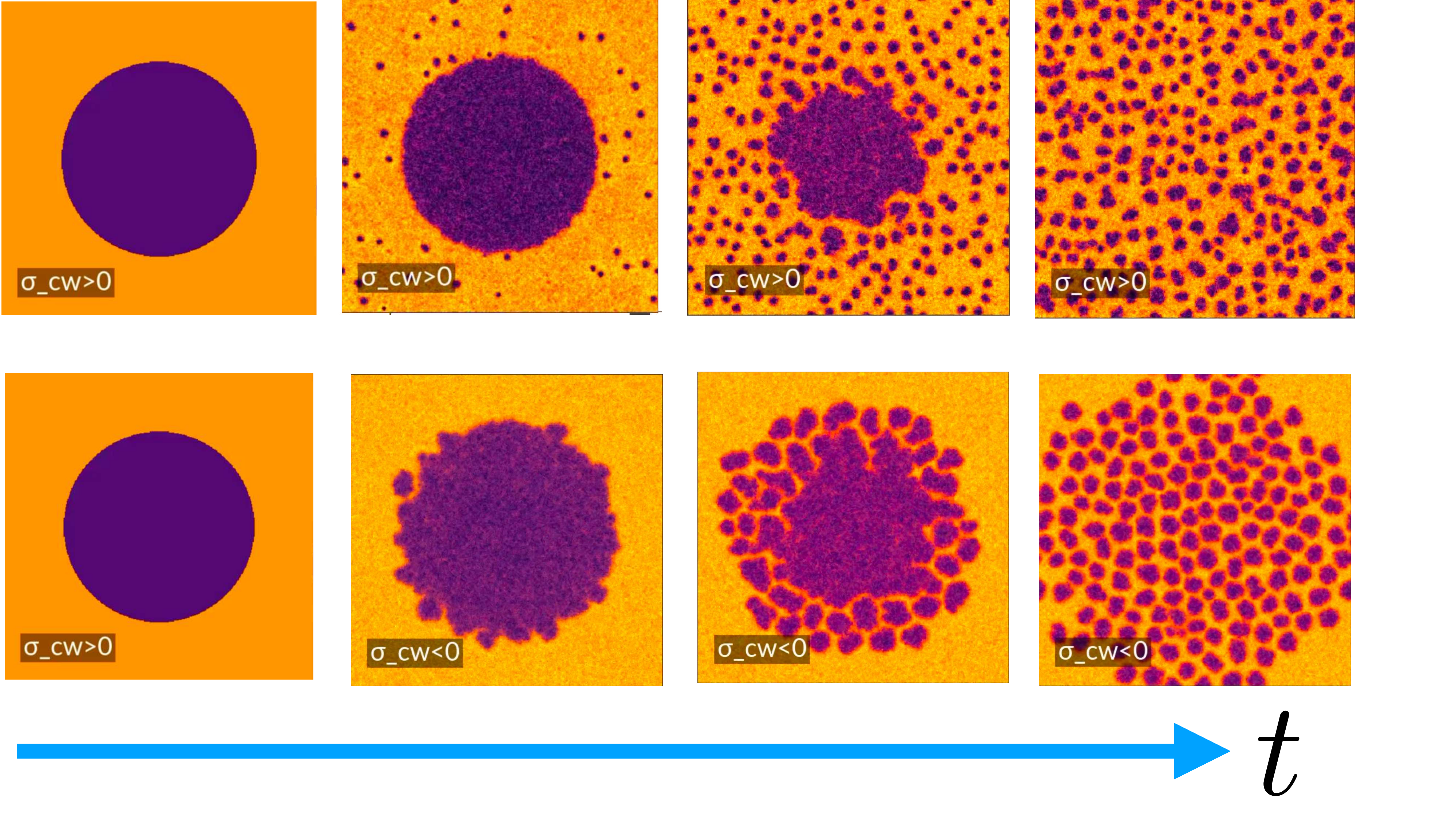}
\par\end{centering}
\caption{Time evolution from an initially fully phase separated state towards microphase separation when (Top) the Ostwald process is reversed ($\sigma_{\rm ne}<0$) but capillary waves are stable ($\sigma_{\rm cw}>0$) or (Bottom) when the latter are unstable ($\sigma_{cw}<0$). In the first case, vapour bubbles are nucleated in the bulk of the liquid and they subsequently grow until the bulk vapour region is destroyed; in the second, bubbles form at the interface between phases via the interfacial instability, and then diffuse away from it into the liquid~\cite{fausti2021capillary}. 
\label{fig:AMB+microphase-time-evolution}}
\end{figure}

The full nonlocal dynamics of the height field is found by an {\em ansatz} of timescale separation akin to \eqref{eq:Ansatz}:
\qq\label{eq:phi-varphi-interface}
\phi(\bfr,t)=\varphi(y-h(\bfx,t))\,,
\qqq
where ${\bf x}$ lies in the plane of the undisturbed interface and the interfacial profile $\varphi$ is again unaffected by curvature to leading order. The  nonlocality is diagonalized by a Fourier transform to give $h(\bfq,t)$, with ${\bfq}$ conjugate to ${\bf x}$. An analysis along lines used above for the Ostwald dynamics yields (with $q = |\bfq|$)
\qq\label{eq:AMB-effectve-eq-h}
\p_t h &=& -\frac{1}{\tau(q)} h +\mathcal{O}( q^2 h^2)\,,\\
\frac{1}{\tau(q)} &=& \frac{2M\sigma_{\rm cw}\,q^3}{(\Delta\phi)^2}\,.\label{eq:AMB-effectve-eq-h-tau}
\qqq
This reduces to standard Model B for $\sigma_{\rm cw} \to \sigma_{\rm eq}$. Direct numerical simulations of AMB+ with an initially imposed height perturbation show excellent agreement with (\ref{eq:AMB-effectve-eq-h},\ref{eq:AMB-effectve-eq-h-tau}) for long wavelengths; this is reported in~\cite{fausti2021capillary} which also gives results for smaller $|\bfq_x|$. 

At large enough activity, we have seen that one or other of the Ostwald tensions $\sigma_{B,D}$  becomes negative. The other does not, so from \eqref{eq:sigma-cw}, $\sigma_{\rm cw}$ initially remains positive as this threshold is passed. Numerical solution of \eqref{eq:AMB+sigma} shows however that at larger activity the negative tension eventually beats the positive one, ands $\sigma_{\rm cw}$ also becomes negative. (See the phase diagram in Fig.~\ref{fig:phi+-}.)  When this happens the interface goes unstable. Yet the system remains phase separated because the active interface remains linearly stable against normal perturbations, as discussed in Sec.~\ref{Sec:interfacial-shape} above.
This behavior has no counterpart in equilibrium Model B where $\sigma_{\rm eq} > 0$ whenever states are demixed.

The mechanism driving the instability to height fluctuations closely resemble the Mullins-Sekerka instability in solidification~\cite{langer1980instabilities}; see Fig.~\ref{fig:CW}.
(In both cases the instability is driven by a single diffusing field: latent heat in crystal growth, and $\phi$ here.) As discussed already in relation to \eqref{eq:sigma-cw}, the diffusive currents on either side of the interface are controlled by separate tensions. Only when the destabilizing current becomes the larger of the two in magnitude does $\sigma_{\rm cw}$ become negative, and the interfacial instability sets in~\cite{fausti2021capillary}.

\subsection{Statistics of active, phase separated states}\label{subsec:statistical}
We now discuss the new stationary states arising at negative tension. We address the case where $\sigma_B<0$ so the Ostwald process is reversed for vapour bubbles. The results then depend on whether $\sigma_{\rm cw}$ is also negative. The corresponding results for droplets with $\sigma_D<0$ follows as usual via the $(\phi,\lambda,\zeta)\to -(\phi,\lambda,\zeta)$ symmetry.

\begin{figure}
\begin{centering}
	\includegraphics[width=1\linewidth]{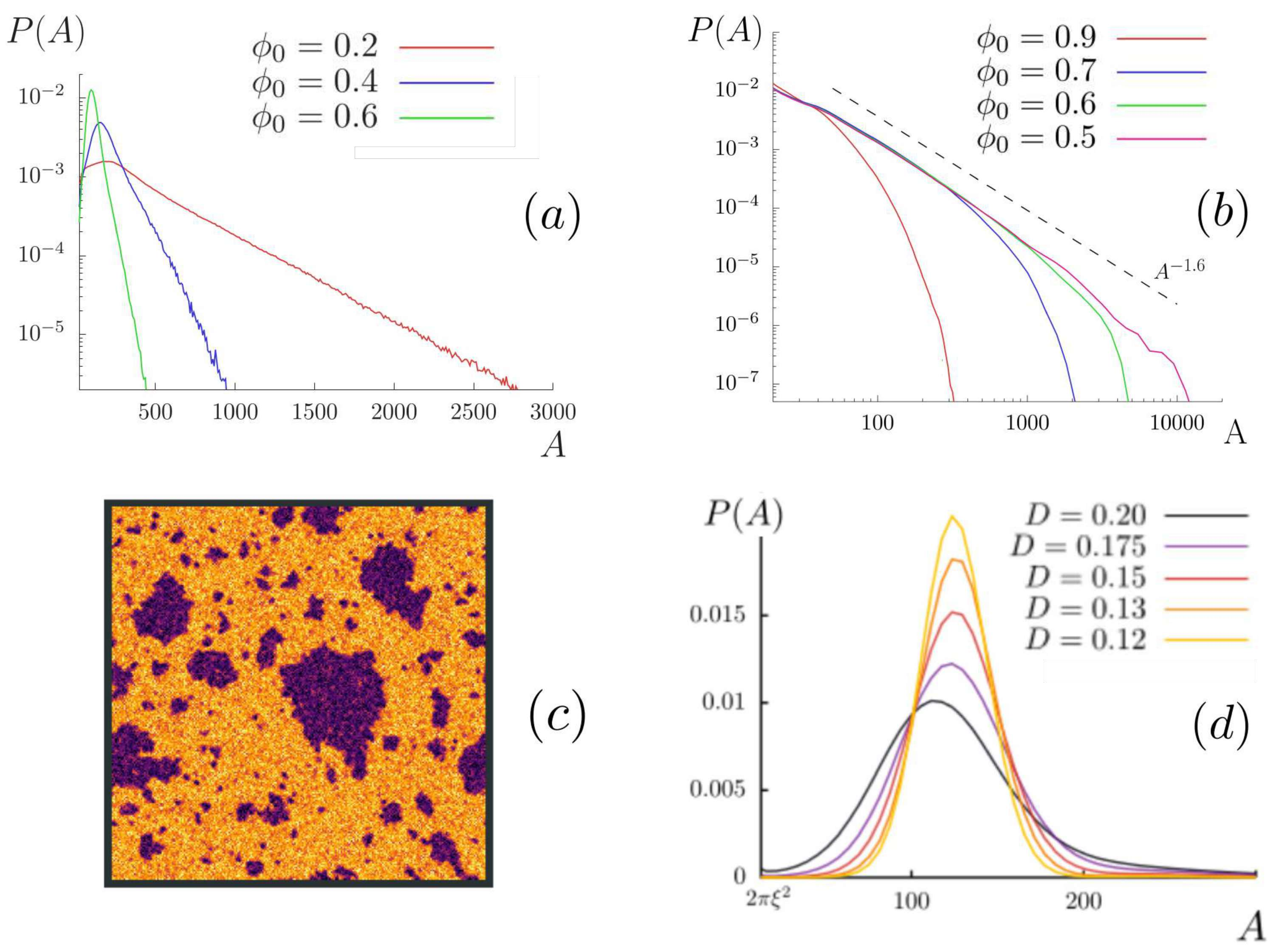}
\par\end{centering}
\caption{(a,b) The distribution of bubble sizes in the microphase separated state at $\sigma_{\rm ne}<0, \sigma_{cw}>0$ broadens when the global density $\phi_0$ decreases towards $\phi_{BL}$, or when decreasing $\sigma_{\rm ne}$ towards zero (not shown)~\cite{fausti2023}. Panel (a) corresponds to parameters for which the reversed Ostwald process dominates over coalescence and nucleation events so that vapour bubbles are almost monodisperse. Panel (b) has the same values of $\sigma_{\rm ne}$ and $\sigma_{\rm cw}$ but coalescence and nucleation have now comparable importance to the Ostwald process: vapour bubbles are broadly distributed and their morphology strongly deviates from a circular shape (Panel (c)). Panel (d) reports the bubble size distribution in the microphase separated state at $\sigma_{\rm cw}<0$ for various noise intensities $D$: the average bubble size is almost independent of $D$, while its variance only slightly increases with $D$. 
\label{fig:statistical}}
\end{figure}

Fig.~\ref{fig:phi+-} shows a phase diagram of AMB+, as a function of $\sigma_B$ and $\sigma_{\rm cw}$, found from simulation with noise $D=0.2$ in $d=2$. Although the analytical predictions presented in this Section hold equally in $d=3$ so that the same phases are to be expected there, no numerical simulations were performed so far in the literature for this case. 
Activity causes four distinct types of phase separation that are absent in passive Model B. When $\sigma_B<0$ but $\sigma_{\rm cw}>0$ we find ($\mu PS_1$) microphase separation and ($BPS$) `bubbly' phase separation (defined below), whereas when $\sigma_B,\sigma_{\rm cw}<0$ we find ($\mu PS_2$) microphase separation (of a different character to $\mu PS_1$), and  ($AF$) an `active foam' state. (Technically, the active foam is also microphase separated, but has distinctive structural features, as described below.) Numerical studies of the statistics of these phases, in $d=2$, can be found in~\cite{fausti2021capillary,fausti2023}; we briefly summarise the results here.

Note first though that the sign of $\sigma_{\rm cw}$ strongly affects kinetic pathways as well as the steady-state statistics~\cite{fausti2021capillary}. As shown in Fig.~\ref{fig:AMB+microphase-time-evolution}, starting from a fully phase separated state, the system breaks into bubbles via different dynamical pathways when $\sigma_{\rm cw}>0$ or $\sigma_{\rm cw}<0$. In the first case, bubbles are nucleated in the bulk and then grow by reverse Ostwald at the expense of the vapour region. In the second, the bubbles emerge from the interface and then diffuse away from it. 

Once the steady state is attained, the bubble size distribution for $\sigma_{\rm cw}>0$ (state $\mu PS_1$ above) is set by a competition between three processes: nucleation, coalescence and reverse Ostwald~\cite{fausti2023}. Nucleation and coalescence both become rarer as the noise amplitude $D$ is decreased, but more so nucleation. Hence the mean size increases upon decreasing the noise strength, diverging in the limit $D\to0$~\cite{tjhung2018cluster}. Furthermore, the average bubble size and its variance increase when decreasing the global density $\phi_0$,  and when decreasing $\sigma_{\rm ne}$ towards zero~\cite{fausti2023}. The size distribution and bubble morphology depends on the relative importance of the three competing processes; see Fig.~\ref{fig:statistical}. When the Ostwald process is dominant, bubbles are almost monodisperse with clear exponential tails in the bubble area distribution $P(A)$. Increasing the rates of nucleation and/or coalescence (both of which increase the width of the size distribution) at fixed $\sigma_B$ and $\sigma_{\rm cw}$, there emerges instead a broad size distribution with non-circular morphology (Figs.~ \ref{fig:statistical}(b,c)). In this regime, an intermediate power-law $P(A)\sim A^{-1.6}$ is observed before a system-size independent cutoff is attained~\cite{fausti2023}.

\begin{figure}
\begin{centering}
	\includegraphics[width=1\linewidth]{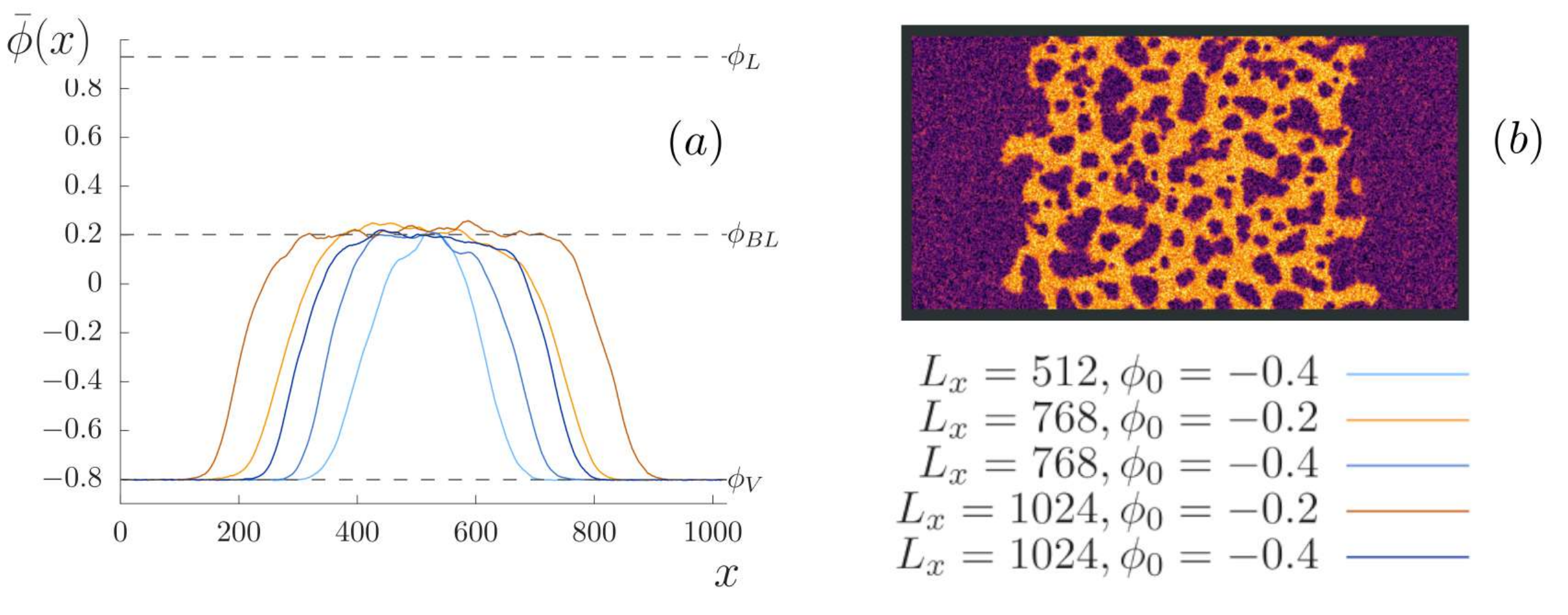}
\par\end{centering}
\caption{Simulations of AMB+ in a rectangular geometry showing bubbly phase separation, which is a coexistence of a uniform vapour phase with a microphase separated region at density $\phi_{BL}$~\cite{fausti2023}. Panel (a) shows the density projected along the vertical direction as a function of the horizontal one, while (b) is a snapshot of the system in steady state ($L_x=2L_y=1024$, $\phi_0=-0.2$, $\phi_{BL}\simeq 0.2$).
\label{fig:statistical-BPS}}
\end{figure}

For the microphase separated state at $\sigma_{\rm cw}<0$, $\mu PS_2$, the average bubble size is almost independent of $D$, while its variance increases with $D$ slightly, see Fig.~\ref{fig:statistical}(d). The typical radius is set here by the fact that the negative capillary tension acquires corrections due to the interface curvature, returning to positive values for sufficiently small bubbles~\cite{fausti2021capillary}, setting a scale for the fragmentation process shown in Fig.~\ref{fig:AMB+microphase-time-evolution}. Unlike state $\mu PS_1$, this type of microphase separation survives in the low-noise limit.

An open question is whether an order parameter can be found that distinguishes between $\mu PS_1$ and $\mu PS_2$ as $\sigma_{\rm cw}$ is varied through zero, perhaps connected with long-ranged spatial ordering of some kind. It is also not clear how closely the bubble statistics of these states can be mimicked by equilibrium microphase separation models~\cite{gompper1994phase}.
It is however clear that the striking breakdown of time-reversal symmetry in the steady state phases of AMB+ cannot be reproduced by any passive theory. For $BPS$ this asymmetry is shown schematically in Fig.~\ref{fig:phi+-}(c): bubbles are nucleated in the liquid region, move to its boundary, and burst to join the vapour. However, similar time-asymmetric life-cycles can be seen in the other phases too. For example, in $\mu PS_1$, bubble coalescence is primarily balanced by diffusive shrinkage. The time reversal of this, {\em i.e.}, spontaneous fission balanced by diffusive growth, is not seen.

The microphase separated state $\mu PS_1$ at  $\sigma_{\rm cw}>0$ can exist only at sufficiently high global densities $\phi_0$, such that $\phi_{BL}<\phi_0\lesssim\phi_2$ (see Fig.~\ref{fig:phi+-}).  The value of $\phi_{BL}$ was found to depend on the noise level $D$, with $\phi_{BL}\gtrsim (\phi_1+\phi_2)/2$ for all simulations so far performed. For $\phi_1<\phi_0<\phi_{BL}$, one observes bubbly phase separation ($BPS$). Here the system develops two macroscopic regions, the first filled by a homogeneous vapour phase and the second with a microphase separated state of type $\mu PS_1$ at mean density $\phi_{BL}$~\cite{tjhung2018cluster,fausti2023}. Simulations performed in a slab geometry have shown that the mean densities within each macroscopic region are independent both of $\phi_0$ and of system size, as is the bubble size distribution in the microphase separated region; see Fig.~\ref{fig:statistical-BPS} and~\cite{fausti2021phase,fausti2023}. Thus state $BPS$ can indeed be viewed as a macroscopic phase separation between $\mu PS_1$ at density $\phi_{BL}$ and excess vapour. (This resembles `emulsification failure' in the passive context~\cite{sear2001emulsification} and/or  the coexistence of periodic and homogeneous solutions in the conservative Swift-Hohenberg equation~\cite{thiele2013localized}).

Turning finally to the active foam state ($AF$), this is the least investigated so far. (For an image, see Fig.~\ref{fig:phi+-}, leftmost panel, bottom row, second column.) While it was shown that the phase boundary between this and state $\mu PS_2$ lies close to $\phi_{BL}\sim (\phi_1+\phi_2)/2$ at the noise levels considered in the simulations, there has been no analysis at the $\mu PS_2 - AF$ boundary to match that summarized above for the $\mu PS_1 - BPS$ boundary. 
It is therefore unknown what is the nature of the transition or crossover between the active foam and the microphase separated state. However this may be linked to a geometric percolation transition of dispersed phase (the vapour in Fig.~\ref{fig:phi+-}). Within a 2D foam, the size distribution of the vapour regions $P(A)$ is scale-free and has $A^{-2}$ tails, limited only by the system-size~\cite{fausti2021capillary}. It remains currently unknown what sets the width of the liquid filaments that make up the foam, 
nor specifically whether this width scales with the interfacial width $\xi$ for the model.

\subsection{Nucleation kinetics}\label{subsec:AMB+nucleation}

A crucial feature of phase-separating systems is homogeneous nucleation~\cite{Bray}. This is the path to phase separation when the uniform phase is metastable (which holds outside of the spinodal instability region, {\em i.e.}, for $|\phi_0|>\phi_s$). In this scenario, a liquid droplet of density close to $\phi_2$ nucleates within a metastable vapour phase of density $\phi_1+\epsilon$, with $\epsilon$ the supersaturation. Droplet growth is then a rare event driven by noise until a critical radius is reached, whereafter the droplet grows spontaneously until macroscopic phase separation is achieved. Nucleation of a vapour bubble in a liquid has a similar description. (Although the same theory covers both cases, we switch language from bubbles to droplets here, the latter being standard in the nucleation literature.) 

In passive fluids, this process is described by Classical Nucleation Theory (CNT)~\cite{debenedetti2021metastable,oxtoby1992homogeneous}. It was proposed by various authors that CNT might be extended to address nucleation in phase-separating active systems~\cite{richard2016nucleation,redner2016classical,levis2017active}. To our knowledge, this idea has so far been fully implemented only for AMB+~\cite{cates2023classical}. 
Below we outline the calculations for $d>2$ noting that logarithmic corrections arise in $d=2$ and that somewhat different physics applies for $d=1$; see~\cite{cates2023classical} for details.

For equilibrium systems CNT addresses the
case where the nucleation barrier is much larger than $k_BT$, in which large deviation theory prevails~\cite{touchette2009large}; in our context, this is best viewed as a weak supersaturation limit. 
The theory therefore assumes that $R_s$ is much larger than the interfacial width and that the critical droplet remains almost spherical. Subleading terms in powers of $R^{-1}$ are thereby suppressed. 

The free energy of a droplet of radius $R$ in $d\ge 2$ dimensions then obeys:
\qq\label{eq:eqlandscape}
U_{\rm eq}(R) = \sigma_{\rm eq} S_d R^{d-1} - \epsilon \Delta\phi f''(\phi_1)\, V_d R^{d} 
\qqq
where $S_d,V_d$ are the area and volume of the unit hypersphere. The first term is interfacial free energy which opposes droplet growth, and the second is the bulk free energy gain which promotes it. The maximum of $U_{\rm eq}$ arises at the critical radius $R_{s,eq}=(d-1)\sigma_{\rm eq}/(\epsilon \Delta\phi f''(\phi_1))$; compare \eqref{eq:Rc}. The corresponding free energy barrier is
\qq\label{eq:DeltaF-eq}
U_{\rm eq}(R_{s,eq})
= \frac{S_d}{d}\sigma_{\rm eq} R_{s,eq}^{d-1} .\qqq

Within equilibrium CNT, the dynamics of $R$ can be expressed as a Langevin equation~\cite{lutsko2012}
\qq\label{eq:R_t-final}
\dot{R}
=
- \mathcal{M}(R) \frac{\partial U_{\rm eq}}{\partial R} +\sqrt{2D{\mathcal{M}}}\,\Lambda
\qqq
where  $\Lambda$ is  zero-mean Gaussian white noise. Here the deterministic term is a special case of \eqref{eq:AMB+Ostwald} above: 
\qq\label{eq:special}
\dot{R} = 
 \frac{(d-2)(d-1)M\sigma_{\rm eq}}{R\,(\Delta\phi)^2}  
\left[ \frac{1}{R_{s,eq}} -\frac{1}{R} \right] 
\qqq
Factorizing this as  $-\mathcal{M}(R)\, {\partial U_{\rm eq}}/{\partial R}$, we find the mobility of the $R$ coordinate obeys
\qq\label{eq:MofR}
\frac{\mathcal{M}(R)}{M} &=& \frac{d-2}{S_d(\Delta\phi)^2 R^d}\,.\label{eq:MR}
\qqq
The noise term in \eqref{eq:R_t-final} then follows by detailed balance. These results still hold in the special case of $d=2$ provided that the factor $(d-2)$ in eq. (\ref{eq:MofR}) is replaced by the logarithmic correction $1/\log(R^+/R)$, where $R^+$ is the upper limit of integration~\cite{cates2023classical}. 

The right hand side of \eqref{eq:MofR} is effectively a Jacobian relating the noise in the $\phi$ dynamics to that for the collective coordinate $R$. This coordinate change is unaffected by the introduction of active terms, except through their effect on $\Delta\phi \equiv \phi_2-\phi_1$. 
Accordingly, in AMB+ the Langevin equation for $R$ is constructed by adding the noise term in \eqref{eq:R_t-final} to the deterministic evolution of \eqref{eq:AMB+Ostwald}. The resulting Langevin equation differs from the equilibrium one \eqref{eq:R_t-final} only through the activity-induced shifts in $\phi_{1,2}$, and the substitution $\sigma_{\rm eq}\to\sigma_{\rm ne}$ in \eqref{eq:eqlandscape} to give a nonequilibrium potential (or quasipotential~\cite{touchette2009large}) $U_{\rm ne}(R)$. A more detailed derivation of these results can be obtained using the ansatz \eqref{eq:Ansatz}~\cite{cates2023classical}. (The factor $(d-2)$ in eq. (\ref{eq:AMB+Ostwald}) and (\ref{eq:MofR}) was omitted in~\cite{cates2023classical}.)

Thus CNT gives for AMB+ a droplet nucleation rate $\mathbb{P}(R_s)\asymp\exp(-U_{\rm ne}(R_s)/D)$, with barrier height
\qq\label{eq:nucleation-barrier-pap}
U_{\rm ne}(R_s)
=
\frac{ S_d}{d }\sigma_D R_s^{d-1},
\qqq
where $R_s$ obeys \eqref{eq:Rc}. 
The theory equally describes nucleation of vapour from liquid: this is achieved by replacing $\phi_1\to\phi_2$ and $\epsilon\to -\epsilon$ (as for equilibrium models), and changing the value of $\sigma_{\rm ne}$ from $\sigma_D$ to $\sigma_B$ throughout.

The reaction pathway given by CNT was confirmed numerically in $d=2$~\cite{zakine2023unveiling} using advanced sampling methods~\cite{zakine2023minimum}.
Intriguingly, the even rarer event of a reverse transition from a stable uniform phase to a metastable one was also studied in~\cite{zakine2023unveiling}, this time for $d=1$. Its pathway is not the time reversal of the forward path, although the two must coincide in equilibrium. This shows that nucleation in AMB+ remains a far-from-equilibrium process even though its analysis via CNT takes a deceptively equilibrium-like form. The latter arises because the CNT droplet dynamics can be written in terms of a one-dimensional variable $R$, for which any Langevin equation can be written in the potential form \eqref{eq:R_t-final}.

The above calculations assume that the relevant value of $\sigma_{\rm ne} = \sigma_{B,D}$ remains positive; otherwise there is no barrier, and the precepts of CNT break down. An open question is to identify the kinetic pathway from a uniform metastable state to one of microphase and/or bubbly phase separation in systems with negative $\sigma_{\rm ne}$. Also open is the same question for the active foam at $\sigma_{\rm cw}<0$.

\subsection{Interfacial roughening}\label{subsec:AMB+roughening}
Phase separation creates a liquid-vapour interface, whose statistical fluctuations and their dynamics have long been studied in passive fluids~\cite{rowlinson2013molecular,safran2018statistical,aarts2004direct}.
There, interface fluctuations are thermal, and described by capillary wave theory. Starting from a flat interface, the system evolves until a stationary structure factor of the interface $S(q) = \lim_{t\to\infty} \langle |h(\bfq,t)|^2\rangle$ is reached; equipartition arguments then require 
\qq\label{eq:CWT-eq}
S(q) \simeq \frac{(2\pi)^{d-1} D }{\sigma_{\rm eq} q^2}\,.
\qqq
This holds for capillary modes whose wavelength is much larger than the interfacial thickness ($q\xi\ll 1$). 

The use of capillary wave theory in phase-separated active systems was first advocated heuristically~\cite{patch2018curvature,bialke2015negative,lee2017interface}, based on replacing $D, \sigma_{\rm eq}$ in (\ref{eq:CWT-eq}) with effective values. Numerical simulations of particle models~\cite{patch2018curvature,bialke2015negative,lee2017interface} lent support to this picture.
Subsequently the capillary wave scenario was fully calculated for AMB+, including the nonlinear corrections~\cite{fausti2021capillary,besse2023interface}.

The resulting theory involves the effective tension $\sigma_{\rm cw}$ introduced in Sec.~\ref{subsec:AMB+capillary} above; we assume this is positive for stability. Moreover, as is usually done for passive systems, we assume that $h({\bfx},t)$ is single valued so that overhangs are negligible. This assumption likely fails in reverse Ostwald regimes where the liquid-vapour interface is disrupted by transport of either vapour bubbles or liquid droplets across it. Therefore we also assume here that $\sigma_{B,D}>0$; the more general case remains open.

To address the capillary fluctuations in AMB+, we must generalize (\ref{eq:AMB-effectve-eq-h}) to allow noise terms. In the small $q$ limit, it was shown in~\cite{fausti2021capillary} that
\qq\label{eq:AMB-effectve-eq-h-noise}
\p_t h &=& -\frac{1}{\tau(q)} h + \xi 
\qqq
where $\langle\xi (\bfq_1,t_1) \xi (\bfq_2,t_2)\rangle = C_{\rm cw}(\bfq_1) \delta(\bfq_1+\bfq_2)\delta(t_1-t_2)$, with 
\qq\label{eq:effective-noise}
C_{\rm cw}(\bfq) =
4(2\pi)^{d-1}\frac{M D }{(\Delta\phi)^2}q\,.
\qqq
For ${\xi q\ll 1}$ this leads to
\qq\label{eq:CWT}
S(q) =
\frac{(2\pi)^{d-1} D}{\sigma_{\rm cw} q^2}\;.
\qqq
Thus, standard capillary wave theory appears to hold for AMB+ up to the replacement $\sigma_{\rm eq}\to\sigma_{\rm cw}$. This prediction was tested against numerical simulations of AMB+ in the low $D$ regime, finding good agreement~\cite{fausti2021capillary}. (As previously noted, our definition of $\sigma_{\rm cw}$ differs from that in~\cite{fausti2021capillary} by a factor $\Delta\phi/\Delta\psi$.)

Within the standard theory, however, (\ref{eq:AMB-effectve-eq-h-noise}) is linearized in $h$. In a renormalization group (RG) context~\cite{tauber2014critical}, this means that the critical exponents for interfacial roughening are given by their mean-field values $z=3$ and $\chi=(z-d)/2$. These exponents govern spatial and temporal correlations scaling, via $\langle \hat{h}(\bfx, t)\hat{h}(\bfx',t)\rangle\sim |\bfx-\bfx'|^{2\chi}$ and $\langle \hat{h}(\bfx, t) \hat{h}(\bfx,t')\rangle\sim |t-t'|^{2\chi/z}$, while the static structure factor $S\sim q^{-d-2\chi+1}$ for small $q$. 

These exponents can be changed by nonlinear corrections to (\ref{eq:AMB-effectve-eq-h-noise}).
For passive systems in $d=2,3$ there are no relevant nonlinearities and the mean-field critical exponents remain exact~\cite{besse2023interface}. Remarkably though, in systems without detailed balance one finds a new nonlinearity~\cite{besse2023interface,caballero2024interface}, relevant for $d<3$. Specifically one must in general add a term in $q F_{\bfx\to \bfq}[|\nabla \hat{h}|^2]$ to the right hand side of (\ref{eq:AMB-effectve-eq-h-noise}), where $F_{\bfx\to \bfq}$ means a Fourier transform. The resulting critical exponents were computed via RG  to one-loop as $z=3-(3-d)/3$ and $\chi=(3-d)/3$, and tested by $d=2$ simulations of the nonlinear counterpart to (\ref{eq:AMB-effectve-eq-h-noise}), which gave $z=2.78$ and $\chi=0.39$
~\cite{besse2023interface}.

So far, these exponents have not been confirmed by direct simulation either of AMB+ or of particle models. Indeed, some particle-based simulations concluded instead that $z\sim 2$~\cite{wysocki2016propagating,patch2018curvature}. One possible explanation is that those models undergo not bulk phase separation, but bubbly phase separation, in which vapour bubbles in the fluid phase merge frequently with the interface. Such events violate a key assumption of capillary-wave theories, namely that $h(\bfx,t)$ describes a dividing surface without overhangs between liquid and vapour regions, whose volumes are fixed. The roughening of the macroscopic interface under conditions of bubbly phase separation thus remains an open problem.

\subsection{Critical behavior}\label{subsec:critical}
The spinodals and binodals meet at the critical point, where $\phi_1=\phi_2$ so that the two coexisting phases become identical. Fluctuations are large here, and characterized by critical exponents that are well known for equilibrium Model B~\cite{Bray,Onuki}. The latter defines a universality class that includes, in the absence of momentum conservation, the vapour-liquid critical point ({\em e.g.,} for interacting passive Brownian particles), and the critical demixing transition for binary fluid mixtures. The corresponding static properties are those of the Ising universality class~\cite{Onuki,goldenfeld2018lectures}. 

There is significant theoretical evidence that this Model B class also controls the behavior of {\em active} systems undergoing critical phase separation between two uniform phases. Specifically, dimensional analysis~\cite{tauber2014critical} shows that for RG purposes active nonlinearities must be irrelevant at the Gaussian fixed point (which controls behavior in $d>4$) and remain so for the Wilson-Fisher fixed point (for $d<4$), at least insofar as the latter can be studied by perturbing in dimensionality at $d = 4-\epsilon$. An explicit one-loop (order $\epsilon$) calculation on AMB+ has confirmed this expectation~\cite{caballero2018bulk}. (A technical challenge, not yet addressed, is to find whether activity is also irrelevant in $d=2$ where the static properties of the Ising class are extremely well characterized via conformal field theory~\cite{belavin1984infinite}.)
Note however that sharing a universality class with equilibrium models does not imply that broken time-reversal symmetry is fully ignorable at the critical point; in general one expects an additional nontrivial exponent describing the entropy production~\cite{caballero2020stealth}. 

Much less is known of any critical points involving non-uniform states. A one-loop analysis has shown that the RG flow of AMB+ in $d<4$ becomes uncontrolled at sufficiently high activity~\cite{caballero2018bulk}, which might indicate a novel universality class linked to the reverse Ostwald regime. Further work is needed to clarify this point; it is not yet known whether there even exists a critical point (rather than first-order transitions) governing activity-induced microphase separation. 

Note that various simulations of phase-separating active particles report exponents matching the passive Model B universality class~\cite{maggi2021universality,partridge2019critical}, while others report quite different exponents~\cite{dittrich2021critical,siebert2018critical}. But, as discussed in Sec.~\ref{subs:micro-bubbly-particles} below, we have as yet little insight into how microscopic parameters map to those at field level. Some of these systems may therefore correspond to cases with $\sigma_{B,D}>0$, for which different critical behavior might arise as just discussed; they are also subject to large finite size effects~\cite{shi2020self}.

\subsection{Coupling to fluid momentum}\label{subsec:AMH}
So far we addressed AMB+, which considers the diffusive time evolution of a conserved scalar field $\phi$ describing, for example, the density of active particles. These typically reside in an incompressible solvent. In many cases the particle+solvent system does not conserve momentum, for instance because it is quasi-two dimensional and exchanges momentum with a supporting solid boundary; such systems are effectively `dry'~\cite{Marchetti2013RMP} and therefore fall within the purview of AMB+. In contrast, for a bulk system of active particles in an incompressible fluid, far from boundaries, the physics of phase separation is strongly altered by momentum conservation. 

\begin{figure}
\begin{centering}
\includegraphics[width=1\columnwidth]{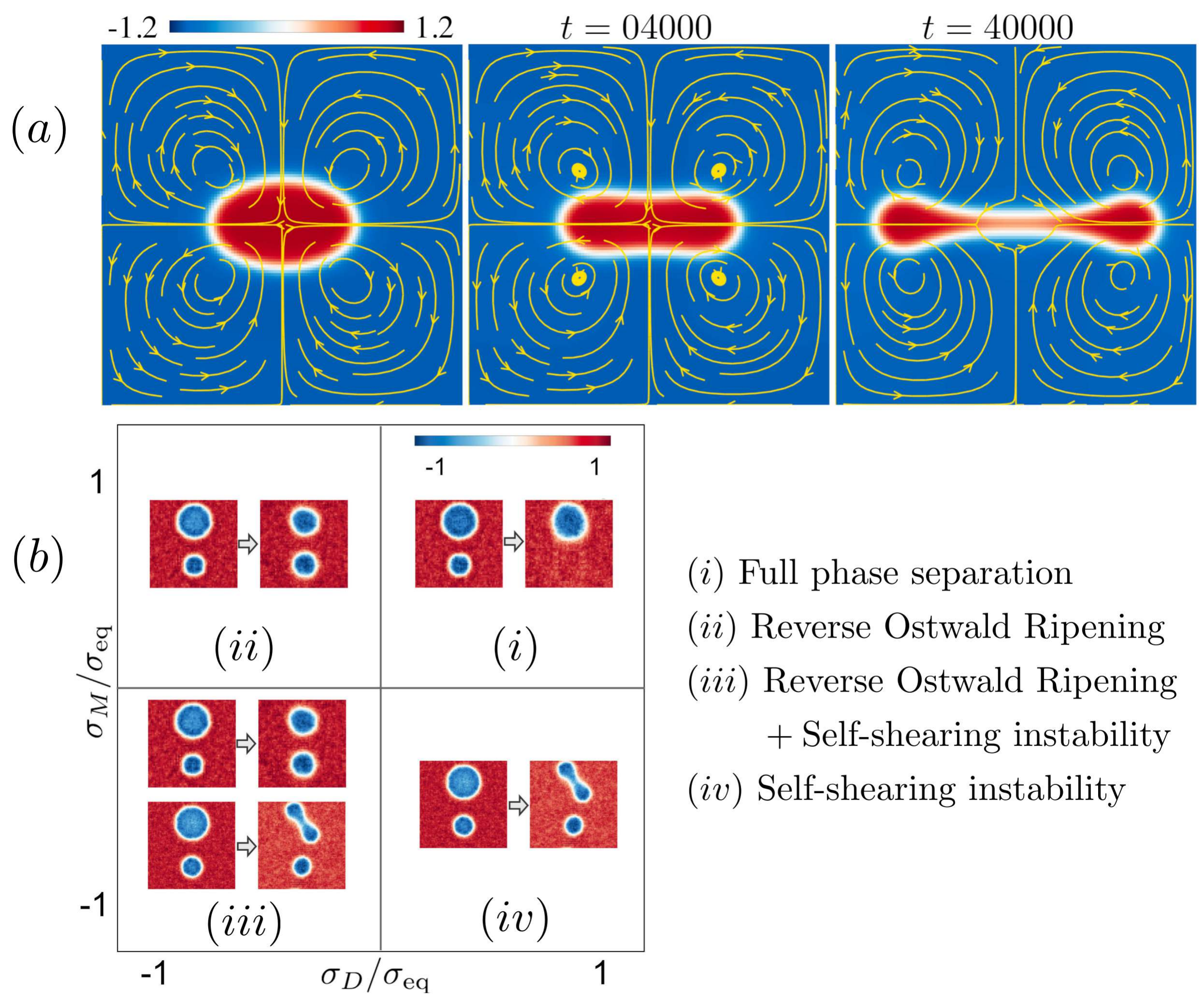}
\par\end{centering}
\caption{(a) Mechanism of droplet instability by self-shearing at $\sigma_{M}<0$. (b) Phase diagram of AMH shown via dynamical diagrams and in words on the $\sigma_M,\sigma_D$ plane. Figure adapted from~\cite{singh2019hydrodynamically}. }\label{fig:AMH}
\end{figure}

For passive systems the minimal theory in this `wet' setting is called Model H~\cite{hohenberg1977theory,Onuki,Bray,CatesJFM:2018}.
Model H couples $\phi$ to the fluid velocity $\boldsymbol{v}$ by adding (i) an advective term $\boldsymbol{v}.\nabla\phi$ to the left side of \eqref{eq:MB}; and (ii) the divergence of a passive order-parameter stress, $\nabla\cdot\boldsymbol{\Sigma}^{\rm eq}$, to the Navier-Stokes equation of a single-component fluid. Also included in that equation, following Landau and Lifshitz~\cite{landau1959fluid}, is a noise stress $\boldsymbol{\Sigma}^{\rm noise}$ whose variance and tensorial structure are fixed by detailed balance.

The order-parameter stress $\boldsymbol{\Sigma}^{\rm eq}$ encodes the facts that changes in the $\phi$ field caused by fluid advection alter the free energy $\mathcal{F}[\phi]$, and conversely that thermodynamic forces acting on $\phi$ to reduce $\mathcal{F}$ can drive fluid motion. It may be shown that
$\nabla\cdot\boldsymbol{\Sigma}^{\rm eq}=-\phi\nabla\mu$,
which can be viewed as the thermodynamic force density acting on a patch of fluid, caused by gradients in the chemical potential $\mu = \delta{\mathcal F}/\delta\phi$~\cite{Bray,CatesJFM:2018}. Tensorially, the order-parameter stress comprises two parts. One is isotropic, and can be absorbed into the pressure field. The latter only acts to enforce the constraint $\nabla\cdot\boldsymbol{v} = 0$. (This is quite unlike the case for interacting particle systems with no solvent; see Sec~\ref{subsec:particles-towards-quantitative-continuum}
 below).  The remaining stress  is traceless (or `deviatoric') and obeys~\cite{Bray,CatesJFM:2018}
\qq
\boldsymbol{\Sigma}^{\rm eq}=-K\boldsymbol{S}\,, \label{eq:eqstress}
\qqq
 where 
 \qq\boldsymbol{S}\equiv({\nabla}\phi)({\nabla}\phi)-\tfrac{1}{d}|{\nabla}\phi|^{2}\boldsymbol{I}\,, \label{eq:eqstressS}\qqq
 with $\boldsymbol{I}$ the unit tensor. 
 
The coupling to fluid flow in Model H has two main effects on phase separation. One is to alter the dynamic exponent near the critical point~\cite{Onuki} and the other is to change the coarsening dynamics whenever the phase-separated domains are bicontinuous~\cite{Bray}. As time proceeds, the curvature-driven diffusion of Model B, with $L\sim (M\sigma t/(\Delta \phi)^2)^{1/3}$ gives way to a `viscous hydrodynamic' regime $L\sim \sigma t/\eta$ in which the interfacial forces are balanced by viscosity. At even larger times, one enters an `inertial hydrodynamic' regime with $L \sim (\sigma/\rho t^2)^{1/3}$ in which the primary balance is between interfacial and inertial terms;  the Reynolds number at scale $L$ is then large~\cite{siggia1979late,CatesJFM:2018}. 
Interestingly, for states comprising spherical droplets these two hydrodynamic scaling regimes are not predicted, because a spherically symmetric stress field cannot drive an incompressible flow. In that case, the main effect of coupling to a (noisy) fluid is to introduce droplet coalescence via Brownian motion which gives $R\sim (D t/\eta)^{1/d}$~\cite{siggia1979late,san1985phase}.
As mentioned in Sec.~\ref{subsec:AMB+coarsening} above, in $d=3$ this has the same power law as Model B, but with an independent coefficient that may be negligible, or may dominate. More complex flow-mediated coarsening routes have also been proposed~\cite{gonnella1999phase,shimizu2015novel}, with particular complications in two dimensions~\cite{wagner1998breakdown}.

The minimal active extension of Model H is called Active Model H (AMH) ~\cite{tiribocchi2015active,singh2019hydrodynamically}, in which
\qq
\dot{\phi} +\boldsymbol{v}\cdot{\nabla}\phi=-{\nabla}\cdot\mathbf{J}.\label{eq:orderPara}
\qqq
Here $\mathbf{J}$ is the diffusive current from AMB+, given by (\ref{eq:AMB+J}). Meanwhile $\boldsymbol{v}$ obeys the Cauchy/Navier-Stokes equation
\qq\label{eq:Navier-Stokes}
\p_t \boldsymbol{v}+\boldsymbol{v}\cdot \nabla \boldsymbol{v} 
&=&\nabla\cdot \boldsymbol{\varsigma} +\nabla\cdot\boldsymbol{\Sigma}\,,\\
\label{eq:NSE-stress}
\bm{\varsigma}&=&-p\boldsymbol{I}+\eta({\nabla}\boldsymbol{v}+({\nabla}\boldsymbol{v})^{T})\,,\\
\label{eq:AMH_stress}
\boldsymbol{\Sigma} &=& \boldsymbol{\Sigma}^{\rm eq} + \boldsymbol{\Sigma}^{a}+\boldsymbol{\Sigma}^{\rm noise}\,.
\qqq
Here  $\bm{\varsigma}$ is the usual Navier-Stokes stress in which $\eta$ is the fluid viscosity, $\boldsymbol I$ is the unit second rank tensor, and the pressure field $p$ enforces incompressibility. The additional deviatoric stress $\boldsymbol{\Sigma}$ contains, alongside the passive order-parameter stress $\boldsymbol{\Sigma}^{\rm eq}$ and the noise stress $\boldsymbol{\Sigma}^{\rm noise}$, an active term $\boldsymbol{\Sigma}^{a}$, discussed below.

As already noted in Sec.~\ref{subsec:AMB+}, one might also break detailed balance via mismatch between the dissipative coefficient $M$ and the noise term in \eqref{eq:AMB+J}. The same opportunity now arises to create a mismatch between the viscosity $\eta$, which is silently taken $\phi$-independent above (and in passive Model H) and the noise stress, $\boldsymbol{\Sigma}^{\rm noise}$. To our knowledge, this opportunity has not yet been taken in the literature. However this case is likely to be relevant only when the solvent itself is active. Another way to break detailed balance would be to have different temperatures in the $\phi$ and $\boldsymbol{v}$ sectors, along lines used for particle mixtures~\cite{grosberg2015nonequilibrium}, see Sec.~\ref{subsec:non-reciprocal}.

In fact, AMH was so far studied only in the noiseless regime $D=0$~\cite{tiribocchi2015active,singh2019hydrodynamically} or for small noise~\cite{singh2019hydrodynamically}. This is legitimate when addressing late-stage phase ordering, which in Model H are controlled by a zero-temperature fixed point (just as they are in Model B)~\cite{Bray}. Another simplification, valid for most active systems, is to ignore inertia: because of the length and time scales at play, the particle-scale Reynold's number is very low (typically of order $10^{-4}$ in bacterial suspensions for example) so that $\nabla\cdot (\boldsymbol{\varsigma} +\boldsymbol{\Sigma})=\boldsymbol{0}$ can safely replace \eqref{eq:Navier-Stokes} on all but the largest length-scales. One advantage of this Stokesian limit is that the equation can be solved in Fourier space, allowing a vastly more efficient numerical scheme~\cite{singh2019hydrodynamically}. 

We now return to the active deviatoric stress, $\boldsymbol{\Sigma}^{a}$.  The lowest order second rank deviatoric tensor constructable from $\phi$ and its gradients is $\boldsymbol{S}$, as defined in \eqref{eq:eqstressS} above. Without detailed balance, the thermodynamic link between free energy and mechanical forces embodied in \eqref{eq:eqstress} is broken. To this order we may formally write  $\boldsymbol{\Sigma}^{a}=-(\tilde{K}-K)\boldsymbol{S}$, or
\begin{gather}
\boldsymbol{\Sigma}^{\rm eq}+\boldsymbol{\Sigma}^{a} = -\tilde{K}\boldsymbol{S}\,\label{eq:activeStress}
\end{gather} 
in \eqref{eq:AMH_stress}. At first sight, it then looks from \eqref{eq:eqstress} as if the active term simply renormalizes $K\to\tilde K$. This is however not true, since $K$ is unchanged in the diffusive current ${\bf J}$ in \eqref{eq:AMB+J}. In other words, even at $\lambda,\zeta=0$, the diffusive current and the order-parameter stress in AMH derive from {\em incompatible} free energies, breaking detailed balance. The $K$ term in ${\bf J}$ is lower order (by one power of $\phi$) than the $\lambda,\zeta$ terms  in \eqref{eq:AMB+J}. Hence the $\tilde K\neq K$ mismatch via \eqref{eq:activeStress} is the leading active term in AMH.

In general, the coefficient $\tilde K - K$ in $\boldsymbol{\Sigma}^{a}$ can have either sign. Therefore, if activity is large enough, so can $\tilde K$ itself (while $K$, as is still present in the equation for ${\bf J}$, remains positive). To exemplify this, note that in a particle picture, discussed further in Sec.~\ref{subsec:nematics},
individual swimming particles with orientation ${\bf u}$ produce a stress field proportional to $\pm({\bf u}{\bf u}-\boldsymbol{I}/d)$ where the sign differs between `contractile' and `extensile' swimmers~\cite{Marchetti2013RMP}. If spherical, these swimmers tend to align normal to  interfaces, in which case $\tilde{K}$ remains positive for
extensile swimmers, but becomes negative, at high enough activity, for contractile
ones~\cite{tiribocchi2015active}. We use this nomenclature below, although the opposite correspondence applies if swimmers instead align {\em parallel} to the interface (see Sec.~\ref{subsec:nematics}).

The gradient expansion implicit in \eqref{eq:activeStress} effectively carries with it a low-order multipole expansion of flow fields created by microswimmers. This cannot capture the effects of their near-field hydrodynamic interactions, which may be considerable and can even reverse the tendency to phase separate~\cite{matas2014hydrodynamic}. Such issues are best addressed in particle-based models in which near-field hydrodynamics is more naturally included; see Sec.~\ref{subsec:hydro-particles}. 

The stress in \eqref{eq:eqstress} creates interfacial forcing through a mechanical tension which in (local) equilibrium obeys $\sigma_{M} = \sigma_{\rm eq}$. This applies wherever $\phi$ lies close to an equilibrium order parameter profile, which is locally relaxed towards by the diffusive current ${\bf J}$. In the case $\lambda= \zeta =0$, the equation for ${\bf J}$, and hence the local profile, is unchanged from the equilibrium case.  To leading order in activity, it follows that the full mechanical forcing term, $\boldsymbol{\Sigma}^{\rm eq}+\boldsymbol{\Sigma}^{a} = -\tilde{K}\boldsymbol{S}$ in (\ref{eq:AMH_stress},\ref{eq:activeStress}), for a given disposition of interfaces, is found from the equilibrium one by the substitition $\sigma_{\rm eq}\to\sigma_{M}=(\tilde{K}/K)\sigma_{\rm eq}$~\cite{tiribocchi2015active}. This is linear in $\tilde K$ because $\nabla\phi$ is independent of this purely mechanical parameter, being fixed instead via local diffusive equilibration of the interfacial profile which thus involves only $K$ not $\tilde K$. The result
agrees with the interfacial tension that determines the relaxation of capillary waves (upon neglecting large-scale diffusive fluxes) as is confirmed by extending to AMH the analysis presented in Sec. II-G~\cite{caballero2022activity,caballero2024interface}. 

So long as the mechanical tension $\sigma_M$ is positive, one can expect relatively mild corrections to the coarsening dynamics described above for passive Model H~\cite{tiribocchi2015active}. (As with AMB+, it is numerically challenging to probe any changes to coarsening exponents, and evidence for such changes remain indecisive~\cite{tiribocchi2015active,padhan2024novel}.)
The required condition is satisfied both for extensile and for weakly contractile microswimmers.
For strongly contractile systems, however, we have $\sigma_{M}<0$. This leads to a self-shearing instability leading to splitting of both droplets and bubbles via the mechanism illustrated in Fig.~\ref{fig:AMH}, whereby shape fluctuations are amplified. In turn this leads to (at least) two types of microphase separation beyond those seen in AMB+~\cite{singh2019hydrodynamically}.

One new type of microphase separation arises when $\sigma_{M}<0$ but the relevant diffusive tension ($\sigma_{D}$, say) is positive. Here the negative mechanical tension causes large droplets to split by self-shearing. This is balanced by (forward) Ostwald ripening: small droplets evapourate while large ones grow until they in turn become unstable. The result is a dynamical steady state of droplet splitting followed, on average, by diffusive growth of one offspring and disappearance of the other(s). Comparing the rates of mass loss by splitting and gain by Ostwald ripening gives a steady-state characteristic droplet size as 
 \qq\label{eq:AMH-scaling-radius}
 \bar R \simeq \left|\frac{\sigma_{D}\eta M}{\sigma_M (\Delta\phi)^2}\right|^{1/2}
 \qqq
 which is confirmed in numerical simulations~\cite{singh2019hydrodynamically}.  
This parameter regime also admits analysis of the relaxation of capillary waves as considered for AMB+ in Sec.~\ref{subsec:AMB+capillary};  the result (\ref{eq:AMB-effectve-eq-h}) still holds, with~\cite{caballero2022activity}
 \qq
\frac{1}{\tau(q)} = \frac{2M\sigma_{\rm cw}\,q^3}{(\Delta\phi)^2}
+\frac{\sigma_M}{4\eta} q \,.
\qqq
Identifying  the wavelength $\ell$ beyond which $\tau(q)$ is negative then recovers  \eqref{eq:AMH-scaling-radius} with $R\to\ell$ and $\sigma_D\to\sigma_{\rm cw}$. 

A second type of microphase separation arises in droplet (resp., bubble) phases when $\sigma_D$ (resp., $\sigma_B$) is negative alongside $\sigma_{M}$. Here, because Ostwald ripening is reversed, were splitting to continue, the number of droplets would increase
forever since small droplets are not removed by evapouration. Accordingly,
for negative $\sigma_D$,
the steady state must comprise almost static droplets with
no splitting. Their characteristic size $\bar R$ then is set by the largest droplet for which the diffusive flux, which now acts to equalize curvature, can stabilize the self-shearing mode. The result is again \eqref{eq:AMH-scaling-radius}, whose $|\sigma_M|^{-1/2}$ scaling was numerically confirmed~\cite{singh2019hydrodynamically}. A third regime of arrested coarsening, in which inertia plays a dominant role, has recently been reported in~\cite{padhan2024novel}.

Capillary waves in AMH were alse investigated at linear~\cite{caballero2022activity} and nonlinear~\cite{caballero2024interface} level showing that, differently from the dry case described in Sec. \ref{subsec:AMB+capillary}, the scaling exponents are given by those of the linear theory. Several other aspects of the AMH phase diagram remain to be fully investigated, such as how negative $\sigma_{M}$ affects the bubbly phase separation and the active foam state arising when $\sigma_{D}$ or $\sigma_{\rm cw}$ are respectively also negative.

\section{Connections with particle models}\label{sec:particle-models}

Typically, the scalar order parameter $\phi({\bf r},t)$ appearing in the stochastic continuum models of Sec.~\ref{sec:AMB+} represents a smooth, coarse-grained density of active particles. This continuum limit emerges at scales beyond the mean interparticle spacing; for the field equations to be local in space as assumed, length-scales of interest must also exceed both the range of interactions and the distance moved by a particle between collisions. Locality in time in turn emerges at time-scales larger than a relaxation time for local structure. In the simplest cases this is the time between those collisions, but for dense systems it is more like a structural (Maxwell) time. These criteria apply for passive and active systems alike, but in the active case one must in addition exceed the persistence length and times of individual self-propelled particles~\cite{stenhammar2013continuum}. 

So long as a conserved scalar order parameter is the only slow variable at large scales, then subject to the above requirements the continuum approach of Sec.~\ref{sec:AMB+} should allow a generic description of phase separation in active systems almost regardless of the specific microscopic physics that causes it.  It is then a separate question to understand which microscopic mechanisms lead to the new phenomenology emerging from the continuum theory as a result of its non-equilibrium character. This research agenda is still in its early stages, and we review in this Section what is known so far. Because the microscopic physics in experimental systems (ranging from bacteria to self-propelled colloids through to interacting social agents) is itself rarely well characterized, much of this knowledge is based on comparing continuum theories such as AMB+ with computational models based on simulation of particles with simple, and fully specified, dynamical rules.

Alongside the usual passive mechanism involving interparticle attractions, there are several microscopic mechanisms that can cause phase separation in active systems. Most strikingly, active phase separation can arise from the fact that particles tend to accumulate where they are slowed down~\cite{schnitzer1993theory}. If they also tend to slow down where the accumulate, then an instability towards phases separation can follow~\cite{Tailleur:08}. This slowdown can be caused by local signals (`quorum sensing', see Sec.~\ref{subsec:QS} below) or by collisions. Even when bare attraction forces are absent, slowdown yields an effective attraction among self-propelled ({\em i.e.,} motile)  particles~\cite{Tailleur:08}, leading to Motility Induced Phase Separation (MIPS)~\cite{Cates:15,fily2012athermal,stenhammar2014phase,redner2013structure,stenhammar2021introduction,o2021introduction}. Importantly, this mechanism is absent in isothermal systems with detailed balance, because in this case local kinetic effects cannot take the steady-state statistics away from the Boltzmann distribution. In contrast, active particles typically accumulate with a steady-state density inversely proportional to their local propulsion speed~\cite{schnitzer1993theory,Tailleur:08,arlt2018painting,frangipane2018dynamic}.

Note also that, within AMB+, the `cause' of phase separation is always the negativity of the quadratic coefficient $a$ of the free energy $F[\phi]$. In the case of MIPS, the effective attractions caused by activity cause $a$ to deviate from the value set by interparticle forces alone. Put differently, the functional $F[\phi]$ in AMB+ is not an equilibrium free energy, as found by coarse-graining a microscopic Hamiltonian -- because there isn't one. The distinction between activity-induced changes to $F[\phi]$ and the {\em explicitly} active terms $\lambda,\zeta$ is that, without the latter, detailed balance would be restored at coarse-grained level despite being absent microscopically.

More generally, several types of actual (rather than effective) attraction can also cause  phase separation among active particles. These include direct particle-particle attractions along passive lines~\cite{Palacci:12,palacci2013living,redner2013reentrant,alarcon2017morphology,ray2023increased}, and forces mediated by chemical fields~\cite{liebchen2015clustering,liebchen2018synthetic}, hydrodynamic flow~\cite{thutupalli2017boundaries,vskultety2023hydrodynamic} or biological mechanisms~\cite{alston2022intermittent,bonazzi2018intermittent,steinberg1963reconstruction,foty2005differential}.  Characterizing these interactions is often an experimental challenge, which is partly why numerical simulation of particle models have proved so important in elucidating the physics of active phase separation.

In simulations~\cite{digregorio2018full,caporusso2020micro,mukhopadhyay2023active,martin2021characterization} and experiments~\cite{Palacci:12,Speck:13,van2019interrupted} in which particles do not vary much in size and shape, local hexatic or crystalline order is often observed in the dense phase, whereas AMB+ assumes that this phase remains a structureless fluid. In other cases, including experiments on microtubule suspensions, the dense phase can have local nematic order~\cite{adkins2022dynamics,tayar2022controlling,lemma2022active}. Such local order can influence phase separation at the mesoscale; see Sec.~\ref{sec:two-order-parameters}.

Below we first address in Sec.~\ref{subsec:QS} the quorum sensing model. This is arguably the simplest active particle model of separation; its coarse-grained description share similarities with AMB+ but lacks its $\zeta$ term and the new physics of negative interfacial tensions. Sec. \ref{subsec:effective-equilibrium} describe attempts at connecting continuum and microscopic descriptions via coarse-graining procedures. Sec.~\ref{subsec:AMB+coarsening2} describes what is known from particle models about  coarsening kinetics. In Sec.~\ref{subs:micro-bubbly-particles} we discuss microphase and bubbly phase separation in models of repulsive self-propelled particles, turning to the attractive case in Sec.~\ref{subs:attraction}. Secs.~\ref{sec:pressure} and ~\ref{sec:pressure2} address the roles of mechanical pressure, swim pressure and hydrodynamics, while \ref{subsec:particles-towards-quantitative-continuum} addresses coarse-graining strategies for mechanical microscopic interactions beyond quorum sensing. Sec.~\ref{subsec:particles-open} discusses some open questions that particle models may address in future.


\subsection{Quorum sensing}\label{subsec:QS}
Quorum-sensing particles (QSPs) comprise a minimal model for bacteria whose behavioural rules depend on their local particle density $\rho({\bf r})$, as sensed via chemical signalling or other means~\cite{miller2001quorum,schnitzer1993theory,Tailleur:08,Cates:15}.
While the natural behaviour is often complicated, genetic engineering can be used to create systems with a self-propulsion speed that decreases at high density in a well-characterized manner~\cite{liu2011sequential}. QSP models with density-dependent swim speed also offer a simplified description of self-propelled synthetic colloidal particles, either interacting via two-body repulsive forces in which crowding hinders forward motion~\cite{Cates:15,stenhammar2013continuum,Speck2014PRL}, or subject to bespoke density-dependent protocols~\cite{bauerle2018self}.

The simplest QSP models assume that a particle at position $\bfr$ with orientation (swim direction) ${\bfu}$ obeys
\qq\label{eq:QS}
\dot{\bfr} = v[\rho](\bfr)\bfu +\sqrt{2D_T}\boldsymbol{\eta}\,.
\qqq
Here $D_T$ is a translational diffusivity which is unrelated to self-propulsion (such as translational Brownian motion) and often negligible, with $\boldsymbol{\eta}$ vectorial unit white noise. The notation $v[\rho](\bfr)$ means that self-propulsion speed  at $\bfr$ and depends on $\rho$ in its neighborhood. The self-propulsion direction $\bfu$ has an autocorrelation time ${\tau}$; its precise dynamics is relatively unimportant~\cite{cates2013active}. Widely studied examples are Active Brownian Particles (ABPs, where $\bfu$ diffuses on the unit sphere), and Run and Tumble Particles (RTPs, where $\bfu$ jumps between values on the unit sphere by a Poisson process).  A third is
Active Ornstein Uhlenbeck particles (AOUPs) where 
\qq\label{eq:AOUP}
\dot{\bfu}= -\bfu/\tau+ \sqrt{2/\tau}\, \boldsymbol{\xi}\qqq
with $\boldsymbol{\xi}$ vectorial unit white noise. In this case, 
$\bfu$ becomes a zero-mean Gaussian random vector with $\langle|{\bf u}|^2\rangle = d$ and autocorrelation time $\tau$. 

We now suppose that $v[\rho](\bfr) = v_0(\tilde\rho(\bfr))$ where $\tilde\rho (\bfr)=\int {\mathbb K}(|\bfr-\bfr'|)\rho(\bfr')d\bfr' $ is a locally averaged density. An anisotropic kernel ${\mathbb K}$, dependent on $\bfu$, can also be included~\cite{solon2018generalized2}. Taylor expanding in density gradients gives~\cite{stenhammar2013continuum,solon2018generalized,solon2018generalized2}
\qq\label{eq:QS-self-propulsion}
v[\rho]= v_0(\rho) + \beta(\rho) |\nabla\rho|^2+
\gamma(\rho) \nabla^2\rho + ...
\qqq
where the function $v_0(\rho)$ and the kernel ${\mathbb K}$ together determine $\beta,\gamma$ (alongside higher order terms). For isotropic kernels, $\beta=0$ so the leading contribution is the $\gamma$ term. 

The coarse-graining of (\ref{eq:QS},{\ref{eq:QS-self-propulsion}) follows standard lines~\cite{dean1996langevin,Tailleur:08,cates2013active,Cates:15,solon2018generalized,tjhung2018cluster,pavliotis2008multiscale}. After eliminating the fast orientation variables,  Dean's equation~\cite{dean1996langevin} gives 
\qq \p_t \rho = -\nabla\cdot ( \mathbf{J} +\sqrt{2 M}\mathbf{\Lambda})\,,
\qqq 
with $\mathbf{J}/M = -\nabla \mu$ where ${\bf \Lambda}$ is unit white vectorial noise,
\qq\label{eq:QS-coarse-grained}
\mu &= &\frac{\delta \mathcal{F_{QS}}}{\delta \rho} + \lambda_{QS}(\rho)|\nabla\rho|^2\,,\\
M[\rho] &=& \rho \tau v^2[\rho] \,,
\\
\label{eq:QS-coarse-grained-f}
\mathcal{F}_{QS} &=& \int \left(f_{QS}(\rho(\bfr)) + \frac{K_{QS}(\rho)}{2}|\nabla\rho|^2\right)d\bfr
\qqq
where we have defined
\qq
\label{eq:QSL}
\lambda_{QS}(\rho) &=& \frac{\beta(\rho)}{v_0(\rho)} +\frac{K'(\rho)}{2}\,,\\
\label{eq:QSK}
K_{QS}(\rho) &=& -\frac{\gamma(\rho)}{v_0(\rho)}\,,\\
f_{QS}(\rho) &=& \rho(\log\rho-1) +\int^\rho ds \log v_0(s)\,.
\label{eq:fQS}
\qqq
In these equations we have set $D_T=0$ for simplicity. Linear stability analysis then shows a spinodal instability for densities such that $v_0'(\rho)/v_0(\rho)<-1/\rho$~\cite{Tailleur:08,Cates:15}. This encodes the central MIPS instability: particles accumulate where they are slow, and slow down where they accumulate. 

We refer to the resulting continuum model (restoring $D_T\neq 0$ if desired) as the Quorum Sensing Model (QSM).
Comparing with AMB+  of Sec.~\ref{sec:AMB+}, we note that in QSM: (i) the mobility $M$ is a functional of the particle density (which is nonlocal unless itself Taylor-expanded);  (ii) $\lambda_{QS}$ also depends (locally) on the density; (iii) the local free energy $f_{QS}$ is not polynomial in $\rho$. Such differences are expected because AMB+ expands everything not only in spatial gradients, but in powers of $\phi = \rho-\rho_c$ about the mean-field critical point. More importantly though, despite its additional layer of microscopic detail, the QSM model has no counterpart of the $\zeta$ term in AMB+. 

On the other hand, the $\lambda$ term in \eqref{eq:QSL} shifts the binodals for the QSM as it does for AMB+, and this effect at mean-field level can be calculated along lines presented in Sec.~\ref{subsec:pseudo-variables}~\cite{solon2018generalized,solon2018generalized2}. Notably, because the coarse-grained mobility $M[\rho]$ enters both the current and the noise in a quasi-equilibrium manner (obeying a fluctuation-dissipation relation at some effective noise temperature) it does not influence that binodal calculation. A transformed density $\psi$ and local potential $g$ are again chosen to solve $\p g/\p\psi = f_{QS}' $ and $K_{QS}\p^2\psi /\p\rho^2 +2\lambda_{QS}\p\psi/\p\rho =0$, giving 
\qq\label{eq:QS-psi-g}
\psi= \int_0^\rho d\rho_2 \,\,e^{-\int_0^{\rho_2} d\rho_1 \,2\lambda_{QS}(\rho_1)/K_{QS}(\rho_1)}
\,.
\qqq
The binodals follow numerically from (\ref{eq:mu-flat},\ref{eq:AMB+pseudo-pressure-equality}) via the common-tangent construction on $g(\psi)$; they compare well to those seen in particle simulations of QSPs with reasonable kernels $K$, at high enough density that the mean-field approach should suffice~\cite{solon2018generalized,solon2018generalized2}; Fig.~\ref{fig:particles-coexisting-predictions}.

 \begin{figure}
\begin{centering}
\includegraphics[width=1.\columnwidth]{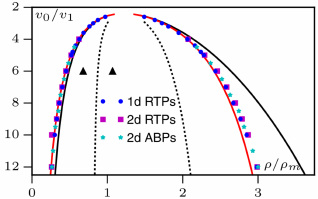}
\par\end{centering}
\caption{Coexisting densities in the QSM with isotropic kernel ($\beta(\rho)=0$) and no translational diffusion ($D_T=0$). The swim speed asymptotes to $v_0$ at low and $v_1$ at high densities; the binodals depend on their ratio and are plotted against normalised density $\rho/\rho_m$. Simulations were performed in the high density regime $\rho_m=200$. Symbols denote the measured binodals in particle-based simulations of QS-RTPs for $d=1,2$ and QS-ABPs for $d=2$. Red lines: mean-field binodals as predicted from $g(\psi)$. Black lines: binodals found from $f_{QS}(\rho)$ by suppressing the detailed-balance breaking term $\lambda_{QS}$. Figure adapted from~\cite{solon2018generalized2}. 
\label{fig:particles-coexisting-predictions}}
\end{figure}

There has been some study of interfacial tensions in the QSM, limited mainly so far to  $\sigma_{\rm ne}$, which alters the binodals (via a Laplace pressure) in the presence of interfacial curvature~\cite{solon2018generalized,solon2018generalized2}.
Note though that, unlike in AMB+, one cannot expect that in QSM the role of $\sigma_{\rm ne}$ in the Ostwald is as simple as (\ref{eq:AMB+Ostwald}); instead, the nontrivial dependence $M[\rho]$ should also enter. Nor were capillary waves in the QSM addressed yet. 

Importantly, the QSM calculation finds $\sigma_{\rm ne} >0$. This suggests that the Ostwald process cannot go into reverse,  so gives only bulk phase separation. This is confirmed in simulations of QSPs~\cite{Cates:15,solon2018generalized} and fully consistent with the absence of any $\zeta$ term within the coarse-grained QSM. Thus, although the QSM is very appealing for its simplicity and bottom-up construction, it captures only part of the new nonequilibrium physics of active phase separation seen in AMB+ and surveyed in Sec.~\ref{sec:AMB+}.

\subsection{Quasi-equilibrium theories}\label{subsec:effective-equilibrium}
Approximate mappings between active phase separation and its equilibrium counterpart have obvious appeal, because the latter case is so well studied. We review these briefly here, although they exclude all of the specifically non-equilibrium phenomenology that is the focus of this Review. There have been two main approaches: coarse-graining of the density, and/or an expansion in small persistence time $\tau$ of self-propelled particles.

The first such mapping was offered in the initial theory of MIPS~\cite{Tailleur:08}, which followed the analysis of Sec.~\ref{subsec:QS} but ignored all gradient terms~\cite{Cates:15}. The result at mean field level is a local free energy density  $f_{QS}(\rho)$ obeying \eqref{eq:fQS} to which standard equilibrium arguments were applied. This theory showed for the first time that liquid-liquid phase separation can arise even in the presence of purely repulsive interactions~\cite{Tailleur:08}. It enables one to locate spinodals (at $f_{QS}''(\rho)=0$) and seemingly also, via a common tangent on $f_{QS}$, the binodals. However, the latter calculation is not correct (see Fig.~\ref{fig:particles-coexisting-predictions}) for reasons explained in Sec.~\ref{subsec:AMB+binodals} above (see \eqref{eq:pressure-eq-equality-AMB+}).

Similarly, it was proposed \cite{Speck2014PRL} that ABPs with pairwise repulsive forces map to Model B at a coarse-grained level, and also predicted that $v_0(\rho)\simeq v_0 - \xi \rho$, with $\xi$ linked to the two-body correlations~\cite{bialke2013microscopic}. Comparison with Sec.~\ref{subsec:QS} shows that the mapping to equilibrium here stems from truncating the gradient expansion before any detailed-balance breaking terms appear.  It is of course better to capture these; see Sec.~\ref{subsec:particles-towards-quantitative-continuum} below.

A related approach, starting from the QSM, is to engineer parameters to restore detailed balance at the coarse-grained level, in effect setting $\lambda_{QS}=0$ in \eqref{eq:QSL}, and then add additional interactions such as a soft repulsion or a competition between repulsion at short distance and attraction at larger distance. The equilibrium mapping was then used to explain the emergence of novel micellar phases and laning states in the resulting model~\cite{o2020lamellar}.

Separately, systematic approximations for the statistics of interacting Active Ornstein-Uhlenbeck particles have been developed, along lines used much earlier in dynamical systems theory to address Langevin equations containing `colored' noise, with a short but finite correlation time $\tau$. (Note that for constant $v$ and $D_T = 0$ in \eqref{eq:QS}, integrating \eqref{eq:AOUP} for AOUPs once with respect to time gives just such a Langevin equation for $\bfr$.) The approaches used are known as the Uniform Coloured Noise Approximation (UCNA)~\cite{haunggi1994colored,cao1993effects,Maggi:15,wittmann2017effective,wittmann2017effectiveII,marconi2015towards} and Fox's theory~\cite{fox1986functional,fox1986uniform,Brader:15}. Although differing at dynamical level, each approach yields an approximate dynamics that restores detailed balance, with the same steady-state probability distribution for particle positions~\cite{Brader:15,rein2016applicability,wittmann2017effective,martin2021statistical}. This is given by (with $D_T=0$)
\qq\label{eq:UCNA-Pss}
P_s(\{\bfr_i\}) = \exp\left(
-\frac{U}{D_{\rm eff}} -\frac{\tau(\nabla_i U)^2}{2D_{\rm eff}}
\right)|\det \mathbb{M}|
\qqq
where $\mathbb{M}_{i,j} = \delta_{ij} +\tau \nabla_i\nabla_j U$, $D_{\rm eff}=v^2\tau$, $v$ is the self-propulsion speed (assumed constant here), and $U$ is the two-body interaction potential. The main effect of finite small $\tau$ is thus to replace this potential by an effective interaction set by the argument of the exponential in \eqref{eq:UCNA-Pss} plus a logarithmic correction from the $|\det \mathbb{M}|$ term. At finite $\tau$, the effective interaction can be attractive at intermediate distances, even if $U$ is purely repulsive, offering a natural quasi-equilibrium explanation of MIPS.

It was however shown, by computing the stationary distribution perturbatively in $\tau$, that only the lowest order in $\tau$ is fully captured by (\ref{eq:UCNA-Pss})~\cite{fodor2016far,martin2021statistical}.  More recently, an extension of these approaches valid to second order in $\tau$ has mapped self-propelled particles to an effective dynamics in which the noise is white in time but non-Gaussian, and detailed balance is explicitly broken~\cite{baek2023markovian}. More generally, quantifying the breakdown of UCNA and Fox's theory beyond the small-$\tau$ limit remains an open problem~\cite{Brader:15,rein2016applicability}. A better understanding might in future allow one to create coarse-grained theories, in the spirit of liquid state theory, using approximations that allow for time-reversal symmetry breaking. A first step in this direction for softly repulsive AOUPs was made in~\cite{li2023towards}. For related approaches based on dynamic density functional theory, see~\cite{bickmann2020predictive,bickmann2020collective,jeggle2020pair,jeggle3d2023,hermann2021phase,hermann2019phase} and Sec.~\ref{subsec:particles-towards-quantitative-continuum} below.

Even in situations where equilibrium mappings give a good account of the stationary probability distribution of particle coordinates (allowing MIPS to be explained in terms of effective attractions), they cannot capture the specific non-equilibrium phenomenology of active phase separation. As shown in Sec.~\ref{sec:AMB+}, this includes circulating currents in steady state, and time-asymmetric correlation functions of the kind arising in bubbly phase separation, where detailed balance is manifestly broken on the meso-scale, in a way disallowed by these mappings.

\subsection{Coarsening kinetics}\label{subsec:AMB+coarsening2}
As described in Sec.~\ref{subsec:AMB+coarsening}, theoretical expectations for phase ordering without momentum conservation favour a quasi-passive, diffusive, $t^{1/3}$ coarsening law whenever the nonequilibrium tension $\sigma_{\rm ne}$ remains positive. In broad agreement with this, 
a  growth exponent close to $1/3$ was observed in particle-based simulations~\cite{stenhammar2013continuum,stenhammar2014phase,shi2020self,caporusso2022dynamics}, although the fitted exponents are often slightly smaller~\cite{stenhammar2013continuum,stenhammar2014phase} (The latter echoes the field-level findings of ~\cite{Wittkowski14,dikshit2024domain,pattanayak2021ordering}.)
For repulsive ABPs, however, it was recently argued that coalescence might in practice modify the Ostwald process at long times, with formation of locally hexatic, fractal clusters also implicated in a complicated scenario that nonetheless gives a  $t^{1/3}$ law~\cite{caporusso2022dynamics}. One reason for this complexity may be that much higher levels of noise are seen in MIPS than in passive phase separation -- noise that becomes diffusive only beyond the persistence length (and time) which exceeds typical interparticle distances. This might make it harder than in passive systems to observe the expected Ostwald regime, where noise scales to zero~\cite{Bray}.

A separate question is whether approximate coarsening laws can be identified for the approach of an active system to a microphase separated state. For example, the particle models studied in~\cite{shi2020self,caporusso2022dynamics} display sufficiently large vapour bubbles in the steady state that the coarsening of these can be studied. Their typical size was shown to increase as $L(t)\sim t^{0.22}$ in~\cite{shi2020self},  similar to the coarsening law  in the relevant regime for AMB+~\cite{fausti2021phase}. The coarsening of hexatic domains in the dense liquid was also studied in~\cite{caporusso2022dynamics}, finding an anomalous $t^{0.13}$ increase of their radius prior to saturation. These power laws for transient coarsening remain unexplained so far. 

So far there are no numerical results reported on the question of hyperuniformity in active particle phase separation as would parallel the work on hyperuniformity at field-theory level that was reviewed in Sec.~\ref{subsec:AMB+coarsening} above.

\subsection{Microphase and bubbly phase separation in particle models}\label{subs:micro-bubbly-particles}
Models of active particles with two-body repulsive interactions have been extensively studied numerically following the seminal works of~\cite{redner2013structure,fily2012athermal,stenhammar2014phase}. These are known to phase separate when the Peclet number $Pe = v_0\tau/\ell$ is sufficiently high, where $\ell$ is an effective particle size set by  the range of the two-body repulsion~\cite{Cates:15}. 
In two-dimensional systems, microphase separation (with a population of bubbles surrounded by liquid) and bubbly phase separation (in which this state coexists with excess vapour) has been observed under varying conditions~\cite{stenhammar2014phase,caporusso2020micro,patch2018curvature,mukhopadhyay2023active,martin2021characterization}; see Fig.~\ref{fig:particles-BPS}. In~\cite{redner2013reentrant}, simulations of {\em attractive} particles were likewise shown to display vapour bubbles in the dense liquid. This suggests that the Ostwald tension $\sigma_B$ for vapour bubbles is negative in these models. However, in three-dimensional simulations of purely repulsive ABPs, no vapour bubbles were reported in the liquid phase. It thus seems possible that the reversal of the Ostwald process depends on specifically two-dimensional physics in the case of repulsive ABPs in which case it is unknown what additional microscopic physics would need to be added to see this in three dimensions~\cite{stenhammar2014phase,omar2021phase,langford2023theory,turci2021phase}.

Many experimental studies of self-propelled colloid show instead cluster phases in which dense droplets exist within a surrounding vapour without growing beyond a certain size~\cite{palacci2013living,Speck:13,van2019interrupted,ginot2018aggregation}. These can be explained if in these systems, whose microscopic interactions are significantly more complicated than any of the computational models, $\sigma_D$ is negative instead. (However we have not seen any reports of coexistence between a cluster phase and a dense fluid.) 

Attempts to characterise the statistical properties of microphase-separated and bubbly phase-separated states in particle models, or understand the link between their reversed apparent Ostwald regimes and microscopic model parameters, remain so far incomplete. We give an overview of what is known here and in Sec.~\ref{subs:attraction} below. 

In the regime where $\sigma_B<0$, AMB+ predicts that the system undergoes microphase separation of vapour bubbles, rather than bulk phase separation, at densities close to but below the liquid binodal, $\phi_{BL}<\phi<\phi_2$ (see Fig.~\ref{fig:phi+-}). For self-propelled repulsive particles, this has been confirmed only recently~\cite{caporusso2020micro,shi2020self}. Caporusso et al.~\cite{caporusso2020micro} performed large-scale off-lattice simulations of self-propelled particles repelling via a short range potential. A uniformly microphase-separated phase was found, comprising vapour bubbles whose size distribution broadens with increasing $Pe$.  On reducing density, an additional bulk vapour phase appeared (giving bubbly phase separation). 
Similar results were reported in~\cite{shi2020self} for a lattice gas model of self-propelled particles with hard-core exclusion; here too, the system is uniformly microphase separated when the global density is close enough to the liquid binodal. (The case an anisotropic mobility was also addressed and found to increase the typical bubble size.) It was observed that the distribution of bubble sizes broadens on moving away from the liquid binodal which, as we have seen in Sec. \ref{subsec:statistical}, matches the AMB+ phenomenology. 

\begin{figure}
\begin{centering}
\includegraphics[width=1\columnwidth]{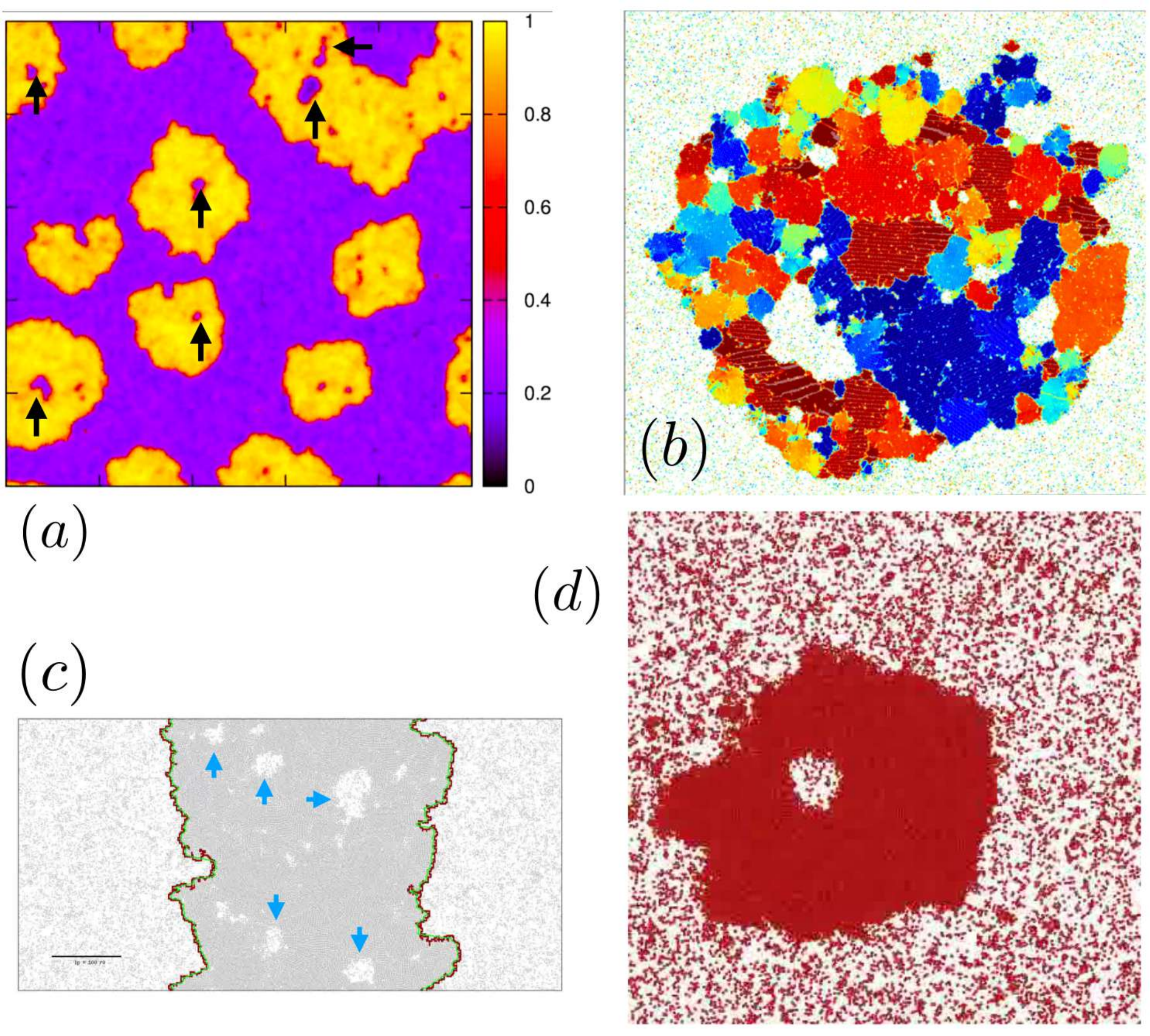}
\par\end{centering}
\caption{Bubbly phase separation in simulations of self-propelled particles. (a,c) Simulations performed with WCA interaction potential for $N\simeq 5\times 10^5$ particles and global density $\rho_0=0.5$, at (a) $Pe=100$  during the coarsening regime (colours indicate the density of particles)~\cite{stenhammar2014phase} and (c) in the steady state in a slab geometry~\cite{patch2018curvature}. Arrows in (a,c) are visual aid for identifying the vapour bubbles in the liquid phase. Panel (b): simulations for the hard-core interaction described in the main text at $Pe=200$, with colours indicating the direction of the hexatic order; vapour bubbles in the dense liquid are clearly visible~\cite{caporusso2020micro}. Panel (d): ABPs with Lennard-Jones potential featuring a short-range hard core and attraction at moderate distance~\cite{redner2013reentrant}. Panels (a), (b, (d) adapted with permission from~\cite{stenhammar2014phase,caporusso2020micro,redner2013reentrant}). Panel (c): Image credit M.C. Marchetti.
\label{fig:particles-BPS} }
\end{figure}

Both of these models show a broad size distribution for the vapour bubbles that becomes independent of system size only for very large systems~\cite{caporusso2020micro,shi2020self}. In practice, convergence with system size is achievable only for global densities up to a few percent below the liquid binodal. Here a decaying power-law distribution $P(A)\sim A^{-1.7}$ for the bubble area $A$ is observed, whose cutoff $A_c$ is independent of system size~\cite{shi2020self,caporusso2020micro}. This echoes the findings of Sec.~\ref{subsec:statistical}: in AMB+, in the regime in which coalescence and nucleation events are frequent, the microphase separated state at $\sigma_{\rm ne}<0$ also displays an intermediate power-law regime with $P(A)\sim A^{-1.6}$, see Fig. \ref{fig:statistical}(c). Intriguingly, the aspherical morphology of the vapour bubbles is also very similar in AMB+ and repulsive ABPs: compare Fig. \ref{fig:particles-microphase}(a) and \ref{fig:statistical}(c).

In~\cite{fausti2023} an approximate theory was introduced for the population dynamics of  vapour bubbles, which should be applicable to both AMB+ and particle-based models. This encodes the fact that reverse Ostwald makes sizes more uniform while coalescence broadens them, so that $P(A)$ is nearly monodisperse when reverse Ostwald dominates but becomes much broader when the coalescence and the Ostwald time scales are comparable~\cite{fausti2023}. Understanding how to control the distribution of bubble sizes (and hence also the rate of the convergence in system size), by varying microscopic parameters, is an open problem. Furthermore, work on particle models is yet  to provided an estimate of $\phi_{BL}$, the density below which repulsive active particles undergo bubbly phase separation rather than microphase separation (see Fig.~\ref{fig:phi+-}).

There have been some preliminary studies on the effect of environmental modulation on MIPS for repulsive ABPs.
First, the role of a periodic external potential (forming a square lattice of period comparable to the particle size) was studied in~\cite{mukhopadhyay2023active} in a setup similar to one that induces crystallization of passive colloids. In the active case, this was shown to induce a transition from a microscopic structure with strong hexatic order to a phase in which particles adapt to the imposed lattice. 
Intriguingly this change of microstructure does not radically alter the mesoscale dynamics of MIPS which continues to show a distribution of voids within the dense phase. This suggests that the reverse Ostwald process is somewhat insensitive to the underlying microstructure  -- as indeed it should be, so long as the tension that controls it remains negative.

\begin{figure}
\begin{centering}
\includegraphics[width=1\columnwidth]{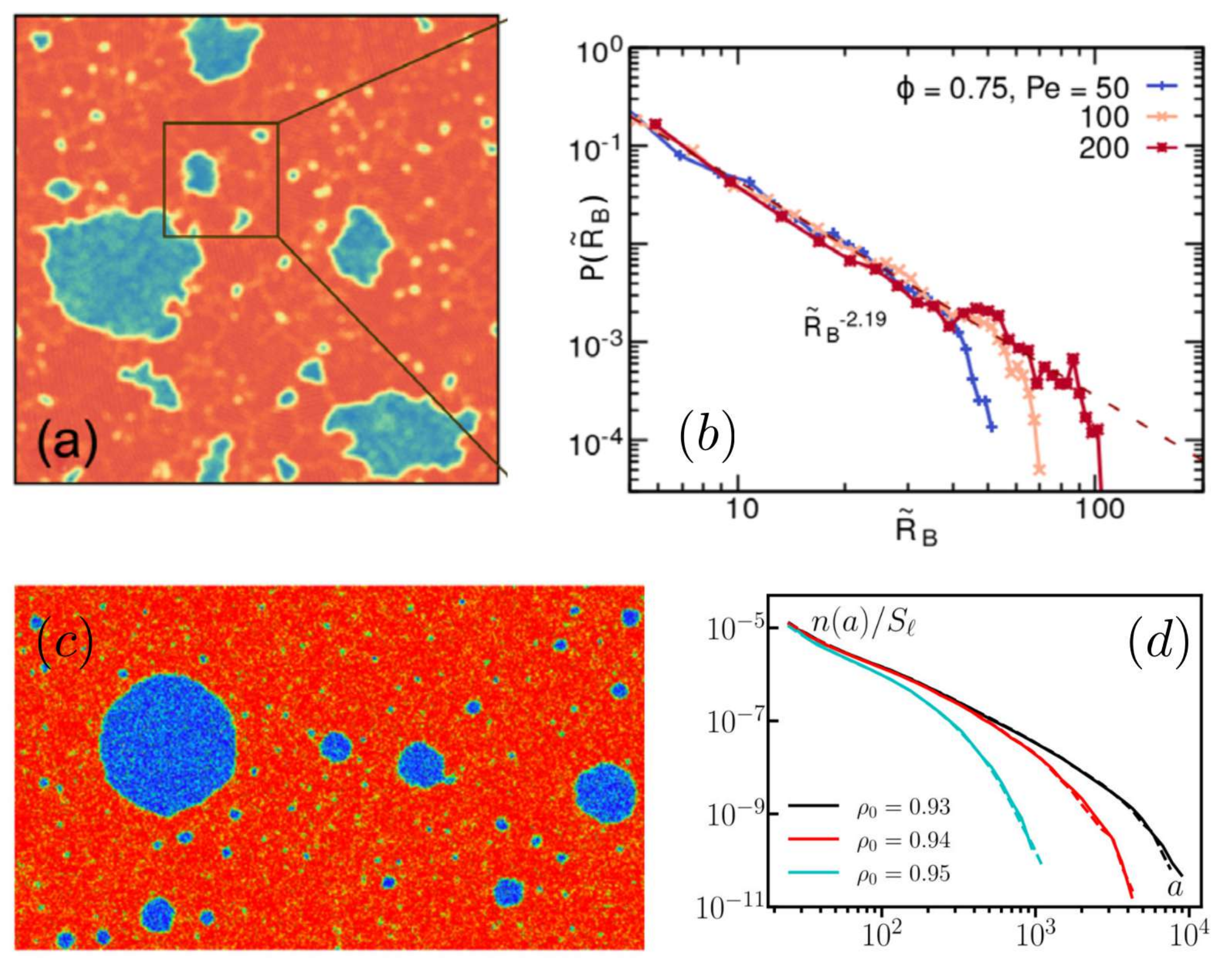}
\par\end{centering}
\caption{Microphase-separated state in simulations of self-propelled repulsive particles showing a population of vapour bubbles within the high density liquid. (a,c) Snapshots in (a) ABPs with hard-core repulsion~\cite{caporusso2020micro} at global density $0.75$ and 
 (c) in a lattice gas model~\cite{shi2020self}. (b,d) Corresponding radii probability distribution $P(R)$ (b) and area probability distribution $P(A)$ (d) of the vapour bubbles as observed in these simulations on varying respectively $Pe$ (b) and density (d). The average size of the bubbles and their variability increases with activity~\cite{caporusso2020micro} and when decreasing the density, in agreement with the predictions of AMB+ in Sec.~\ref{subsec:statistical}. 
Panels (a) and (b) adapted with permission from~\cite{caporusso2020micro}. 
\label{fig:particles-microphase}
\label{fig:particles-microphase}}
\end{figure}

Second, uniaxial anisotropy was addressed in~\cite{nakano2023universal} by adding a potential that orients self-propelled particles along a preferred axis. Both for the particle model and for an anisotropic extension of AMB+, it was found that at sufficiently strong anisotropy the system reverts from reverse to forward Ostwald dynamics~\cite{nakano2023universal}. This may be because, as discussed in Sec.~\ref{subsec:AMB+}, reverse Ostwald requires the $\zeta$ term in AMB+, which is effectively absent in one dimension, whereas strong enough anisotropy should restore quasi 1D behaviour. Anisotropy in the swimming direction was also shown to lead to dense regions with shapes that are programmable via microscopic parameters and contain steady circulating currents~\cite{broker2023orientation}.

\subsection{Role of attractive interactions}\label{subs:attraction}
Attractive pairwise interactions among self-propelled particles can of course be present at the bare level~\cite{redner2013reentrant,alarcon2017morphology,mognetti2013living,ray2023increased}, or arise by integrating out passive sub-colloid degrees of freedom. An example of the latter is the depletion force between particles suspended in a polymer solution~\cite{poon2002physics} whose effect on active particles (bacteria) was studied experimentally~\cite{schwarz2012phase} and theoretically~\cite{harder2014role}. In active systems, attractive interactions can also emerge by integrating out some of the degrees of freedom of the underlying irreversible or dissipative dynamics, including chemical fields~\cite{liebchen2015clustering,liebchen2018synthetic,marsden2014chemotactic}, hydrodynamics~\cite{thutupalli2017boundaries,vskultety2023hydrodynamic,mallory2017self}, and various specifically biological mechanisms~\cite{alston2022intermittent,bonazzi2018intermittent}. 

A natural question is then how activity perturbs the standard phase diagram of attractive colloids. For example, experiments on self-propelled colloids showed clustering at low volume fractions where collision-induced slowdown is too weak to cause MIPS, and particle-particle attraction was suggested as a possible explanation~\cite{Palacci:12,palacci2013living}.

The interplay between attraction and self-propulsion in active phase separation has  been studied computationally both in two~\cite{redner2013reentrant,Brader:15,prymidis2015self,rein2016applicability} and three dimensional systems~\cite{mognetti2013living,prymidis2016vapour}. 
A standard model for passive particles of radius $\simeq r_0$ uses a Lennard-Jones potential $U(r)=4\epsilon \left[(r_0/r)^{12}-(r_0/r)^6\right]$ and shows phase separation at high density for sufficiently large non-dimensional attraction strength $\epsilon/(k_B T)$. 
With self-propulsion present, significantly stronger attraction than in passive fluids is needed to induce phase separation~\cite{schwarz2012phase,redner2013reentrant,mognetti2013living,prymidis2016vapour,Brader:15,ray2023increased}. Interestingly, at low volume fractions, these systems are not bulk phase-separated but show a microphase separated state formed of dense clusters~\cite{redner2013reentrant,mognetti2013living,prymidis2016vapour}. At high volume fraction, particles form a highly dynamic percolating network state~\cite{prymidis2016vapour,redner2013reentrant} which might  relate to the active foam state described in Sec.~\ref{subsec:AMB+capillary}.
Semi-analytical results based on a a quasi-equilibrium approximation (see Sec.~\ref{subsec:effective-equilibrium} above) and use of liquid-state theory predict an effective intermediate-range repulsion to emerge, causing microphase separation at large $Pe$~\cite{Brader:15}. A notable aspect of all these results, alongside earlier ones on MIPS,  is that self-propulsion not only cause attractive interactions among repulsive particles, but also {\em vice versa}.

The impact of activity has been studied computationally also in systems that undergo microphase separation in the absence of activity~\cite{mani2015effect,pohl2014dynamic,tung2016micro}; these comprise particles in which the interparticle potential has a hard core, is attractive at moderate separations, and repulsive at large separations~\cite{sear1999microphase,archer2007phase}.

\begin{figure}
\begin{centering}
\includegraphics[width=1\columnwidth]{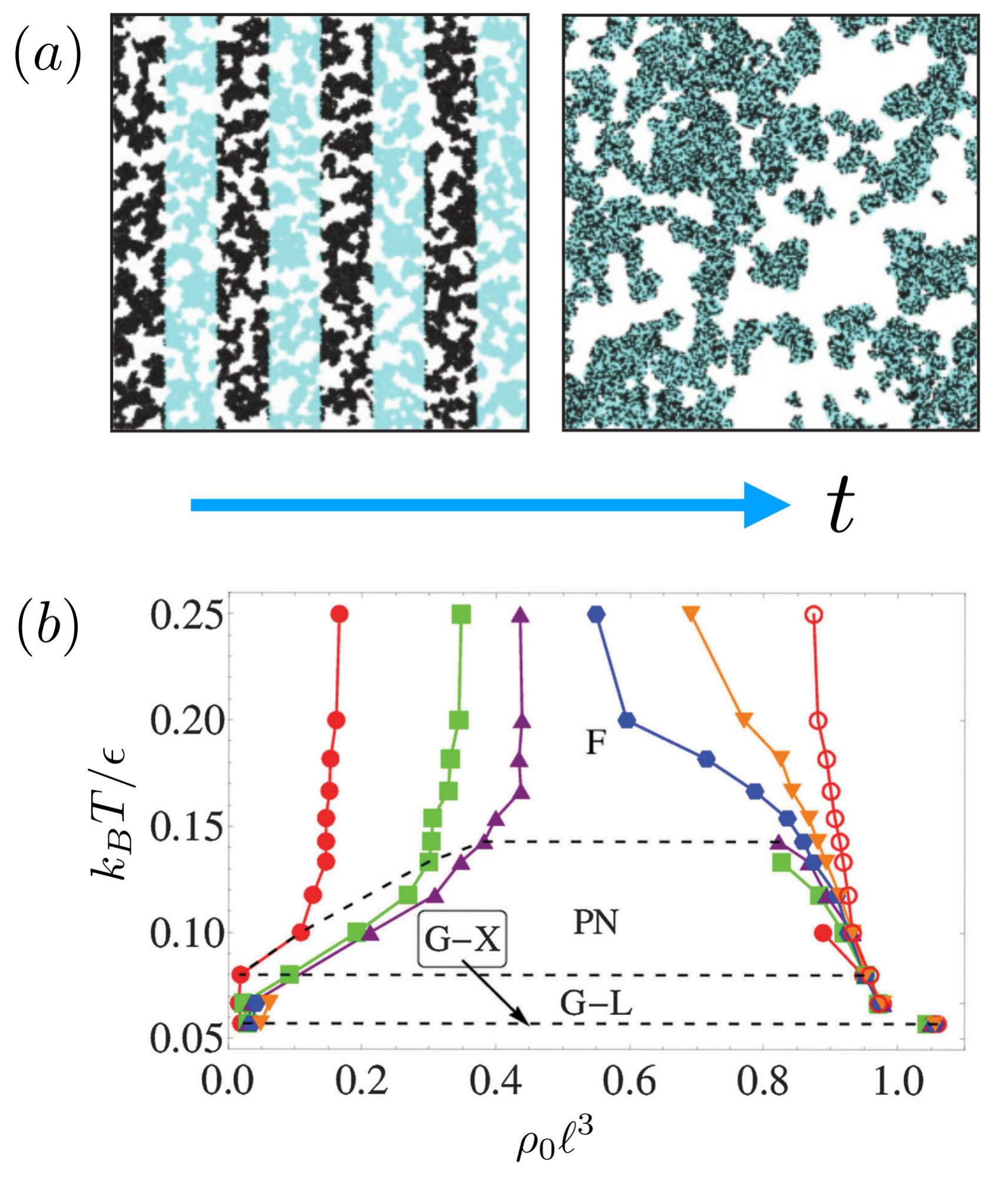}
\par\end{centering}
\caption{Percolating networks of dense particles were observed in simulations of 2D~\cite{redner2013reentrant} and 3D~\cite{prymidis2016vapour} self-propelled particles with attractive (Lennard-Jones) interactions at high volume fraction. Panel (a): Figure adapted with permission from~\cite{redner2013reentrant} showing the time evolution of an initially disordered state at $Pe=40$, global volume fraction $\rho_0r_0^3=0.4$ and attraction strength $\epsilon=40k_BT$ showing the formation of a well mixed percolating cluster (colours denote initial particle positions as shown). Panel (b): Figure adapted with permission from~\cite{prymidis2015self}. Phase diagram at large persistence time of the 3D system as a function of volume fraction and $k_BT/\epsilon$; dots denote the estimated liquid and vapour densities at different volume fractions. G-X, G-L, PN, F respectively denote  vapour-solid coexistence, vapour-liquid coexistence, percolating network, and homogeneous fluid. 
\label{fig:particles-attractive}}
\end{figure}

Self-propulsion is not the only kind of activity that can be combined with attractive interactions so as to break detailed balance in a single component system. For example, a model 
where particles do not self-propel, but are driven out of equilibrium by intermittent attractive interactions, was introduced in~\cite{alston2022intermittent}. This model was inspired by the role of pili attachment in the dynamics of  bacterial colonies~\cite{bonazzi2018intermittent}, but non-reversible attractions may play a wider role in biological systems~\cite{kuan2021continuum}. (A more elaborate continuum description including a nematic tensor for pili  was also developed in~\cite{kuan2021continuum}.)
Such intermittently attractive particles were shown to undergo microphase separation into liquid-like droplets at low global density when the on-off rate of the attractions is slower than particle diffusion. An explicit coarse-graining was used map this microscopic model to AMB+, showing that microphase separation at slow switching corresponds to a reverse Ostwald regime~\cite{alston2022intermittent}.  
 
\subsection{Role of mechanical pressure}\label{sec:pressure}
For active particle systems, the pressure can be defined mechanically as the force per unit area exerted on a wall by the particles. (In the presence of a solvent, this particle contribution is termed the osmotic pressure, and must be distinguished from the total pressure governing fluid flow in, {\em e.g.}, Active Model H.) Unlike in equilibrium, this pressure is generally not a state function, but depends on the wall-particle interactions~\cite{solon2015pressure}.

The pressure is, however, a state function for the specific case in which active particles interact only via mechanical two-body forces, without torques or `alignment interactions'~\cite{solon2015pressureII}. (For this state function to equate to the force density on a wall,  the particle-wall interaction must also be torque-free
~\cite{takatori2014swim,solon2015pressure,solon2015pressureII,takatori2015towards}.) For systems meeting this condition, which excludes quorum sensing and other non-mechanical, many-body interactions, the mechanical pressure is equal in coexisting phases. However the equilibrium Maxwell construction does not hold: this implies that if an `active chemical potential' is constructed from the equation of state using equilibrium arguments~\cite{takatori2015towards}, then it is generally unequal in coexisting phases~\cite{solon2015pressureII}. 

This might seem at odds with our discussion in Sec.~\ref{sec:AMB+} of AMB+, whose local chemical potential $f'(\phi)$ is equal in coexisting phases, with the equal pressure condition replaced by \eqref{eq:AMB+pseudo-pressure-equality} which instead asserts equality of the `pseudo-pressure' $\mu\psi-g$. This paradox is partly caused by a mismatch of nomenclature between particle and continuum levels. In nonequilibrium systems, $\mu\psi-g$ cannot be interpreted as a mechanical pressure; and there is no agreed particle-level definition of chemical potential. 

Nonetheless, it is a good check of any coarse-grained description of active particles that the equilibrium relationships between chemical potential and pressure should be recovered in the limit of vanishing activity. To address this we now discuss the relation between the pseudo-variables and the mechanical pressure within a Landau-Ginzburg description, and the role of the collective mobility in this relationship. 
Consider a model that reduces in the passive limit to Brownian dynamics with pairwise interparticle forces and constant particle mobility $\bar m$. In this case, working in one dimension for simplicity, we note that the particle current obeys not only $J = - M(\rho)\nabla\mu$, with $M=\rho\bar m$ the collective mobility and $\mu$ the chemical potential, but also $J = -\bar m\nabla p$ where $p$ is the (osmotic) pressure. The latter form arises because in a subvolume $V$, the force density is $-\nabla p = V^{-1}\sum_{\alpha\in V}{f}_\alpha$, and each particle $\alpha$ responds with a velocity $\bar m{f}_\alpha$ to  the force ${f}_\alpha$ upon it. 

For Brownian dynamics (setting $k_BT = 1$) one has the free energy density $\mathbb{F}(\rho) = \rho(\ln\rho-1) +f_{\rm int}[\rho]$, where $f_{\rm int}$ is the interaction term.
We ignore for now the gradient terms found by expanding $f_{\rm int}$ for finite-ranged interactions, treating it as a local function $f_{\rm int}(\rho)$. Then we have $\mu = \ln\rho + f_{\rm int}'(\rho)$, and $\nabla\mu = (\rho^{-1}+f_{\rm int}'')\nabla\rho$. Meanwhile, $p = \rho\mu- \mathbb{F}  = \rho -f_{\rm int} +\rho f_{\rm int}'$, giving
\qq
-{J} = M(\rho)\nabla\mu =  \bar m\nabla p =  \bar m(1+\rho f_{\rm int}'')\nabla\rho + ... \label{eq:muPi}
\qqq
with $(...)$ denoting higher gradient terms.  To get from this to Model B, one must expand in $\phi = \rho- \rho_c$ about the critical density $\rho_c$ near which 
$M = \bar m(\rho_c+\phi)$ and
\qq
\mathbb{F} = \mathbb{F}(\rho_c)
 + \frac{a}{2}\phi^2+\frac{b}{4}\phi^4 + ... \nonumber
\qqq
 It is then easy to confirm that $M(\rho)\nabla\mu =  \bar m\nabla p$ still holds when the full mobility $M(\rho) = \bar m(\rho_c+\phi)$ is used, but not when approximating the mobility by a constant, $M(\rho) = M(\rho_c) \equiv M$. While harmless within the chemical potential picture, this standard Model-B approximation breaks the mechanical link to ABPs. Coarse-graining ABP dynamics must therefore gives theories more complicated than AMB+, albeit with broadly equivalent physics; and among these, the relative roles of chemical potential and pressure can vary from one theory to another, depending on the precise coarse-graining strategy.
  
Note also that when a square-gradient term $K(\nabla\phi)^2/2$ is added to $F[\phi]$, $\mu=\delta F/\delta\phi$ acquires a term $-K\nabla^2\phi$ while $p = \mu\phi-\mathbb{F}+K(\nabla\phi)^2$ 
acquires $K((\nabla\phi)^2-\phi\nabla^2\phi)$~\cite{kendon2001inertial}. At this point the statements $\nabla\mu=0$ and $\nabla p=0$ cease to be equivalent, though both remain true in thermal equilibrium. Active terms can violate one or both of these conditions in steady state, as described in Sec.~\ref{subsec:pseudo-variables} and in~\cite{solon2018generalized}.
On the other hand, active particle models are defined through their equations of motion alone, and we are free to choose definitions of $\mu$ and $p$ to suit ourselves, so long as these recover thermodynamic consistency in the zero-activity limit. One such choice is to define $\mu$ such that $\nabla\mu = -J/(\rho \bar m)$ and another choice is to define $p$ such that  $\nabla p= -J/\bar m$. For these choices in turn, $\mu$ and $p$ respectively must be equal among coexisting phases. However, only for equilibrium systems can one expect $\mu$ and $p$ to {\em both} be equal, unless this is chosen as a defining property of at least one of them. (For instance, if $\mu$ is defined such that ${\bf J}/M = -\nabla\mu$, one could use the pseudo-pressure $\mu\psi-g$ to define $p$; but there is no micro-mechanical formula for this $p$.)

\subsection{Momentum conservation, swim pressure, and hydrodynamic interactions}\label{subsec:hydro-particles}\label{sec:pressure2}
One simplification of the ABP paradigm for active particles is that the self propulsive forcing enters as an external force on each particle. This is justifiable only in dry systems. For wet systems, one should consider instead particles whose self propulsion is force-free, moving within a fluid that conserves momentum. The dry and wet cases can however both be addressed using the statistical mechanical concept of `swim pressure'~\cite{takatori2014swim,yan2015swim}. For ABPs with pairwise interaction forces, this can be used to link the pressure (which is a state function in that case as detailed above) to bulk correlation functions, along lines used in liquid state theory ~\cite{solon2015pressureII}. One must then account for an external force density in any region of nonzero alignment (such as near a wall~\cite{solon2015pressure} or at the edge of a phase-separated cluster~\cite{fily2014freezing,redner2013structure}). In contrast, for force-free swimmers there is no such external force density, but in calculating the force on a wall or interface one must allow solvent-mediated momentum fluxes~\cite{omar2020microscopic,yan2018curved}.

The effect of hydrodynamic interactions among active particles, mediated by solvent flow, can be studied at particle level with models that address the local dynamics of swimming. Of these, the most widely studied model is that of so-called squirmers: these are hard-core spherical particles with a specific flow pattern prescribed on their surface~\cite{lighthill1952squirming}. In squirmer suspensions, hydrodynamic interactions were argued to suppress MIPS, because they cause reorientation of particles on collision, so there is no regime where the reorientation time greatly exceeds the time between collisions~\cite{matas2014hydrodynamic,yoshinaga2017hydrodynamic}. Nonetheless, contractile squirmers were found numerically to microphase separate (with a broad distribution of cluster sizes)~\cite{theers2018clustering} -- a result also found also in attractive squirmers~\cite{alarcon2017morphology}.
Meanwhile it was also reported that hydrodynamics can enhance MIPS~\cite{zottl2014hydrodynamics}, but this finding may be due to fluid compressibility in the particular simulation scheme used. More generally, velocity alignment interactions do certainly promote MIPS~\cite{sese2018velocity}, and near-field hydrodynamics can create such interactions for certain types of swimmer~\cite{yoshinaga2017hydrodynamic}.

The natural setting for a continuum model of hydrodynamically interacting particles is Active Model H (AMH); see Section~\ref{subsec:AMH} above. As explained there, this model has a scalar density field with an active stress expanded in density gradients; orientational dynamics is assumed fast compared to density relaxation. Thus there is  no simple way to connect AMH interaction parameters with the local physics of self-propulsion, nor the fact that both near-field and far-field hydrodynamic interactions can rotate other swimmers. In principle, the latter could be partly addressed via models that allow for an additional polar order parameter (see Sec.~\ref{subsec:flocking} below), but these have almost exclusively been used to address the onset of polar ordering (flocking) rather than try to create an improved theory of isotropic MIPS in wet systems.

A somewhat different mechanism for phase separation among contractile swimmers is predicted when particles are interfacially confined~\cite{thutupalli2017boundaries}. Here, out-of-plane fluid motion can create an effective in-plane compressibility, allowing a distinct type of effective interparticle attraction to arise~\cite{vskultety2023hydrodynamic}. This leads to clustering~\cite{bardfalvy2023collective} which may be closer to passive (attraction-mediated) phase separation than to MIPS.   We are not aware of continuum theories for such systems so far.
 
\subsection{Coarse-graining of models with pairwise forces}\label{subsec:particles-towards-quantitative-continuum}
The coarse-graining of the quorum sensing model (QSM) in Sec.~\ref{subsec:QS} was particularly simple because the model lacks hard-core interactions. In contrast, coarse-graining of particles interacting via pairwise forces (including hard-core) is challenging even in the passive limit, where liquid-state theory is needed to find the equation of state (relating pressure to density) in terms of two-body correlations~\cite{hansen2013theory}. Moreover, as explained in Sec.~\ref{sec:pressure}, among active systems  an equation of state only exists in favorable cases.

In one such case, a partial correspondence between pairwise repulsive active Brownian particles and AMB+ has been obtained by a non-equilibrium extension of dynamical density functional theory in~\cite{bickmann2020predictive,bickmann2020collective,teVrugtDDFTinpreparation}. (See~\cite{te2020classical} for a general review of this theory and~\cite{te2023derive} for a self-contained overview focused on active systems.) This approach uses numerical estimates for the pair-distribution function \cite{jeggle2020pair,jeggle3d2023}, and gives equations of motion for the density field at deterministic level. So far it does not predict noise terms, precluding a full correspondence with AMB+. Moreover the current theory predicts spinodals rather than a full phase diagram, and makes no analysis of interfacial tensions. Nonetheless, this approach establishes valuable connections between parameters of a credibly derived continuum theory and the underlying microsopic interactions. A distinct theoretical approach based on a Enskog-like kinetic theory valid at low-density was recently introduced in~\cite{soto2024kinetic} but it is also, so far, limited at predicting the spinodals. 

An alternative route, similar in spirit to dynamical density functional theory, was recently proposed in~\cite{omar2023mechanical,langford2023theory} for active particles interacting solely via repulsive two-body forces, building on results in~\cite{solon2015pressureII}. This starts from the hierarchy of evolution equations for the density $\rho$, polarisation $\bfm= \sum_i \bfu_i \delta(\bfr-\bfr_i)$ and nematic order ${\bf Q}=\sum_i \bfu_i \bfu_j \delta(\bfr-\bfr_i)$. Several approximations are then used. First, it is assumed that the effect of interactions on the evolution of the polarisation and nematic order only results in renormalising the self-propulsion by inducing a local density-dependent swim speed $v_s\to v(\rho(\bfr))$. This leads to the approximations
\qq
\bfm &\simeq& \frac{\tau}{d-1} \nabla \cdot(v_0(\rho) {\bf Q})\\
{\bf Q} &\simeq& \frac{\rho}{d}\boldsymbol{I} -\frac{\tau}{2d(d+2)} \nabla\cdot [v(\rho) \mathbb{I}\cdot \bfm]
\qqq
where $\boldsymbol{I}$ is the rank-four isotropic tensor. It is then assumed that the density is almost stationary, which gives (setting to unity the particle mobility) $\bfm\simeq -v_s^{-1} \nabla\cdot \boldsymbol{\sigma}^C$, where $\boldsymbol{\sigma}^C$ is the standard passive stress tensor generated by inter-particle forces. The latter is in turn approximated as a local, isotropic function of density,  $\boldsymbol{\sigma}^C(\bfr) \simeq - p^C(\rho)\boldsymbol{I}$, with $\boldsymbol{\sigma}$ the unit tensor. 

The resulting theory closely resembles AMB+ but without reliance on a Taylor expansion in $\phi = \rho-\rho_c$. At zeroth order in gradients this approach recovers \eqref{eq:muPi} above, while, in AMB+ notation, the square-gradient and activity coefficients $K,\zeta,\lambda$ (alongside the mobility $M$ and the corresponding noise) become $\rho$-dependent. This approach requires estimation from particle simulation data of both the mean propulsion speed $v(\rho)$, and the pressure contribution $p_C(\rho)$, from which the full (state-function) pressure is found by adding a swim pressure $p_S(\rho)\propto v_s v(\rho)\rho$ as in~\cite{solon2015pressureII}. 

The theory of~\cite{omar2023mechanical,langford2023theory}, as summarized above, was used to obtain quantitative binodals in repulsive ABPs~\cite{omar2023mechanical} and further deployed to create a theory of capillarity that parallels that for AMB+ as reviewed in Secs.~\ref{subsec:AMB+capillary} and \ref{subsec:AMB+roughening} above~\cite{langford2023theory}. (So far the latter predictions have not been extensively tested against particle simulations.) Moreover, the emergence of a  $\zeta$ term promises a potential microscopic explanation of microphase and bubbly phase separation as were observed in repulsive ABP simulations (as noted in Sec.~\ref{subs:micro-bubbly-particles} above). We however note that the $\rho$-dependent expressions for parameters $K,\lambda,\zeta$  do not allow for easily assessing their signs. Moreover, the presence of several uncontrolled approximations (such as the neglect of gradient corrections to the passive stress tensor) means that the domain of validity of this approach remains somewhat unclear. One open question is whether it is can be accurately used under non-stationary conditions such as govern capillary wave dynamics. 

A distinct coarse-graining strategy, bypassing the need for an equation of state, was developed in~\cite{tjhung2018cluster} to address the case of active particles with {\em both} quorum sensing {\em and} repulsive pairwise forces. For weak enough two-body interactions, the approach outlined in Sec.~\ref{subsec:QS} remains valid. In AMB+ language, it was shown that with both interactions, $\zeta$ and $\kappa$ are each nonzero~\cite{tjhung2018cluster}. (This contrasts with the pure QSM of Sec.~\ref{subsec:QS} where $\zeta=0$ was found.) One thus expects that repulsive quorum sensing particles  can display microphase separation, bubbly phase separation, and possibly an active foam phase. 

 \subsection{Interfacial tensions from microscopic models}\label{subsec:particles-tension}
In passive systems, there are various equivalent definitions of the 
interfacial tension. These include the mechanical work per unit area to 
enlarge the interface; the thermodynamic free energy excess per unit 
area relative to the bulk phases with no interface present; and various 
definitions involving specific macroscopic processes, such as the 
amplitudes or relaxation rates of capillary waves, or the 
curvature-dependent fluxes that control the Ostwald process. Definitions 
of the latter type were used in Sec.~\ref{sec:AMB+} for defining 
$\sigma_{\rm B,D}$, $\sigma_{\rm cw}$ and $\sigma_{M}$.

The realization that, unlike in passive systems, different macroscopic 
phenomena are determined by different interfacial tensions (see 
Sec.~\ref{sec:AMB+})~\cite{fausti2021capillary,bialke2015negative} 
negates any debate about which of these is `the' correct definition in 
the active case.
Yet, for each of them, one would like to understand how the tension in 
question depend on microscopic parameters, where possible via direct 
microscopic formulae that allow a value to be computed or measured.

An early proposal~\cite{bialke2015negative,patch2018curvature} for 
defining a mechanical interfacial tension $\sigma_{\rm m}$ in models of 
self-propelled particles with two-body forces was to adopt the 
equilibrium relation of Kirkwood and 
Buff~\cite{kirkwood1949statistical}, which equates $\sigma$ to an 
integral of the stress anisotropy (the difference between parallel and 
perpendicular normal stresses) on passing through the interface:
\qq\label{eq:tension-particles-mech}
\sigma_{\rm m} = \int dx [\Sigma_\parallel(x)-\Sigma_\perp]\,,
\qqq
This was measured to be sharply negative in simulations of repulsive 
ABPs~\cite{bialke2015negative,das2020morphological,patch2018curvature}. 
Furthermore, unlike in passive systems, $\sigma_{\rm m}$ does not 
capture the mechanical work needed to increase interface 
area~\cite{omar2020microscopic,li2023surface}; computing the latter 
directly in numerics gives a positive value~\cite{li2023surface}. 
Another proposal involves subtracting a swim-pressure contribution from 
the stresses in 
\eqref{eq:tension-particles-mech}~\cite{omar2020microscopic,lauersdorf2021phase,yan2015swim,epstein2019statistical,chacon2022intrinsic,paliwal2017non}, 
giving a nearly vanishing outcome~\cite{omar2020microscopic}. Further 
approaches based on power-functional theory ~\cite{hermann2021phase} or 
detailed-balance restoring approximations such as 
UCNA~\cite{marconi2016pressure} have also been considered.

More work is needed to understand which macroscopic phenomena are 
described by these various mechanically defined tensions. It was shown 
that, in ABPs interacting with pairwise 
forces (for which the pressure is a state function), $\sigma_{\rm m}$ in \eqref{eq:tension-particles-mech} captures the 
Laplace pressure~\cite{solon2018generalized2} and it can be measured numerically via 
the force exerted on a solid surface~\cite{zhao2024active}. However, there can be no direct connection between 
$\sigma_{\rm m}$ which is defined in \eqref{eq:tension-particles-mech} 
for a flat interface, and $\sigma_{\rm B,D}$ which depend on the sign of 
interfacial curvature~\cite{speck2021coexistence}, nor between it and 
$\sigma_{\rm cw}$ which have opposite signs because capillary 
fluctuations are 
stable~\cite{bialke2015negative,das2020morphological,patch2018curvature}.

As mentioned in Sec.~\ref{subsec:AMB+Ostwald}, $\sigma_{\rm B,D}$ 
determine the deviations of coexisting densities in a liquid droplet or 
vapour bubbles from the binodal values. The relevant tension in particle 
systems was found for the QSM~\cite{solon2018generalized2} by first 
coarse-graining it to give (\ref{eq:QS-coarse-grained}-\ref{eq:fQS}) and 
then using similar techniques as those employed for 
AMB+~\cite{tjhung2018cluster}. These results were tested against 
numerical simulations with excellent 
agreement~\cite{solon2018generalized2}.
Additionally, using a derivation along lines presented for AMB+ in 
Sec.~\ref{subsec:AMB+nucleation}, it was shown that $\sigma_{\rm B,D}$ 
determines the nucleation dynamics in ABPs~\cite{langford2024mechanics}. 
Under the approximation that interactions only result in a swimming 
speed that locally depends on the density, analogous interfacial 
tensions were proposed for ABPs with repulsive 
interactions~\cite{speck2020collective,speck2021coexistence}.

A tension associated with the width of a fluctuating interface 
(sometimes called `stiffness') has also been analysed in particle 
models, either by fitting the density profile to a known 
function~\cite{paliwal2017non} or by analysing height fluctuations 
directly~\cite{bialke2015negative,patch2018curvature}. Despite 
having opposite signs, $\sigma_{\rm m}$ and $\sigma_{\rm cw}$ were found 
to be proportional to each numerically for repulsive 
ABPs~\cite{paliwal2017non}.
Whether in this case $\sigma_{\rm cw}$  coincides with any of the other 
tensions considered above remains an open question.
A derivation of it, using the theory discussed in 
Sec.~\ref{subsec:particles-towards-quantitative-continuum} was proposed 
in~\cite{langford2023theory}. Numerical verifications of the ensuing 
predictions remain incomplete: they do not definitely confirm the 
expected scaling with wavenumber of relaxation times (see 
\eqref{eq:AMB-effectve-eq-h} and Sec.~\ref{subsec:AMB+roughening}). A 
difficulty is that if the system undergoes bubbly phase separation, the
assumptions underlying the derivation may be invalidated (as already 
discussed in Sec.~\ref{subsec:AMB+roughening}). Second, accessing the 
long length- and time-scales associated with the capillary dynamics 
becomes computationally very expensive~\cite{yue2024scale}.

A further area with many open questions is the influence of activity on 
wetting phenomena, for instance whether tensions can be identified that 
preserve the Young equation relating the three tensions at the 
liquid/liquid/solid interface to the contact 
angle~\cite{zakine2020surface,turci2021wetting}. So far it has been 
argued that the Young equation holds for tensions defined via droplet 
shape fluctuations~\cite{turci2024partial}, whereas it is 
violated~\cite{zhao2024active} for tensions defined via 
\eqref{eq:tension-particles-mech} (with~\cite{zhao2024active} also 
arguing that a contact angle cannot be defined for large system sizes).

\subsection{Links to particle models: further open questions}\label{subsec:particles-open}
In Secs.~\ref{subsec:QS}-\ref{subsec:particles-tension}  we have surveyed the links between  phase separation as modelled in AMB+ with particle-based models of active systems, identifying {\em en route} a number of open questions. Here we add some more.

First, it seems fairly clear (if not indisputable) that several particle models display the reverse Ostwald process, and thus microphase and bubbly phase separation, including the case of repulsive self-propelled particles in two dimensions. However, it remains unclear how to steer among these phases (and find their counterparts involving clusters or droplets rather than bubbles) by tuning the microscopic parameters in particle models. 

For example, several computational studies of repulsive ABPs have shown the presence of vapour bubbles in the dense phase (Sec.~\ref{subs:micro-bubbly-particles}). However, similar simulations for repulsive active dumbbells do not show this~\cite{caporusso2024phase}, perhaps because of the resulting interparticle torques (compare Sec.~\ref{subsec:hydro-particles}). And, while cluster phases are widely seen experimentally~\cite{palacci2013living,Speck:13,van2019interrupted,ginot2018aggregation}, whether multiple system-specific mechanisms causes them, or they can instead be explained within a unified mechanistic picture, is still an open question.


Another open question is whether local order in the dense phase can significantly affect the character of active phase separation. In two dimensions, strong local hexatic order is often observed in the dense phase both in simulations of repulsive active particles~\cite{digregorio2018full,caporusso2020micro,mukhopadhyay2023active,klamser2018thermodynamic} and in experiments on self-propelled colloids~\cite{Palacci:12,Speck:13,van2019interrupted}, although in other experiments, presumably with different interparticle forces, there is no hexatic order even at local level~\cite{zhang2021active}. Hexatic patches of tens or hundreds of particles can be separated by defect lines~\cite{digregorio2018full,caporusso2020micro}, which then act as nucleation sites for vapour bubbles, although a quantification of this effect is missing so far. Meanwhile in three dimensions, a theory of crystallisation for repulsive ABPs was recently proposed by extending the equation of state to depend on a crystallinity order parameter
~\cite{evans2023theory}; it was shown numerically that the dense liquid phase is metastable to the crystalline structure over a large part of the coexistence region~\cite{omar2021phase}. In two-dimensions, a phase-field model to describe dense clusters with crystalline order was recently proposed~\cite{holl2024motility}. 

Another unresolved issue is the role of inertia. While particle-scale inertia is negligible in most mesoscopic situations, understanding its effects remains fundamental to active systems composed of macroscopic particles including large multicellular organisms.  For repulsive particles, inertia causes a reentrant phase diagram, with the homogeneous state returning at  high $Pe$~\cite{su2021inertia,mandal2019motility,omar2023tuning}. This is because collisions are ineffective in slowing down particles at high density once an inertial time-scale, given by the ratio $\tau_m = m/\gamma$ between particle mass $m$ and friction constant $\gamma = k_BT/D_T$,  exceeds the orientational relaxation time $\tau$ (compare \eqref{eq:QS} and \eqref{eq:AOUP}). It was further argued that domain growth is significantly slowed down in the presence of inertia with a $t^{1/5}$ coarsening law reported~\cite{mandal2019motility}. Meanwhile, {\em rotational} (rather than translational) inertia was instead found to favour (rather than suppress) MIPS~\cite{caprini2022role}.

Computational and analytical studies have shown that self-propulsion causes particle-scale velocity correlations that are significantly longer-ranged than in overdamped passive systems~\cite{szamel2021long,caprini2020hidden,keta2023emerging,marconi2021hydrodynamics,marconi2016velocity}. Without changing the ultimate hydrodynamic structure of the theory, such emergent long scales (alongside the persistence time $\tau$ and length  $v_0\tau$~\cite{stenhammar2013continuum}) helps explain why active particle models often show strong density fluctuations and large finite-size effects ~\cite{caporusso2022dynamics,shi2020self}. More generally, the strong density fluctuations within coexisting active phases deserve closer study; it is not clear for example whether these mask, or instead promote, the critical regime (see Sec.~\ref{subsec:critical}).

Most of our discussion has focused on systems in the absence of imposed disorder such as random external fields, nonuniform swim parameters, or randomly placed obstacles to motion. The latter are known to cause long-range modulations of the density field in active systems~\cite{baek2018generic,poncet2021pair}. One interesting implication is that imposing disorder on the boundaries of an active system, but not in its bulk, is enough to destroy macroscopic phase separation at large length-scales~\cite{ro2021disorder}.  There are no studies yet of the impact of disorder on microphase separated and/or active foam states; however since these already have finite correlation lengths, it seems likely that a finite threshold of bulk disorder must be reached before they change in character.

There are many open questions concerning phase separation in systems of nonspherical particles, even when none of the coexisting phases involves long-range orientational order other additional symmetry breaking. In particular, chiral symmetry is often broken at microscopic level: many bacterial species are chiral due to the helical shape of their flagella~\cite{grober2023unconventional}, and in some cases active particles have intrinsic spin. Recent experiments on self-spinning colloids activated by an external magnetic field~\cite{soni2019odd}, and on spinning starfish embryos~\cite{tan2022odd} show rich new phase-separated regimes, and recent theoretical works have started to address such cases~\cite{bickmann2022analytical,ma2022dynamical,caporusso2023phase,semwal2022macro,liao2018clustering,kalz2023field}.

\section{Active phase separation: beyond a single conserved scalar}\label{sec:two-order-parameters}
So far in this Review, we have addressed in detail the non-equilibrium phenomenology arising in the simplest type of active phase separation, involving a single conserved particle density. We included a discussion of momentum conservation (Sec.~\ref{subsec:AMH}), and of hexatic and/or crystalline order in the dense phase (Sec.~\ref{subsec:particles-open}).  
We survey next (at field theory or continuum level)
some further non-equilibrium phenomena arising from active phase separation. These emerge when more than one conserved density is present; when a scalar has both conserved and non-conserved dynamics simultaneously; and when there is a non-conserved orientational order parameter alongside the conserved density.
 
Multiple conserved scalar fields are needed to describe mixtures of particles with different properties so that more than one density can become nonuniform. In computer models, the particles are typically treated in vacuo; in colloidal experiments they would reside in a common solvent. Activity breaks detailed balance as usual, and generically this happens at lower order in the Landau-Ginzburg expansion than for a single density. Specifically, the presence of more than one scalar admits terms representing {\em non-reciprocal interactions} between species, leading often to travelling-wave patterns. We discuss this case in Sec.~\ref{subsec:non-reciprocal}. 

Processes causing the birth and death of active particles, or chemical reactions among the various species present,  instead violate the conservation law(s) for particle density. In combination with active phase separation, the conserved and non-conserved dynamics can move the system in competing directions. (This is not possible in equilibrium where both act to minimize the same free energy.)
Such competition might be at play in the self-organisation of some of the membraneless organelles that shape the inner structure of cells~\cite{hyman2014liquid, berry2018physical, Jacobs2017biophysical, Weber2019review, hyman2011, banani2017, brangwynne2015, nott2015phase}, and also in some types of pattern formation of bacterial colonies~\cite{catesPNAS2010,curatolo2020cooperative}. In Sec.~\ref{sec:mass-non-conservation} we briefly review a minimal model for such situations, involving a single scalar order parameter with antagonistic conserved and nonconserved dynamics. 

For active particles with aligning interactions, such as self propelled rods, polar or nematic liquid-crystalline order is expected at high density. The onset of orientational ordering interacts with active (micro-)phase separation in subtle ways: phase separation in active liquid crystals arguably requires a separate review article. Aiming only to give entry points into the large literature, we briefly discuss these aspects in Secs.~\ref{subsec:flocking} and \ref{subsec:nematics}.

\subsection{Multiple species}\label{subsec:non-reciprocal} 
A mixture of two species (in vacuo or surrounded by a passive solvent)  is described by two order parameters $\phi_{i}$ with $i=A,B$. We assume that momentum is not conserved, so that the minimal description in the passive case is a two-component Model B. For simplicity, following~\cite{saha2020scalar,you2020nonreciprocity}, we include coupling only through the free energy $\mathcal{F}$ rather than via the mobility (which is in general a matrix $M_{ij}$). Setting $M_{ij}=M\delta_{ij}$ we have
\qq
\p_t\phi_i &=& -\nabla \cdot \left( \bfJ_i+\sqrt{2D M}\mathbf{\Lambda}_i \right)\label{eq:-2-components-Model-B}\\
\bfJ_i &=&
-M\nabla\mu_i
=-M\nabla\frac{\delta \mathcal{F}}{\delta \phi_i}\;.\label{eq:-2-components-Model-B-J}
\qqq
The free-energy $\mathcal{F}$ is of two-component Landau-Ginzburg form (not written here). In general such an $\mathcal{F}$ can exhibit two linearly independent instabilities towards phase separation; by a suitable change of coordinates in $\phi_{i}$-space (if needed), one can choose these to be a liquid-vapour mode (dense A+B liquid coexists with dilute A+B vapour) and a demixing mode (A separates from B with the sum of their densities remaining uniform). 

\begin{figure}
\begin{centering}
\includegraphics[width=1\columnwidth]{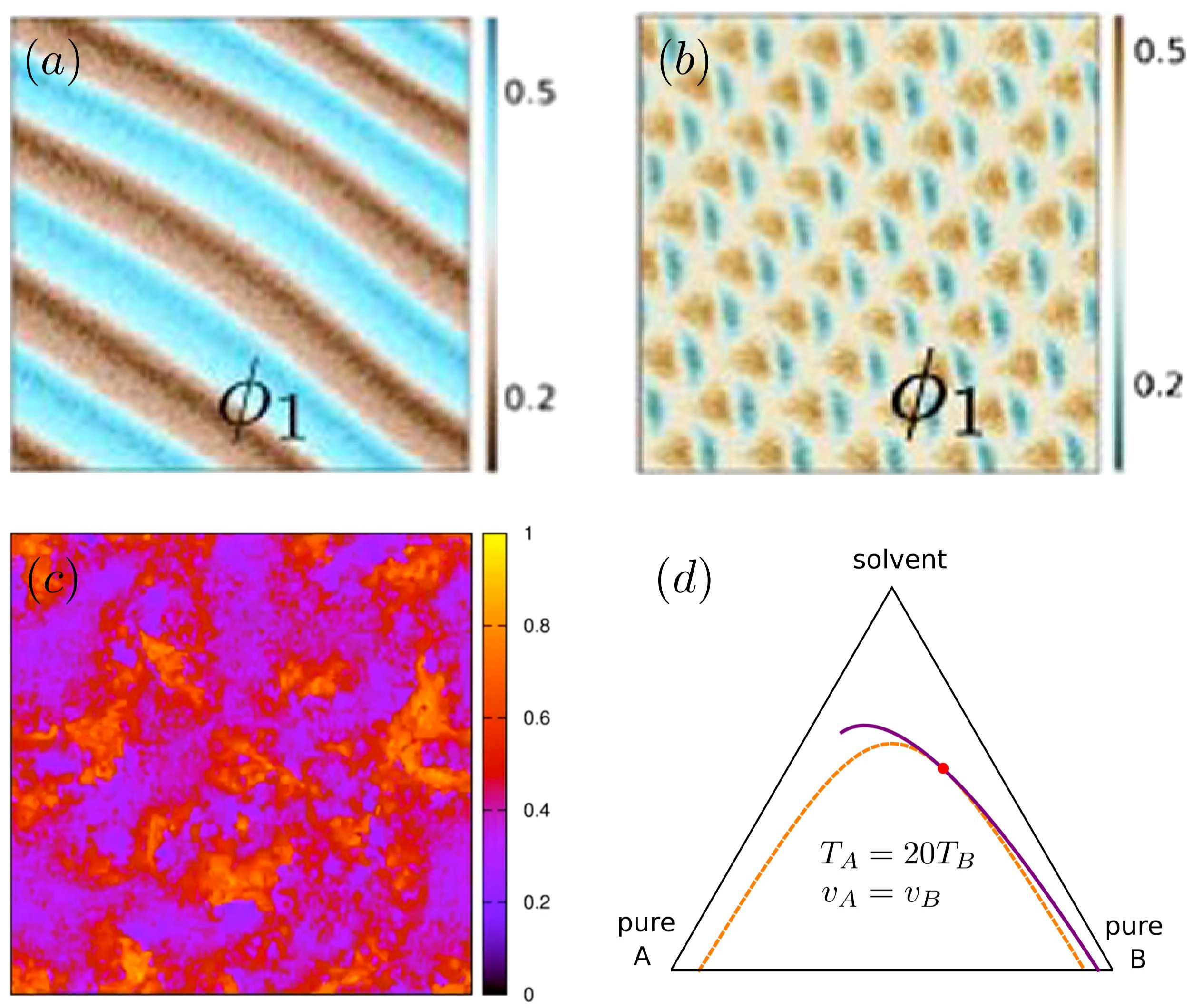}
\par\end{centering}
\caption{Travelling patterns found numerically in the NRCHM; only one of the two densities $\phi_{i=1}$ is shown~\cite{saha2020scalar}. (a) travelling quasi-smectic pattern; (b) another moving microphase-separated state. Figure adapted from~\cite{saha2020scalar}. 
(c) Demixing in mixtures of active and passive particles: the density field of the passive component as obtained from particle simulations of a 50-50 mixture during coarsening from an initially uniform initial condition. Figure adapted from~\cite{stenhammar2015activity}. (d) Phase diagram of a binary mixtures of repulsive Brownian particles with equal size ($v_A=v_B$) and different temperatures $(T_A > T_B)$ as predicted by a generalisation of the virial expansion in the low density limit. Orange line, spinodal; purple line, binodal; red dot, critical point. `Solvent' can equally be vacuum. The fraction of each component at a given point in the triangle is the perpendicular distance from the opposite edge. Figure adapted from~\cite{ilker2020phase}. }
\label{fig:2-Cahn-Hilliard}
\end{figure}

Taking two functional derivatives of $\mathcal{F}$, Schwarz's theorem then implies 
\qq\label{eq:-2-components-Model-B-eq-condition}
\frac{\delta \mu_i(\bfr_1)}{\delta\phi_j(\bfr_2)} = \frac{\delta \mu_j(\bfr_2)}{\delta\phi_i(\bfr_1)}\,
\qqq
for any pair $i,j$, and all pairs of spatial coordinates, $\bfr_1, \bfr_2$. 
This induces a strong dynamical constraint on passive phase separation: the linear operator describing the dynamics close to any homogeneous state is symmetric and therefore has real eigenvalues. Either of the spinodal instabilities mentioned above therefore entails (in the linear regime) a fixed spatial pattern of growing amplitude, rather than (say) the emergence of a travelling wave pattern, as can arise when an eigenvalue is complex.

Although static patterns can be found in multicomponent active systems~\cite{dinelli2022self}, the generic situation is that activity violates \eqref{eq:-2-components-Model-B-eq-condition}. The linear dynamics is then not symmetric, and can have complex eigenvalues at the onset of instability~\cite{saha2020scalar,you2020nonreciprocity,frohoff2023non,frohoff2021suppression}. This leading order effect of activity arises already at order $\nabla^0$ in $\bfJ_i$, in contrast to order $\nabla^3$ for the single component case (see Sec.~\ref{subsec:AMB+}). Indeed, the lowest order violation of \eqref{eq:-2-components-Model-B-eq-condition}, within an expansion in $\phi_i$, is given by setting in \eqref{eq:-2-components-Model-B-J}
\qq
\mu_i = \frac{\delta\mathcal{F}}{\delta \phi_i} +  A_{ij}\phi_j \label{eq:nonrecmu}
\qqq
with $A_{ij}= - A_{ji}$~\cite{saha2020scalar,you2020nonreciprocity}.

The antisymmetric linear term in \eqref{eq:nonrecmu} complements a symmetric one that is, generically, already present in $\delta \mathcal{F}/\delta\phi_i$. It arises, for example, when interactions break Newton's third law (say, species A attracts B but B repels A). More generally, any coupling that breaks detailed balance by violating
\eqref{eq:-2-components-Model-B-eq-condition} can be considered non-reciprocal. Models comprising (\ref{eq:-2-components-Model-B},\ref{eq:-2-components-Model-B-J},\ref{eq:nonrecmu}) are called non-reciprocal Cahn-Hilliard Models or NRCHMs
~\cite{saha2020scalar,you2020nonreciprocity}. Some of them can be viewed as conservative analogues of the complex Ginzburg-Landau equation~\cite{aranson2002world}.

The phase diagram  of NRCHMs include travelling patterns such as self-propelled smectic patterns, moving microphase separated states, and spiral waves as shown in Fig.~\ref{fig:2-Cahn-Hilliard}; their precise topology depends on parameters and spatial dimension (see ~\cite{you2020nonreciprocity} for the $d=1$ case and ~\cite{saha2020scalar,frohoff2023non,dinelli2022self,rana2023defect,duan2023dynamical,brauns2024nonreciprocal} for $d=2$). The general origin of travelling patterns is clear; if A attracts B but B repels A, then B-rich domains will chase A-rich ones which in turn try to evade the pursuit. Full phase separation would represent success for the chased and failure for the chaser, so in many parameter regimes a `fairer' outcome, involving finite-scale patterns, is expected. It is then clear from the chaser/chased dynamics that generically these patterns will be in motion. This however is not the whole story; the parameter space of NRCHMs is quite large, and subject to continuing study.

Arguably the first active matter system to be studied using this type of theory was a mixture of self-propelled and passive particles. (An experimental realisation of this was recently investigated using vibrated rods and passive beads~\cite{kant2024bulk}.) It was shown by explicit construction of a coarse-grained model, along lines described in Sec.~\ref{subsec:particles-towards-quantitative-continuum} above, that there are indeed two spinodal instabilities; a liquid-vapour mode (broadly similar to MIPS) and a demixing mode with a complex eigenvalue~\cite{wittkowski2017nonequilibrium}. The latter causes the onset of travelling waves, which, it was argued, help to explain various dynamical patterns reported earlier in particle-based simulations~\cite{stenhammar2015activity}. These include amoeboid motions of clumps of passive particles being pushed around by flocks of active ones. 

Although foreshadowed in part by this more specific work, the introduction of NRCHMs in~\cite{saha2020scalar,you2020nonreciprocity} has created a powerful framework to address generic features of active mixtures, at a similar level to AMB+ for a single density. This has allowed study of entropy production in travelling wave states~\cite{suchanek2023entropy,alston2023irreversibility}, and also their critical behaviour, which is likely to be ruled by a strong-coupling fixed point~\cite{hanai2020critical,zelle2023universal}. Partial prediction of the binodals has been recently achieved for an NRCHM~\cite{saha2024phase}. The effect of $\mathcal{O}(\nabla^3)$ terms (analogous to those appearing in AMB+) has been investigated in~\cite{chiu2024theory}.

Another type of active mixture is a passively demixing binary fluid to which self-propelled particles are added.  Simulations show the emergence of microphase separation by a simple mechanism~\cite{diaz2023activity}: active particles tend to localise at the interface of fluid droplets, and to point inward. Coarsening of the passive droplets is then arrested when the pressure exerted by active particles balances the passive Laplace pressure of the droplets. This argument allows one to give an estimate of the radius of the passive droplets in terms of the surface tension of the passive fluid and the self-propulsion speed~\cite{diaz2023activity}. 

Self-propelled colloids have a variety of complicated interactions, at least some of which are non-reciprocal at microscopic level and may remain so upon coarse graining. Indeed these systems are predicted and/or seen to show non-stationary patterns of various types such as travelling waves, and self-propelled clusters~\cite{liebchen2015clustering,liebchen2018synthetic,agudo2019active,grauer2020swarm,ouazan2021non,zhao2023chemotactic,saha2022effervescent} (see also~\cite{liebchen2021interactions} for a review). Quincke rollers with heterogeneous self-propulsion~\cite{maity2023spontaneous} or mixtures of self-propelled spherical particles and dumbells~\cite{tung2016micro} also interact non-reciprocally, as do activated oil droplets in micellar
surfactant solutions~\cite{meredith2020predator}. Finally, bacterial suspensions can be genetically engineered to have non-reciprocal interactions and indeed display travelling wave patterns, although so far these were studied experimentally only in the regime where their birth and death crucially affect the outcome~\cite{curatolo2020cooperative}. 

A final example of non-reciprocal particles, closely related to the active-passive mixtures referred to above,  is a mixtures of two species (each individually quasi-passive) whose dynamics is governed by different noise temperatures, such that the $i$'th particle of type $a\in (A,B)$ obeys $\dot{\bfr}_i^a = -\mu^a {\bf f}_i^a + \sqrt{2\mu^a T^a}{\boldsymbol{\eta}}_i^a$, with ${\bf f}_i^a$  the total interaction force and ${\boldsymbol{\eta}}_i^a$ unit white noise. Choosing $T^A\geq T^B$, one may redefine the mobility of the $B$ species as $\tilde{\mu}^B=\mu^B T^B/T^A$ in which case the forces become nonreciprocal. This model might be relevant to mixtures of photo-activated colloidal particles~\cite{weber2016binary}, irradiated alloys~\cite{zarkadoula2016effects}, and the structure of chromatin~\cite{ganai2014chromosome,agrawal2017chromatin,papale2021nanorheology,smrek2017small}. Interestingly, it was shown in numerical and analytical work that such a system will demix into a cold dense region and a hot dilute one, even for purely repulsive particles~\cite{weber2016binary,grosberg2015nonequilibrium,ilker2020phase,smrek2017small,mccarthy2023demixing}.  The analytic results exploit an ansatz for the pair correlation function $g^{ab}(\bfr_i-\bfr_j) \simeq \exp(-U(\bfr_1-\bfr_2)/T^{ab})$, where $T^{ab}=(\mu^a T^a+\mu^b T^b)/(\mu^a+\mu^b)$, which becomes exact at low densities. This ansatz was shown to accurately predict the onset of phase separation in two-temperature mixtures of soft colloids~\cite{mccarthy2023demixing,dumont2024}.

\subsection{Phase separation with a non-conserved density}\label{sec:mass-non-conservation}
In many active systems, the total mass of a given chemical or biological species is not conserved at long times, but the non-conserving dynamics is slow enough that  phase separation remains a dominant feature of the phenomenology. In passive models with any nonconserved dynamics, slow or fast, the final equilibrium state is not phase-separated but instead is the single, homogeneous phase of lowest free energy. We shall see that in the active case, steady states can be more complicated and generically involve microphase separation.

Examples range from bacteria, whose time-scale for birth and death is typically several hours,~\cite{catesPNAS2010,curatolo2020cooperative}, to biomolecular condensates within cells~\cite{hyman2014liquid, berry2018physical, Jacobs2017biophysical, Weber2019review, hyman2011, banani2017, brangwynne2015, nott2015phase, jacobs2023theory}, where chemical species have a turnover time set by cellular metabolism. As noted in Sec.~\ref{sec:intro}, membrane-less organelles are often viewed as phase separated liquid-liquid mixtures and some authors have argued that chemical reactions are crucial for their phenomenology~\cite{Weber2019review,zwicker2017growth}.  Another class of active systems in which phase separation arises without long-term mass conservation is compartmentalisation in biological tissues. Indeed, models similar to the one described below arise in the description of tumors, although these models often explicitly take into account fluid flows and nutrient uptake~\cite{hoshino2019pattern,chatelain2011emergence,wise2008three}. Models of (micro-)phase separation have also been used to describe ecological systems such as (say) forest formation. Here the diffusive scalar underlying phase separation can be a nutrient density or similar underlying quantity rather than the density of (say) trees~\cite{liu2016phase}.

\begin{figure}
\begin{centering}
\includegraphics[width=1\columnwidth]{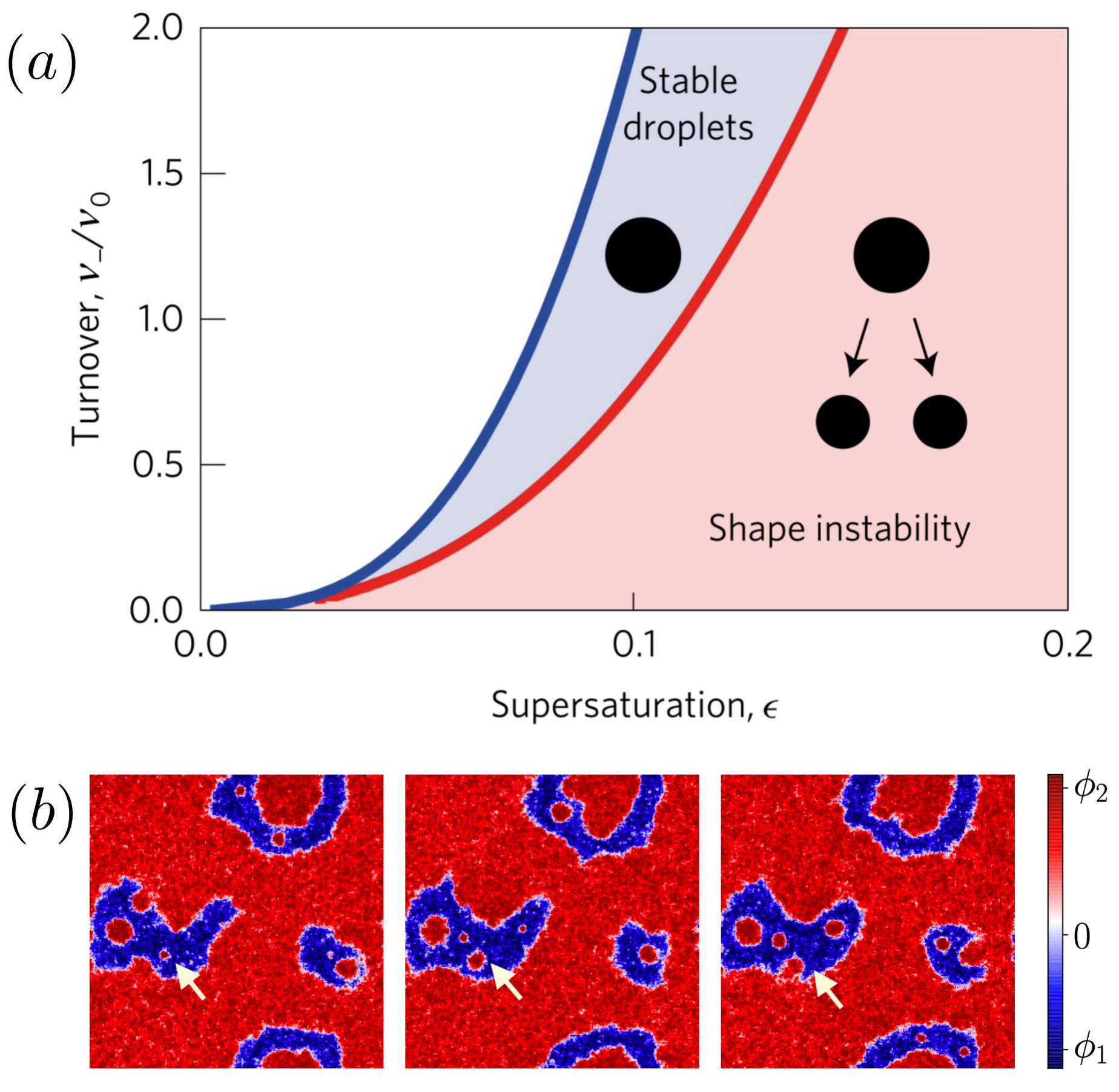}
\par\end{centering}
\caption{Panel (a): Phase diagram of a model of phase separation with chemical reactions (similar but not identical to Model AB+) showing regions where liquid droplets undergo the reversed Ostwald process due to chemical reactions (blue) and undergo a shape instability (red). Figure adapted with permission from~\cite{zwicker2017growth}. In both regimes, even for $\lambda=\zeta=0$, the system is microphase separated. Panel (b): Temporal evolution of Model AB+ within the regime of bubbly microphase separation. Figure adapted from~\cite{li2021hierarchical}}
\label{fig:Model-AB}
\end{figure}

In what follows, for the non-conservative part of the dynamics, we use the language of chemical reactions interchangeably with that of birth and death~\cite{catesPNAS2010,grafke2017spatiotemporal}. 
The emergence of stationary microphase-separated states in chemically reacting binary fluids was first predicted in the 1990s~\cite{glotzer1994monte,glotzer1995reaction,glotzer1994self,motoyama1996morphology,motoyama1997morphology,puri1994segregation,puri1998phase}. It was however soon clarified that the models employed break detailed balance, whereas in equilibrium, as mentioned above, the steady state is always uniform~\cite{lefever1995comment}. These investigations were revived more recently for describing condensates~\cite{hyman2014liquid, berry2018physical, Jacobs2017biophysical, Weber2019review, hyman2011, banani2017, brangwynne2015} and bacterial colonies~\cite{catesPNAS2010,curatolo2020cooperative}. Here we discuss a minimal version of the models employed for describing these situations, termed Model AB+~\cite{li2020non,li2021hierarchical}.

Model AB+ augments AMB+ with leading-order terms to break $\phi$ conservation. It replaces \eqref{eq:AMB+} with
\qq
\partial_t \phi &=& - \nabla \cdot \mathbf{J}- M_{\rm A} \mu_{\rm A} + \sqrt{2 D} \tilde\Lambda
\label{eq:phi_dot_ab}\\
\mu_{\rm A} &=& u (\phi-  \phi_{a})  (\phi - \phi_{t}) \equiv
- \frac{\delta \mathcal{F}_A}{\delta\phi}
\label{eq:AB-muA}
\qqq
where $\bfJ$ obeys \eqref{eq:AMB+J}. This dynamics is the sum of conserved (AMB+) and  nonconserved (Model A-type) contributions. 
The noise in \eqref{eq:phi_dot_ab} combines the noises from these two dynamical sectors; the resulting correlator is
\qq
\langle \tilde\Lambda(\bfr_1,t_1) \tilde\Lambda(\bfr_2,t_2) \rangle 
= 
(M_A -M \nabla^2)\delta(\bfr_1-\bfr_2)\delta(t_1-t_2)\,.\nonumber
\qqq
The nonconserved dynamics is governed by $\mathcal{F}_A = \int f_A(\phi) d\bfr$ where $f_A$ is a polynomial (gradient terms could be added but are inessential). In \eqref{eq:AB-muA}, $\phi_t$ is the stable fixed point or `target density' for this dynamics: which, on its own,
would relax the system to a uniform state of $\phi = \phi_t$.
Meanwhile $\phi_a<\phi_t$ is an unstable fixed point setting the minimum value that $\phi$ can attain. (Choosing $\phi_a = -\rho_c$ then imposes that the physical $\rho = \rho_c+\phi$ stays positive.)  
A partial derivation of Model AB+ by coarse-graining a microscopic bacterial colony model is presented in~\cite{grafke2017spatiotemporal,li2020non}, while the interpretation of continuum model parameters in terms of microscopic ones for biophysical condensates can be found in~\cite{Weber2019review,zwicker2017growth}.

Detailed balance is broken in Model AB+ in two separate ways. One is inherited from AMB+, whose active nonlinearities $\lambda$ and $\zeta$ were described in Sec.~\ref{subsec:AMB+}. 
Second, even when these terms are set to zero,  $\mu_A = \delta\mathcal{F}_A/\delta\phi \neq \delta\mathcal{F}/\delta\phi$, so that the `equilibrium' chemical potentials in the conserved and nonconserved sectors do not match. Then one has a model (called Model AB~\cite{li2020non}) whose detailed balance breaking resides solely in the fact that the
two sectors are governed by different free energy landscapes. This is the lowest-order way in which detailed balance can be broken; in this sense the $\lambda$ and $\zeta$ terms are subdominant (though they do give additional effects, noted below).

For a special subset of parameters, Model AB+ has been shown to map exactly onto an equilibrium model with long-range attractions~\cite{li2020non,ziethen2022nucleation}. Crucially, the mapping holds for the stationary probability measure of the density $\phi$, but not of its conservative current ${\bf J}$, which is generically nonzero at stationarity, creating a finite entropy production~\cite{li2020non,li2021steady}. 

The interesting regime of Model AB+ has $\phi_1<\phi_t<\phi_2$ so that the target density of the nonconserved dynamics lies within the miscibility gap of the conserved dynamics. This rules out both a uniformly homogeneous state (which would phase separate) and bulk phase separation (whose coexisting densities would each be moving torwards $\phi_t$). The result can include nonstationary states such as limit cycles~\cite{grafke2017spatiotemporal,li2020non}, but Model AB+ also supports stationary microphase separation. In birth/death language, liquid regions ($\phi_2 > \phi_t$) form in which particles continually die off while in nearby vapour regions ($\phi_1 < \phi_t$) they are continually being born. A steady conservative flux from vapour to liquid maintains the state, whose length-scale is large so long as the chemical reactions are slow. For $\phi_t$ close to zero the steady state is lamellar but for strongly asymmetric situations it involves droplets of one phase in the other, as first seen in a more detailed model of bacterial colonies~\cite{catesPNAS2010}. 

Physically, the situation somewhat resembles the task of sweeping snow off a parking lot while it is still snowing. Any attempt to sweep across large distances into one big pile (analogous to bulk phase separation) allows too much snowfall between successive sweeps, and in steady state the lot is almost covered in snow. However it is possible to keep the lot almost clear by continually sweeping snow over shorter distances, into a number of smaller `microphase-separated' piles distributed across it. The analogy is of course imperfect: in Model AB+ the option of a `uniformly snow-covered lot' does not really exist, because when its density moves above $\phi_s$ it will undergo spinodal decomposition.

The emergent length-scale in this type of steady state, and the regime in which it arises, can be obtained from dimensional analysis of (\ref{eq:phi_dot_ab}) and (\ref{eq:AB-muA}). For phase separation to arise, it is clear that the conservative part of the dynamics must be fast with respect to chemical reactions; this happens at scales smaller than $\ell_s=\sqrt{M/M_A}$. This scale criterion is satisfied in typical subcellular conditions where relatively rapid thermal diffusion of chemical species is accompanied by a relatively slow reaction-driven turnover time. 

Experiments on bacteria often initialize the system with a single dense droplet rather than studying the instabilities of an initially uniform bacterial carpet. Here concentric rings can be formed at early times which then themselves may split into separate droplets and lead to a microphase-separated stationary pattern~\cite{shapiro1995significances,budrene1991complex,woodward1995spatio,ben2000cooperative,curatolo2020cooperative}.
While historically most models of these patterns have assumed specific chemotactic interactions, they can also be explained using Model AB+ and/or closely related coarse-grained models~\cite{li2020non,catesPNAS2010,curatolo2020cooperative,tyson1999minimal,brenner1998physical}. 

Model AB+ supports three distinct mechanisms that can each cause microphase separation. One is the conservative reverse Ostwald process of AMB+, discussed in Sec. \ref{subsec:AMB+Ostwald} above. Two more stem from chemical reactions (and are present even when $\lambda=\zeta=0$). The first is that without detailed balance, chemical reactions can separately drive the Ostwald process into reverse hence stabilising finite-size droplets~\cite{zwicker2015suppression,li2021hierarchical}. The second is that such reactions can destabilize large spherical droplets, causing these to split~\cite{zwicker2017growth,li2021hierarchical}, see Fig.~\ref{fig:Model-AB}. The precise interplay between these mechanisms is yet to be fully explored, and it remains uncertain which, if any, are relevant to the emergence of membraneless organelles, of self-limiting size, within the cell interior~\cite{hyman2014liquid, berry2018physical, Jacobs2017biophysical, Weber2019review, hyman2011, banani2017, brangwynne2015}.

An intriguing new phenomenology arises by when microphase separation is favoured {\em both} by the conservative and by the non-conservative parts of the Model AB+ dynamics. Specifically, if one sets $\phi_t$ close to the liquid binodal (so that the vapour undergoes microphase separation), while the AMB+ part of the dynamics displays reversed Ostwald process, the system settles in a new kind of ``bubbly microphase separated'' state~\cite{li2021hierarchical}, see Fig.~\ref{fig:Model-AB}.
Recall that bubbly phase separation in AMB+ comprises a single domain of liquid containing vapour bubbles, coexisting with excess vapour. In bubbly microphase separation this motif repeats locally across numerous separate liquid domains; the organization of these domains itself is a microphase separation, but at a larger scale. The resulting hierarchical structure, which may or may not be of biological interest, stems from the presence of {\em two distinct} microphase-separation mechanisms that act at different length scales; one (reverse Ostwald) is intrinsic to the conservative dynamics, and the other is controlled by the competition between conservative currents and birth/death dynamics~\cite{li2021hierarchical}.

The full phase diagram of Model AB+ is yet to be investigated and there is every likelihood of finding further new phenomenology, perhaps related to additional surface tensions~\cite{cho2023nonequilibrium}. For example, it would be interesting to see what is the interplay between reverse Ostwald dynamics, and the limit cycles~\cite{grafke2017spatiotemporal,li2020non} found in Model AB+. Furthermore, most of the studies of how chemical reactions affect active phase separation were so far conducted at mean-field level. The noise strength is known to play an important role in determining the distribution of cluster sizes in AMB+ and it is safe to speculate that this is true in Model AB+ as well. Whether noise might control the broad size distributions observed experimentally in (some types of) membraneless organelles is an open question~\cite{nott2015phase}. Noise is obviously crucial also in driving nucleation~\cite{Jacobs2022}. It was recently shown experimentally~\cite{shimobayashi2021nucleation}, and explained theoretically~\cite{ziethen2022nucleation}, that nucleation is facilitated by active chemical reactions. However a full  theory of reaction-mediated nucleation rates, analogous to Classical Nucleation Theory for AMB+ (Sec.~\ref{subsec:AMB+nucleation} above), remains to be developed.

\subsection{Phase separation in flocking}\label{subsec:flocking}

Active particles generically have either a dipolar or quadrupolar axis attached to each particle. For self-propelled particles this axis is typically the swimming direction; an interaction that tends to align the swimming axis of neighbouring particles is capable of creating a flock that moves persistently in one direction. Flocking is indeed one of the first proposed, and most studied, phenomena of active matter~\cite{vicsek1995novel,toner1995long,chate2019dry}. Although some flocking models impose a homogeneous spatial density~\cite{chen2015critical}, in the compressible cases more normally studied, density fronts resembling shock-waves can form close to the flocking transition~\cite{gregoire2004onset,caussin2014emergent,chate2019dry}. The result is a microphase-separated state: regions of disordered vapour coexist with  domains of flocked particles.  The latter are in motion while the vapour is not, so that particles switch between phases as patterns move; the moving flocks often have marked fore-aft asymmetry. These flocks can organize in several ways including a lamellar state~\cite{Solon:2015:PRL} and a so-called `cross-sea' phase~\cite{kursten2020dry}. 
Such microphase-separated regimes arise in systems having short-range interactions and spontaneous breaking of a continuous rotational symmetry~\cite{Solon:2015:PRL}; see Fig.~\ref{fig:flocking}. On the other hand, bulk phase separation can be found when the order parameter has discrete rotational symmetry (in lattice models) instead~\cite{solon2015flocking,solon2022flying}. In both cases, on further increasing the alignment interactions, a spatially uniform polar flock emerges.

\begin{figure}
\begin{centering}
\includegraphics[width=1\columnwidth]{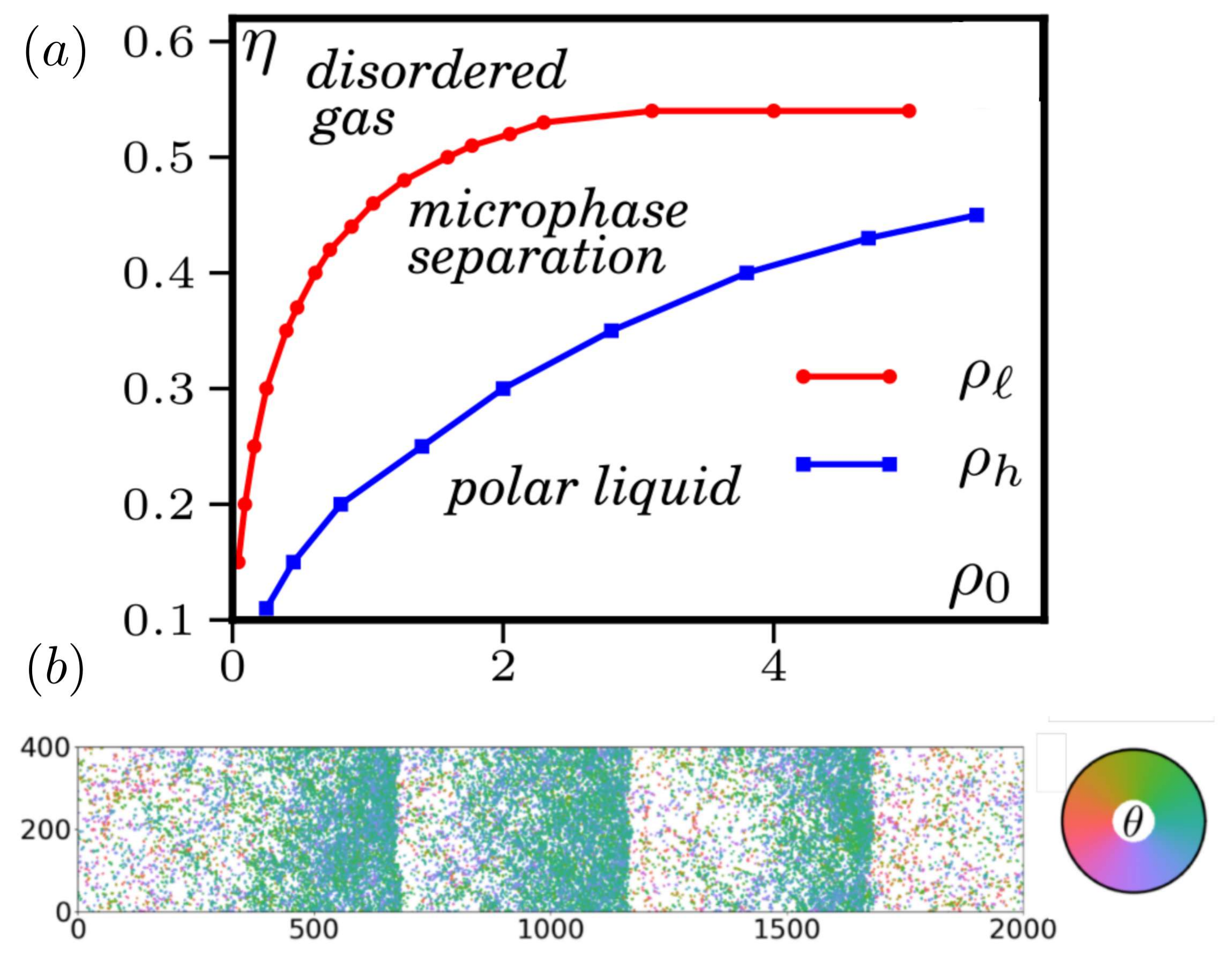}
\par\end{centering}
\caption{Panel (a): Phase diagram of systems in which a transition to polar order (breaking continuous rotational symmetry) is coupled to a density field. 
When the polar order has discrete symmetry, microphase separation is replaced by bulk phase separation~\cite{solon2022flying}. Figure adapted with permission from~\cite{Solon:2015:PRL}. 
Panel (b): Microphase separated state in a particle model of flocking with topological interactions~\cite{martin2021fluctuation}. }
\label{fig:flocking}
\end{figure}

A simple analytical argument~\cite{martin2021fluctuation,solon2015flocking,martin2024fluctuation}, reviewed below, demonstrates that phase separation should generically emerge close to the flocking transition, without directly explaining why this usually takes the form of microphase- rather than bulk phase separation. Moreover, multiple solutions of travelling-wave type were found at mean-field level whose selection mechanism remains unknown~\cite{bertin2009hydrodynamic,caussin2014emergent,ihle2013invasion}. A further complication is that, at least in systems with a discrete rotational order parameter, rare nucleation of differently oriented sub-flocks destroys long range order in the polar liquid phase ~\cite{benvegnen2023metastability}. It is not yet known if the same applies for continuous rotational symmetry; if so the phase diagram in Fig.~\ref{fig:flocking} holds only up to a certain system size.

The minimal model for describing active systems with polar order is that of Toner-Tu theory~\cite{toner1995long,toner2005hydrodynamics}. As with AMB+, it is helpful to gather all terms compatible with a free energy structure into a functional $\mathcal{F}_{\bfp}$; doing so, one finds~\cite{Marchetti2013RMP}:
\qq\label{eq:TT}
&&\p_t \rho + v_0\nabla\cdot (\rho \bfp)= -\nabla\cdot \left(
-M \frac{\delta \mathcal{F}_\bfp}{\delta\rho} +\mathbf{\Lambda}_\rho
\right)\\
&&\p_t \bfp 
+\lambda_1 \bfp\cdot \nabla \bfp
+\lambda_2 \bfp \nabla\cdot \bfp 
=-M_\bfp \frac{\delta \mathcal{F}_\bfp}{\delta\bfp} + \mathbf{\Lambda}_\bfp\;,
\qqq
where $\mathbf{\Lambda}_\rho$, $\mathbf{\Lambda}_\bfp$ the associated noises.
The effective free energy $\mathcal{F}_{\bfp}$ is a standard one for compressible polar liquid
 (writing $\phi=\rho-\rho_0$): 
  \qq \label{eq:effTTF}
\mathcal{F}_\bfp&=&\int d\bfr 
\left(-\frac{\alpha(\rho)}{2}+\frac{\beta}{4}|\bfp|^2\right)|\bfp|^2
+\frac{K}{2}(\p_\alpha p_\beta)(\p_\alpha p_\beta)\nonumber\\
&+&\frac{w_0}{2}|\bfp|^2\nabla\cdot\bfp -w_1 \frac{\phi}{\rho_0} \nabla\cdot \bfp 
+\frac{A}{2}\left(\frac{\phi}{\rho_0}\right)^2
\qqq

Toner-Tu theory famously predicts the emergence of true long-range polar order even in two-dimensional systems where in equilibrium it is forbidden~\cite{toner1995long}. Although originally derived from symmetry arguments and gradient expansion, \eqref{eq:TT} was also obtained by coarse-graining microscopic models~\cite{bertin2006boltzmann,bertin2009hydrodynamic,ihle2011kinetic,chate2019dry}, although uncertainty remains over the noise terms~\cite{bertin2013mesoscopic,feliachi2022fluctuating}. 

 At mean-field level, with $\rho$-independent $\alpha$, the theory predicts a continuous transition from a disordered fluid at $\alpha<0$, to a flocked state (with $\langle \bfp\rangle\neq0$) for $\alpha>0$. When $\alpha$ depends on the density $\rho$ instead, the transition generically becomes first order because there is a range of $\alpha(\rho_0)$ and global density $\rho_0$ where the uniform states with and without polar order are both linearly unstable~\cite{mishra2010fluctuations,gopinath2012dynamical,bertin2009hydrodynamic,peshkov2012nonlinear,solon2015flocking,chate2019dry}. Moreover such dependence can be generically expected, as it is not forbidden by symmetry or related arguments~\cite{toner2005hydrodynamics}. Indeed, a $\rho$-dependent $\alpha$ is often found when obtaining \eqref{eq:TT} from coarse-graining of microscopic models~\cite{bertin2006boltzmann,mishra2010fluctuations,solon2013revisiting,bricard2013emergence,bertin2009hydrodynamic,ihle2016chapman}. Moreover, it was shown that fluctuations reinstate a $\rho$-dependence in $\alpha$ even for models in that lack such dependence at mean-field level~\cite{solon2015flocking,martin2021fluctuation,martin2024fluctuation}. To summarize then, just as in passive systems, a first order orientational transition coupled to a density generically leads to a miscibility gap and hence a phase separation~\cite{vroege1992phase}. 
 
The arguments summarised above exclude the case of mean-field and other long-ranged interactions~\cite{martin2021fluctuation,martin2024fluctuation}, including when mean-field-like models are obtained by judicious coarse-graining of microscopic interactions~\cite{agranov2024thermodynamically}. (There may also be an exception for active particles with broadly-enough distributed L\'evy flight dynamics ~\cite{cairoli2019hydrodynamics}.) In such cases the flocking transition is second order without phase coexistence, as also arises in cases where the density is decoupled from the aligning mechanism~\cite{martin2021fluctuation,chen2015critical}. 
 Models in which particles align with a set of neighbours defined topologically, rather than those within a fixed interaction range, were initially believed to also display a second-order flocking transition ~\cite{peshkov2012continuous,ginelli2010relevance,chou2012kinetic}. It was however recently shown that, at least when the number of topological neighbours is fixed, the transition is first order. Here a density-dependent $\alpha$ is absent at mean-field level but reinstated by fluctuations~\cite{martin2021fluctuation,martin2024fluctuation}. 

Above we focused on phase separation between a dilute disordered fluid and dense polar domains. Several other phase separation mechanisms are however possible~\cite{bertrand2022diversity,jentsch2023critical}: these can lead to coexistence of two ordered phases of different density~\cite{miller2024phase}; or even coexistence of  a dense disordered phase and a dilute ordered one. The latter case was recently investigated experimentally~\cite{geyer2019freezing} and theoretically~\cite{schnyder2017collective,geyer2019freezing,bertrand2022diversity}.

Another distinct phase separation mechanism is when interparticle attractions, quorum-sensing~\cite{degond2016self}, or other alignment-independent terms create a miscibility gap via a double welled structure in the density sector of $\mathcal{F}_\bfp$. This can be modelled by replacing the final term in \eqref{eq:effTTF} with a Model B-like free energy density. Hydrodynamics can also be incorporated by using a Model H structure in the same way. In such cases, rather than the ordering transition driving the phase separation as in Fig.~\ref{fig:flocking},  the reverse holds. Since the polar order is essentially a spectator in this case, it is somewhat less interesting from a strict phase-separation perspective. However, under conditions where a dense polar droplet coexists with a surrounding isotropic fluid, 
such models have proven useful as a minimal description of motile cells ~\cite{tjhung2012spontaneous,tjhung2015minimal} and (for the multi-droplet case) of active emulsions~\cite{carenza2020multiscale}.

\subsection{Phase separation in active nematics}\label{subsec:nematics}

Nematic ordering commonly arises in rod-shaped particles whose interactions favour a common orientation of the rod axis without regard for which end of the rod is which. Even in self-propelled particles (which do have a front and a back), it can be the dominant effect, giving orientational order described not by a vector ${\bf p}$ but by a traceless symmetric second rank tensor ${\bf Q}$. For passive systems in 3D, the isotropic-nematic transition is generically first order, and therefore accompanied by phase separation unless the particle density is incompressible.  In 2D it is instead generically second order because the free energy is then even in ${\bf Q}$ with no cubic term. (This holds because changing ${\bf Q}\to -{\bf Q}$ can be absorbed into a global rotation of ${\bf Q}$ in two dimensions only~\cite{chaikin2000principles}.)

In dry active systems, close to the onset of the nematic order, there exist a region of parameters where neither the homogeneous isotropic nor the homogeneous nematic state  is stable~\cite{ngo2014large,maryshev2020pattern}. This happens even in two dimensions, where the system undergoes phase separation broadly similarly to that described in Sec.~\ref{subsec:flocking} for the polar case. Here, however, the dense nematic bands can themselves be unstable to bending or to breakup in transverse segments~\cite{chate2019dry,maryshev2020pattern}, in both cases leading to microphase-separated bands displaying a large-scale chaotic dynamics~\cite{maryshev2020pattern,ngo2014large,chate2019dry}. 

Wet active nematics likewise present novel physics close to the transition and in the well-separated (active emulsion) regime~\cite{mueller2021phase,doostmohammadi2018active,CatesJFM:2018}. Uniform contractile and extensile nematics are unstable to splay and bend distortions respectively~\cite{simha2002hydrodynamic}. Due to coupling between splay and density, a circular contractile nematic droplet can become asymmetric and then motile~\cite{tjhung2012spontaneous,giomi2014spontaneous,CatesJFM:2018}. Furthermore, contractile droplets can split by self-stirring~\cite{giomi2014spontaneous} with a dynamics qualitatively similar to that seen in Active Model H (see Sec.~\ref{subsec:AMH}), while extensile droplets can become elongated and then destabilized by bending~\cite{blow2014biphasic}. 

\begin{figure}
\begin{centering}
\includegraphics[width=1.\columnwidth]{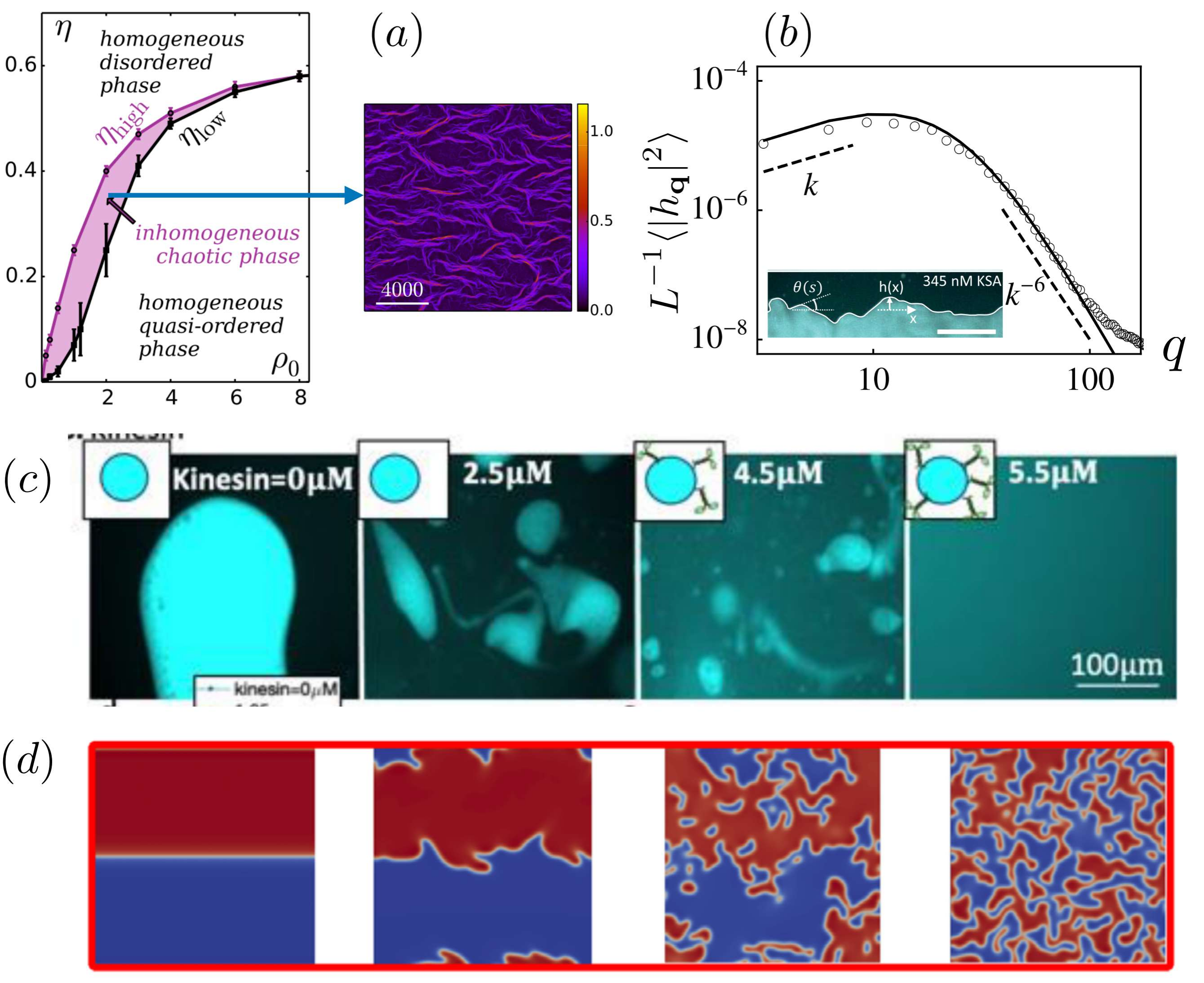}
\par\end{centering}
\caption{  Panel (a): Two-dimensional dry active nematics present a region of the phase diagram (here in terms of alignment noise $\eta$ and global density $\rho_0$) in which uniform disordered and nematic states are both unstable, causing microphase separated in bands which, being themselves unstable, undergo a chaotic dynamics. Figure adapted with permission from~\cite{ngo2014large}. Panel (b): Capillary wave spectrum in experiments on wet active nematics. Figure adapted with permission from~\cite{adkins2022dynamics}.
Panel (c): Suppression of phase separation in experiments onextensile active nematics; increasing Kinesin concentration increases activity (left to right).  Figure adapted with permission from~\cite{tayar2022controlling}. 
Panel (d): Numerical simulation of unstable capillary waves at a bulk nematic-isotropic interface leading to a microphase separated state. Figure adapted with permission from~\cite{caballero2022activity}. }\label{fig:nematics}
\end{figure}

Demixing dynamics in an incompressible binary mixture of an isotropic passive fluid and an active nematic one has recently been studied experimentally~\cite{adkins2022dynamics,tayar2022controlling,zhao2024asymmetric} and theoretically~\cite{tayar2022controlling,giomi2014spontaneous,caballero2022activity,blow2014biphasic,maryshev2020pattern,coelho2022dispersion}. In these systems the active nematic phase is generally turbulent, and thus isotropic at sufficiently long length-scales. At these scales it should arguably be addressable using the scalar models of Sec.~\ref{sec:AMB+}, but the required scale separation is not reached in the experiments so far. Thus the appropriate theory has an active ${\bf Q}$ coupled to a Model H type description of the phase separation. 

The theory then comprises the following elements~\cite{giomi2014spontaneous,blow2014biphasic,maryshev2020pattern,tayar2022controlling,caballero2022activity,coelho2022dispersion}. (i) The equations of Active Model H, (\ref{eq:orderPara}, \ref{eq:Navier-Stokes}, \ref{eq:NSE-stress}, \ref{eq:AMH_stress}) in which the diffusive current {\bf J} obeys (\ref{eq:AMB+J}, \ref{eq:AMB+L}), with $\zeta = \lambda = 0$ so that the only source of activity is the nematic one, and where an additional chemical potential contribution arises from the equilibrium coupling of $\phi$ to the nematic order via a nematic free energy ${\mathcal{F}}_{\bf Q}[{\bf Q},\phi]$; (ii) An additional nematic contribution to the passive stress ${\bf \Sigma}^{\rm eq}$ in \eqref{eq:AMH_stress} that derives from the same nematic free energy ${\mathcal{F}}_{\bf Q}$. (iii) An active stress
in \eqref{eq:AMH_stress} given by ${\bf \Sigma}^{a} = \varsigma\tilde\phi{\bf Q}$ where a function $\tilde{\phi}(\phi)$ optionally encodes the fact that only the dense phase is nematically ordered~\cite{tayar2022controlling,caballero2022activity}. (iv) An advective-relaxational equation of motion for the nematic order parameter that, like $\mathcal{F}_{\bf Q}$, takes a standard form inherited from the passive limit, as reviewed {\em e.g.}, in~\cite{CatesJFM:2018}.

\if{
Complementing AMH with stresses due to the nematic order amounts at replacing (\ref{eq:AMH-stresses}) with~\cite{tayar2022controlling,caballero2022activity}
 \qq\label{eq:AMQ-stresses}
\boldsymbol{\nabla}\cdot\boldsymbol{\sigma}=-\boldsymbol{f}\qquad 
\boldsymbol{f}=\boldsymbol{\nabla}\cdot(\boldsymbol{\Sigma}^{A}+\boldsymbol{\Sigma}^{E}+\boldsymbol{\Sigma}^Q)
\qqq
where $\Sigma^Q = \alpha \tilde{\phi} Q$ is the familiar active nematic stress~\cite{Marchetti2013RMP} (extensile for $\alpha<0$ and contractile for $\alpha>0$), and the coupling with $\tilde{\phi} = (1+\phi/\phi_2)/2$ ensures that the nematic order is non-vanishing only in the dense phase while the dilute phase remains isotropic. The nematic order follows the standard active nematic evolution
\qq\label{eq:dt-Q}
\p_t Q +\bfv\cdot \nabla Q  = \lambda D + Q\cdot \omega -\omega\cdot Q -\frac{1}{\gamma}\frac{\delta \mathcal{F}_Q}{\delta Q}
\qqq
with rate of strain $D=(\nabla \bfv + (\nabla\bfv)^T)/2$, vorticity $\omega = (\nabla \bfv - (\nabla\bfv)^T)/2$, and relaxation controlled by the Landau-de Gennes free energy $\mathcal{F}_Q$ in the single Frank constant approximation
\qq\label{eq:nematic-FQ}
\mathcal{F}_Q=\frac{1}{2}\int d\bfr [-\tilde{\phi}  + \textrm{Tr}(Q^2)]\textrm{Tr}(Q^2)+ K_Q (\nabla Q)^2\,,
\qqq
(observe that eq. (\ref{eq:nematic-FQ}) assumes nematic ordering even in the absence of activity). The final model given by eq. (\ref{eq:orderPara}), (\ref{eq:AMB+J}), (\ref{eq:AMQ-stresses}), (\ref{eq:dt-Q}) was studied in the case where $\lambda=\zeta=0$ and $\tilde{\kappa}=0$, thus retaining the strength of active nematic stresses $\alpha$ as the only activity parameter~\cite{tayar2022controlling,caballero2022activity}, for extensile stresses ($\alpha<0$).
}\fi

The resulting theory (see Fig.~\ref{fig:nematics}) has $\varsigma$ as its sole activity parameter; this is positive (negative) for extensile (contractile) nematics. 
It was used to show
that extensile activity acts to suppress phase separation~\cite{caballero2022activity}, which agrees with experimental observations~\cite{tayar2022controlling,caballero2022activity}. Furthermore, in the phase separated region, the system is not bulk phase separated, but rather microphase separated in a dynamical steady state in which droplets are continuously stirred and broken up (Fig.~\ref{fig:nematics}). The dynamics in the steady state is broadly similar to the one described by AMH when $\sigma_M<0$ but $\sigma_{\rm eq}>0$ (see Sec.~\ref{subsec:AMH}). 

The same theoretical framework can be used to address capillary waves at the interface between an active nematic and an isotropic fluid~\cite{caballero2022activity,gulati2024traveling}, a situation also studied experimentally for the extensile case~\cite{adkins2022dynamics}. This analysis links microphase separation  to an instability of these capillary waves, whose theoretical description extends that for AMB+ in Sec.~\ref{subsec:AMB+capillary} to incorporate both the fluid velocity and the nematic order as additional dynamical fields. (This requires an ansatz for ${\bf Q}({\bf r},t)$ that parallels \eqref{eq:phi-varphi-interface} for the density.) It is then found that extensile stresses ($\varsigma<0$) induce unstable capillary waves at any activity level, for modes $q<\sqrt{|\sigma_M |/(M\sigma_{\rm eq}\eta)}$~\cite{caballero2022activity}.
This means that (even for $\sigma_M<0$), in small enough systems capillary waves remain stable.  However in the presence of active rather than thermal fluctuations, they are predicted to get significantly amplified, as found experimentally~\cite{adkins2022dynamics}. In the stable regime at low activity, the static capillary fluctuation spectrum $S(q)=\langle|h_\bfq|^2\rangle$ was also measured; at wavelengths near the nematic correlation length, $S(q)\sim q$. This strongly deviates from the passive case where (with gravity) $S(q)\sim q^0$~\cite{adkins2022dynamics}. 

Several more complex phenomenologies are also possible for phase-separated active nematic systems. For example, foam-like states and aster-like structures were both experimentally observed~\cite{lemma2022active} and theoretically described~\cite{maryshev2019dry,gulati2024active}, but their relation to phase separation and to the phases described in Sec. \ref{sec:AMB+} is yet to be investigated. Further scenarios could arise from having activity also in the isotropic phase; from different types of anchoring condition at the isotropic-nematic interface; or from the effect of friction with a supporting substrate (leading back to the dry limit). This combinatoric parameter space is yet to be explored systematically in either theory or experiment.

\section{Concluding remarks}\label{sec:Conclusions}
Phase separation in the absence of detailed balance at microscopic level
is widespread across scales and physical systems. The understanding of
its properties paves the way for applications as diverse as creating new
routes to self-assembly; engineering materials with novel properties at
rheological or mechanical level; and perhaps controlling how living
systems responds to varying external conditions -- a crucial question
for systems as diverse as tumors and vegetation patterns.

Perhaps even more important, from the viewpoint of the Review, is that
active phase separation offers a testing ground for broader efforts to
extend the familiar toolbox of equilibrium statistical mechanics to
systems that are far from equilibrium -- not just at large scales with respect to those of 
external forcing (say), but in their fundamental microscopic dynamics.
The principle of detailed balance, or equivalently that of microscopic
reversibility, is so deeply embedded in the foundations of equilibrium
statistical mechanics that all too frequently it gets forgotten about,
especially in elementary courses on the subject.

Take for instance the fact that in equilibrium the pressure $P$ of a
fluid on a confining wall does not depend on the fluid-wall interaction
(because pressure is a state function) whereas in active matter it can
do so~\cite{solon2015pressure}. The equilibrium result follows from $P =
-\langle \partial H/\partial V\rangle = -\partial F/\partial V$, where
$H(x,p)$ is the Hamiltonian, $x$ a list of generalized coordinates
(including $V$) and $p$ their canonical momenta; $\langle\cdot\rangle$
is an average over the Boltzmann probability distribution $e^{-\beta
H}/Z$; and $\beta F = -\ln Z$. The pressure $P$, defined as the ensemble
average of a force, is thus equal to a thermodynamic derivative {\em
only} because the microscopic dynamics has a Hamiltonian structure, which is
reversible. (A separate argument, based on momentum conservation, does
not generalize to other observables and is anyway not applicable to dry
systems.) In active matter, there is no
Hamiltonian, so the stationary probability measure cannot be written
down by inspection of $H$ but must be constructed from the dynamical
equations. This task is generically intractable; but more importantly,
even knowing the measure {\em a priori}, one cannot calculate $P$ (say)
without also knowing those equations including the particle-wall
interactions.

Likewise, as we have outlined in this Review, statistical mechanics has
in recent years undergone extensive (but not yet complete) rebuilding to
address active phase separation. While we have emphasized the emergence
of new phenomenology such as multiple interfacial tensions, in one
important sense it is remarkable that the concept of phase separation
survives at all without detailed balance. Specifically, in equilibrium
it goes without saying that when a system of global density $\phi_0$ phase
separates into two coexisting states of densities $\phi_{1,2}$ (with
$\phi_1<\phi_0<\phi_2$) these binodal densities {\em do not} depend on
$\phi_0$. This precept is violated in other types of density modulation,
such as Turing patterns, as discussed in the Introduction. For active
matter, the very existence of phase separation, in this sense of
well-defined binodals, is not obvious and requires explicit
confirmation. The theoretical basis for that confirmation was provided
above in Secs.~\ref{subsec:AMB+binodals}, \ref{subsec:pseudo-variables}.

More generally, thanks to a concerted effort between theoretical
investigations, numerical simulations, observations of natural systems,
and laboratory experiments, it is now clear that active phase separation
can be both observed and, to some extent, controlled. As we have
emphasized, it generically presents novel and more complex phenomenology
compared to the passive case. Among other phenomena, broken detailed
balance naturally leads to microphase separation in the presence of only
local interactions; circulating currents and macroscopically detectable
entropy production in the steady state; travelling patterns; and
competition with other dynamical instabilities (including the Turing
instability). While there is observational support for several of these
new phenomenologies,  the main focus of this Review has been on
theoretical work that can account for such intriguing physical outcomes.
Although the variety and subtlety of those outcomes can easily defeat an
intuition trained mainly on equilibrium concepts, significant progress
can be made by extending such concepts rather than abandoning them
entirely. As we have reported, much can be understood by generalizing
the concept of interfacial tension into a setting-specific quantity. In
abandoning the equilibrium concept of a uniquely defined tension,
na\"ive intuition may fail, but informed reasoning does succeed, in
certain cases at least.

One possible reason for the power of such quasi-equilibrium reasoning,
at least at the level of continuum theories, is that these are heavily
constrained by symmetries and conservation laws, especially at low
orders in any Landau-Ginzburg type expansion. In equilibrium, such
expansions are often justified by appeal to universality close to the
liquid-vapour critical point (where higher order terms are irrelevant in
a renormalization-group sense~\cite{chaikin2000principles}). But they
also account for generic behaviour elsewhere on the phase diagram, such
as late stage phase ordering which is governed by its own
low-temperature fixed point~\cite{Bray}. Moreover such low-order
theories deeply inform much of soft-matter physics, especially in
systems with liquid-crystalline order~\cite{Marchetti2013RMP}.
Accordingly, notwithstanding the diversity of physical systems
displaying active phase separation, it is natural to hope that generic
field theories such as those described in Sec.~\ref{sec:AMB+} of this
Review provide a unified framework for predicting and understanding its
phenomenology.

A major drawback of such a top-down continuum approach is that the
predicted phenomena remain unconnected to specific mechanisms at
microscopic level. To make those connections, top-down work must be
complemented by careful study of the same phenomena bottom-up, starting
from particle-level descriptions or experiments. While much has been done along these
lines, as reviewed in Sec.~\ref{sec:particle-models}, the actual
building of connections between top-down and bottom-up levels remains
very much in its infancy.

In general terms, however,  the search for a bespoke mechanism in a
given system -- say to explain a microphase separation pattern --  could
be misguided if the phenomena being explained are actually widespread
and general. (This holds also in equilibrium, where one should not try
to explain bulk phase separation, as such, by appeal to specific
features of the pair potential, although these do control the critical
density.) Also, a putative microscopic explanation may be virtually
untestable if the generic scenario admits a host of alternative causes.
On the other hand, if a generic model predicts some outcome only under a
subset of conditions ({\em e.g.} when an active coupling has a specific
sign and/or exceeds a certain threshold), it certainly makes sense to
try to explain microscopically why those conditions are or are not met
in particular cases. Also, if one observes behaviour that is not
predicted at all by a seemingly suitable generic theory, it is
appropriate to ask why, in microscopic terms, the system does not match
the precepts of that theory.

A closely related question that arises in many settings, is as follows.
If the phenomenology observed in an active system is among those also
seen in equilibrium, is the activity crucial in causing the observed
effects or is it just a detail? At continuum level, a distinction exists
between activity causing changes in quasi-passive parameters of the
theory, such as those defining $\mathcal{F}[\phi]$ in \eqref{eq:F}, and
giving rise to new, active-only parameters such as $\lambda,\zeta$ in
(\ref{eq:AMB+J},\ref{eq:AMB+L}). In the case of motility-induced phase
separation, both outcomes occur. But the situation is less clear in many
more complex cases --  such as biomolecular condensates in cells,
ecosystems, and the formation of patterns in social dynamics. While in
such cases one should expect the active-only parameters to be present
unless proved otherwise, the top-down theory is silent concerning their
magnitude and hence their importance. To address this we need to know
more, in mechanistic terms, about what separates activity-specific from
quasi-passive phenomena in any given setting.

Overall then, much interesting work remains to be done in the field of
active phase separation. In Sections
\ref{sec:AMB+}--\ref{sec:two-order-parameters} we identified at least a
dozen open questions and problems ranging from narrowly technical to
broadly conceptual. Each was considered in context, so we do not repeat 
them here.  In general terms, though, future work needs both to deepen
our understanding of the phenomenology predicted from continuum models,
and to better characterize the connections between these models and
their microscopic (particle-level) counterparts. Crucially we also need
much better connections with experimental observations.

One promising avenue is offered by coarse-graining techniques such as
kinetic theory and density functional theory which can directly connect
microscopic and continuum descriptions. This has long been exploited for
atomic and molecular systems in equilibrium, but by comparison with
those cases, the microscopic models used so far in active matter are
often extremely stylized rather than realistic. Examples include
Run-and-Tumble Particles or Active Brownian Particles, in vacuo with
pairwise interactions, when used respectively to model bacteria or
autophoretic colloids suspended in a solvent.
It may be illusory to consider such simple microscopic models as more
fundamental descriptions than continuum theories.

Arguably the hallmark of a realistic microscopic model is that it can be
{\em progressively refined} by adding details without removing what went
before, rather than having to be {\em successively replaced} by more
complex models that each erase the previous attempt. Progressive
refinability is not yet an established principle  for either of the
cases just mentioned. This is partly because there is no strategy in
place to measure quantitatively the parameters that would enable
successive refinements, and partly because the resulting predictions
might even then become too system-specific to be mechanistically
convincing.  This all means that, for active phase separation, the
relation between microscopic and macroscopic modelling may always remain
less sharply delineated than the corresponding relation for phase
separation in equilibrium.

As often attributed to John Maynard Keynes, `it is better to be roughly right than
precisely wrong'~\footnote{The quote was not attributed to J.M. Keynes until after his death. The original quote comes from Carveth Read~\cite{CarvethRead} and is `It is better to be vaguely right than exactly wrong'.}.  A `roughly right' approach, avoiding direct coarse
graining of models that are themselves questionable, could entail a
search for robust observables whose measurements could fix or at least
constrain the unspecified active and passive couplings that arise within
continuum models. This applies whether the latter are formulated at the
level of fluctuating density fields (as in AMB+) or in even more
macroscopic terms (such as interfacial coordinates). This route of
course has precedents for equilibrium models: the surface tension of a
fluid can be measured directly with much higher precision and simplicity
than inferring it from attempted measurements of the particle-particle
interactions and then applying density functional theory; and the
measured tension can then be used, for example, in an Ostwald ripening
calculation. In active matter the situation is more complicated because
there are more parameters involved and not all of their influences are
yet understood. However, the same principles apply. A similar comparison
can be made in relation to recent progress using machine-learning
techniques to fit model parameters at continuum (or indeed microscopic)
level; this promising avenue has begun to be applied to several active
matter problems albeit not so far those addressed in this Review~\cite{cichos2020machine,joshi2022data}.

{\em Acknowledgements}. We thank all the colleagues and collaborators, too numerous to mention individually, with whom we have worked on or discussed these topics in the past two decades. The authors would like to thank the Isaac Newton Institute for Mathematical Sciences, Cambridge, for support and hospitality during the programmes (i) New statistical physics in living matter: non equilibrium states under adaptive control; and (ii) Anti-diffusive dynamics: from sub-cellular to astrophysical scales, where work on this paper was undertaken. This work was supported by EPSRC grant EP/R014604/1. CN  acknowledges the support of the ANR grant ``PSAM'', the support of the INP-IRP grant ``IFAM'' and from the Simons Foundation.


%
%
%
%
%

\bibliographystyle{iopart-num.bst}
\bibliography{./biblio.bib}
\end{document}